\documentclass[journal]{new-aiaa}

\usepackage[english]{babel}

\usepackage{textcomp}
\usepackage[super]{nth}
\usepackage{float}
\usepackage{subfloat}
\usepackage{subcaption}
\usepackage{multirow}
\usepackage{makecell}
\usepackage{longtable,tabularx}
\usepackage{lipsum}
\usepackage{pgfplots}
\usepackage{tikz}
\usetikzlibrary{positioning,external,patterns}
\usepackage{algorithm}
\usepackage{algpseudocode}
\usepackage{amsmath}
\usepackage{amstext}
\usepackage{amsfonts}
\usepackage{bm}
\usepackage{mathtools}
\usepackage[bb=boondox]{mathalfa}

\usepackage[nopostdot,acronym,nogroupskip,nonumberlist]{glossaries}
\usepackage[capitalize]{cleveref}
\usepackage{apptools}
\usepackage{graphicx,import}
\usepackage{lastpage}
\usepackage{fancyhdr}
\usepackage{ifthen}
\usepackage[]{siunitx}
\usepackage{algorithm}
\usepackage{algpseudocode}
\usepackage{xcolor}
\usepackage{bm}

\let\oldhat\hat
\renewcommand{\vec}[1]{\boldsymbol{#1}}
\renewcommand{\hat}[1]{\oldhat{\boldsymbol{#1}}}

\newcommand*{\doublerule}{\hrule width \hsize height 1pt \kern 0.5mm \hrule width \hsize height 2pt}
\newcommand\doublerulefill{\leavevmode\leaders\vbox{\hrule width .1pt\kern1pt\hrule}\hfill\kern0pt }
\usetikzlibrary{plotmarks}
\pgfplotsset{compat=newest,
every plot/.append style={color=black, mark=none},
every axis/.append style={
label style={font=\fontsize{9pt}{1em}\color{white!15!black}\selectfont},
tick style={font=\fontsize{9pt}{1em}\selectfont, color=black, line cap=round},
ticklabel style= {font=\fontsize{9pt}{1em}\selectfont},
legend style={legend cell align=left, font=\fontsize{9pt}{1em}\selectfont, align=left, draw=white!15!black},
/pgf/number format/1000 sep={},
yticklabel style={
        /pgf/number format/fixed,
        /pgf/number format/precision=5
},
scaled y ticks=false
}}
\newcommand{\smd}{d_m^2}
\newcommand{\ipc}{P_{IC}}
\newcommand{\pc}{P_C}
\newcommand{\ipci}{P_{IC,i}}
\newcommand{\smdlim}{\overline{d}_m^2}
\newcommand{\ipclim}{\overline{P}_{IC}}
\newcommand{\pclim}{\overline{P}_C}
\newcommand{\ipcmax}{P_{IC,m}}
\newcommand{\ipcmaxlim}{\overline{P}_{IC,m}}
\newcommand{\dmiss}{d_{sep}}
\renewcommand{\transp}{^\mathrm{T}}
\newcommand{\dv}{\Delta v}
\newcommand\blfootnote[1]{%
  \begingroup
  \renewcommand\thefootnote{}\footnote{#1}%
  \addtocounter{footnote}{-1}%
  \endgroup
}

\newacronym{leo}{LEO}{low Earth orbit}
\newacronym{geo}{GEO}{geostationary Earth orbit}
\newacronym{smd}{SMD}{squared Mahalanobis distance}
\newacronym{ssa}{SSA}{Space Situational Awareness}
\newacronym{adr}{ADR}{Active Debris Removal}
\newacronym{sk}{SK}{station keeping}
\newacronym{ca}{CA}{collision avoidance}
\newacronym{ipc}{IPoC}{instantaneous probability of collision}
\newacronym{cpc}{CPoC}{cumulative probability of collision}
\newacronym{pc}{PoC}{probability of collision}
\newacronym{cam}{CAM}{collision avoidance maneuver}
\newacronym{scvx}{SCvx}{successive convexification}
\newacronym{erf}{erf}{error function}
\newacronym{da}{DA}{differential algebra}
\newacronym{eci}{ECI}{Earth Centered Inertial}
\newacronym{ecef}{ECEF}{Earth Centered Earth Fixed}
\newacronym{lvlh}{LVLH}{Local Vertical Local Horizon}
\newacronym{hbr}{HBR}{hard body radius}
\newacronym{pdf}{PDF}{probability density function}
\newacronym{nlp}{NLP}{non-linear program}
\newacronym{srp}{SRP}{solar radiation pressure}
\newacronym{mpbvp}{MPBVP}{multi-point boundary value problem}
\newacronym{cw}{CW}{Clohessy-Wiltshire}
\newacronym{ns}{N-S}{North-South}
\newacronym{sdp}{SDP}{Semidefinite program}
\newacronym{heo}{HEO}{highly elliptic orbit}
\newacronym{qcqp}{QCQP}{Quadratically constrained quadratic program}
\newacronym{qp}{QP}{quadratic program}
\newacronym{zoh}{ZOH}{zeroth-order hold}
\newacronym{ocp}{OCP}{optimal control problem}
\newacronym{ew}{E-W}{East-West}
\newacronym{stm}{STM}{state transition matrix}
\newacronym{rtn}{RTN}{radial, along-track, cross-track}
\newacronym{iss}{ISS}{International Space Station}
\newacronym{lp}{LP}{linear program}
\newacronym{gnc}{GNC}{guidance, navigation, and control}
\newacronym{socp}{SOCP}{second-order cone program}
\newacronym{cp}{CP}{convex program}
\newacronym{mrv}{MRV}{multivariate random variable}
\newacronym{gmm}{GMM}{Gaussian mixture model}
\newacronym{rv}{RV}{random variable}
\newacronym{scp}{SCP}{sequential convex program}
\newacronym{rso}{RSO}{resident space object}
\newacronym{mc}{MC}{Monte Carlo}
\newacronym{mss}{MSS}{Mahalanobis Shell Sampling}
\newacronym{nli}{NLI}{nonlinearity index}
\newacronym{rk}{RK}{Runge-Kutta}
\newacronym{trr}{TRR}{trust region radius}
\newacronym{tca}{TCA}{time of closest approach}
\newacronym{gto}{GTO}{Geostationary Transfer Orbit}
\newacronym{aida}{AIDA}{Accurate Integrator for Debris Analysis}
\newacronym{koz}{KOZ}{keep-out-zone}

\title{Long-Term Fuel-Optimal Collision Avoidance Maneuvers with Station-Keeping Constraints}

\begin{document}

\author{Zeno Pavanello\footnote{PhD candidate, Te P\=unaha \=Atea - Space Institute, 20 Symonds Street, Auckland Central, Auckland 1010, New Zealand; zpav176@aucklanduni.ac.nz (Corresponding Author).}, Laura Pirovano\footnote{Research Fellow, Te P\=unaha \=Atea - Space Institute, 20 Symonds Street, Auckland Central, Auckland 1010, New Zealand; laura.pirovano@auckland.ac.nz.} and Roberto Armellin\footnote{Professor, Te P\=unaha \=Atea - Space Institute, 20 Symonds Street, Auckland Central, Auckland 1010, New Zealand; roberto.armellin@auckland.ac.nz. Member AIAA.}}
\affil{The University of Auckland, Auckland, 1010, New Zealand}

\maketitle

\begin{abstract}
This work presents a \acrlong{scp}  method to compute fuel-optimal \acrlong{cam}s for long-term encounters. 
 The low-thrust acceleration model is used to account for the control, but the method can compute high-thrust maneuvers by increasing the maximum available acceleration.  
Dealing with the long-term conjunction poses additional challenges compared to the short-term problem because the encounter is not instantaneous. Thus,  under the assumption of Gaussian statistics, the \acrlong{pc} is replaced by a simpler metric, the \acrfull{ipc} and a keep-out zone constraint  is formulated as a continuous condition to be respected throughout the time frame of interest. The robustness of the solution is improved by introducing a constraint on the sensitivity of the \acrshort{ipc}.
Furthermore, the \acrlong{ca} problem is coupled with the classical station-keeping requirement for \acrlong{geo} satellites and with a return to the nominal orbit condition for \acrlong{leo} satellites.
Even though no guarantee is given for the recovery of the global optimum solution, numerical simulations in different orbital regimes show that the proposed approach can yield a local fuel-optimal solution with a run-time suitable for autonomous applications\blfootnote{A part of this work was presented at the 33rd AAS/AIAA Space Flight Mechanics Meeting, held in Austin, Texas, January 15-19 2023. Paper 23-104 was entitled "Long-Term Encounters Collision Avoidance Maneuver Optimization with a Convex Formulation".}.

\end{abstract}

\section{Introduction}
The space population is rapidly increasing. As of the end of 2022, more than 32,000 objects larger than $10$ \si{cm} are being tracked in space, and the active space population has doubled in the last four years \cite{ESA2023}. The increase in the number of operational spacecraft is mainly due to miniaturization (e.g., CubeSats) and the launch of mega-constellations (e.g., Starlink). In parallel, the number of tracked space debris is growing thanks to improved monitoring systems (e.g., the space fence system). This situation brings new challenges in space situational awareness and spacecraft operations. 

One of these is the risk of collision between space objects. The probability of a conjunction event occurring increases with the increasing density of the orbital regime \cite{Letizia2015}, as indicated by research conducted in both \gls{leo} \cite{Zhang2022} and \gls{geo} \cite{Zhang2022Analysis}.
In recent times, there has been a significant increase in research  focusing on \gls{ca} .
In particular, the development of autonomous and fuel-optimized \glspl{cam} aims to effectively decrease operational expenses, minimize reliance on human intervention, and extend the lifespan of missions by mitigating propellant wastage.
Conjunction events are typically considered too risky when the \acrfull{pc} is greater than an arbitrary threshold. NASA identifies two such thresholds, corresponding respectively to the yellow and the red warning: the former is $10^{-5}$, and the latter is $10^{-4}$ \cite{Browns2010}. When these limits are exceeded, a \gls{cam} is designed to lower the value of \gls{pc} while respecting operational constraints, if required.

Conjunctions can be divided into two major types: short-term and long-term. The former is the most common and has been investigated more thoroughly. In this circumstance, the relative velocity involved is high, and the collision is almost instantaneous: the conjunction dynamics can be approximated as linear without loss of accuracy, and the event is studied on the two-dimensional B-plane \cite{Hernando-Ayuso2020, Armellin2021, DeVittori2022}. In long-term conjunctions, the two objects move on similar orbits, and the event cannot be considered instantaneous \cite{NunezGarzon2022}.  Patera \cite{Patera2003Satellite} shows that when the relative trajectory of the primary spacecraft with respect to the secondary is far from a straight line inside the combined covariance ellipsoid, the encounter can be classified as long-term and a different model must be used. Despite many recent attempts at finding ways to efficiently compute \gls{pc} for long-term encounters \cite{Coppola2012,AlfanoSpace,Alfano2014,Xu2011Research,Wen2022}, non-rectilinear relative trajectories inside the uncertainty ellipsoid render the risk quantification and, consequently, the design of long-term \glspl{cam} very complex. In a recent publication \cite{NunezGarzon2022}, Núñez Garzón and Lightsey have suggested that \gls{ipc} is a valuable proxy for \gls{pc}: in general, an increase in \gls{pc} is accompanied by a high value of \gls{ipc}, so controlling the evolution of \gls{ipc} allows to indirectly control the growth of \gls{pc}. For this reason, in agreement with a good portion of the state-of-the-art \cite{chan2008Spacecraft,Jones2013Satellite,Adurthi2015Conjugate,Zhang2020,NunezGarzon2022},  we use \gls{ipc} as a risk metric rather than the more complex \gls{pc}.

\gls{ca} strategies have been developed for long-term encounters in terms of \gls{lp} \cite{Mueller2009} and deterministic disjunctive \gls{lp} \cite{Serra2015}. In reference \cite{Mueller2009}, Mueller simplifies the relative motion using a linear time-varying system and converts the \gls{ca} problem into an \gls{lp}. To this end, the ellipsoidal \gls{koz} constraint is linearized into a hemispace. Moreover, to keep the problem linear, the $l^1$ norm of the control is minimized, which can result in higher propellant consumption if compared with the $l^2$ norm. Notably, Mueller's approach does not take the uncertainty of the states into account; instead, the \gls{koz} region is arbitrarily defined by the author. In this work, we will demonstrate that the evolution of the uncertainty is a key aspect of the definition of the \gls{koz} in the long-term encounter, which was first addressed by Serra \textit{et al.} \cite{Serra2015}. They formulate a deterministic disjunctive \gls{lp}. They linearize the dynamics using the Yamanaka-Ankersen \gls{stm}, which is also used to propagate the covariance in a linear fashion. In this approach, since a relative motion model is used, the combined covariance is propagated from the \gls{tca} to the desired final time. This may lead to inaccuracies because the difference between the orbits of the two spacecraft, which results in different \glspl{stm}, is not considered. Like in our approach, \gls{pc} is substituted by \gls{ipc}, but this variable is not directly controlled. Instead, a polyhedral \gls{koz} constraint is employed, and \gls{ipc} is only checked a posteriori in the validation process. Furthermore, the same $l^1$ norm of the control is used, as in Mueller's approach.

This paper investigates a convex formulation of the problem to mitigate current limitations. Under the hypothesis that the Gaussian states are known at \gls{tca} (from a conjunction data message), we assume that the uncertainties evolve linearly and independently from the control action. With respect to the previously cited approaches, linear propagation of the covariance is carried out separately for the two spacecraft, using the \glspl{stm} of an accurate dynamics model that includes all the relevant orbital perturbations.

The use of convex optimization to solve trajectory design problems is particularly appealing as it has been proven to be efficient for many aerospace applications \cite{Liu2017}, from short-term CAMs \cite{Armellin2021,Dutta2022} to asteroid landing \cite{Pinson2018}, drone formation flying \cite{Alonso-Mora2019}, or spacecraft maneuver estimation \cite{Pirovano2024}. 
These contributions have shown that it is convenient to use a \gls{socp} to formulate the convexified problem because this enables the minimization of the $l^2$ norm of the control history, which is a fundamental requirement in spacecraft trajectory optimization problems. To frame the problem into a \gls{socp}, all of the non-convex parts of the original \gls{ocp} must be convexified \cite{Boyd2}, most notably the fuel-optimal objective function, the nonlinear dynamics constraint, and the \gls{pc} constraint. In the proposed method, additional control magnitude variables and a lossless relaxation are introduced to make the objective function linear and transform the constraint on control magnitude into a second-order cone. The nonlinear dynamics  are automatically linearized using \gls{da}, and the associated \glspl{stm} are used to propagate the covariances of the spacecraft.

The convexification approach for the \gls{pc} constraint proposed in reference \cite{Armellin2021} for short-term encounters is extended to long-term encounters. In reference  \cite{Armellin2021}, the author transcribes the \gls{pc} constraint using the \gls{smd}, effectively turning it into a \gls{koz} condition on the B-plane; this \gls{koz} is convexified by means of a projection and linearization algorithm. We propose an analogous approach where we define a three-dimensional \gls{smd} based on \gls{ipc},  and we iteratively approximate the \gls{koz} constraint into a succession of hemispaces.  Unlike in the short-term encounter case,  the \gls{smd} constraint is  applied to the whole window of interest and not only at the nominal conjunction time. To guarantee a robust solution, a novel constraint is introduced to limit the sensitivity of \gls{ipc} to state uncertainties. As a result, safety is ensured even in the presence of thrust misalignment or \acrlong{gnc} errors. 
The initial stochastic problem is therefore reduced into a deterministic one. This approach bears some similarity to works on robust trajectory optimization (for example, see \cite{Lew2020}, \cite{Ridderhof2021}, and \cite{Benedikter2022Covariance, Benedikter2022Convex}), in which chance constraints are converted into a deterministic form suitable for convex optimization. However, unlike these works, in the \gls{ca} scenario, it is not advantageous to include a feedback component to control the covariance of the primary since most of the uncertainty is generally associated with the uncontrolled secondary. On the other hand, the state of the primary is well-known, and its uncertainty does not have sufficient time to grow significantly.


Post-\gls{cam} \gls{sk} constraints are introduced both for \gls{geo} and for \gls{leo} scenarios. In the first case, recalling the work from Mueller \textit{et al.} \cite{Mueller2013}, a linearised keep-in-box constraint bounds the spacecraft to respect a latitude-longitude requirement; in \gls{leo} a constraint on the final state forces the spacecraft to return to its nominal orbit. 

A trust region constraint based on references \cite{Malyuta2021Tutorial,Losacco2023,Bernardini2023} to limit the solution space and favor the convergence of the algorithm is adopted.
The linearization of the dynamics and \gls{ipc} constraints prevents the optimizer from finding the optimal solution with a single \gls{socp} run. For this reason, a \gls{scp} is built as a succession of \glspl{socp}, where the solution from the previous \gls{socp} is taken as the linearization point for the new problem. This process can take a discretely large number of iterations before converging to the optimal solution of the original \gls{ocp}. This iterative approach allows for updating the evolution of the covariance history, which might lead to changes in the shape of the \gls{koz} that can be accounted for in the next iteration. This aspect has been neglected in the literature but might have relevance for cases in which the \gls{ipc} is highly sensitive to variation in the covariance.

The article is organized into five sections. In \cref{sec:dyn}, the dynamics of the problem are presented; in \cref{sec:socpprob}, the base \gls{cam} optimization problem is posed as a \gls{scp}, and the sensitivity constraint on the collision metric is introduced. The \gls{sk} constraints are introduced in \cref{sec:sk}, leading to the final formulation of the convex problem in \cref{sec:finalize}. The algorithm is then applied to realistic \gls{leo}, \gls{geo}  and \gls{heo}  test cases and conclusions are drawn in \cref{sec:results,sec:conclusion} respectively.

\section{Formulation of the Long-Term Collision Avoidance Problem}
\label{sec:dyn}

Let the states of two spacecraft (primary and secondary) be described at time $t_0$ by two uncorrelated Gaussian multivariate random variables.  The states are represented using any arbitrary set of elements, e.g. Cartesian, Keplerian or generalized equinoctial elements. 
\begin{equation}
\begin{aligned}
      \quad &  \vec{x}_p(t_0)\sim \mathcal{N}(\vec{\zeta}_p(t_0),\vec{C}_p(t_0)) \, \mbox{ and } \,
      \vec{x}_s(t_0)\sim \mathcal{N}(\vec{\zeta}_s(t_0),\vec{C}_s(t_0)),
\end{aligned}
\label{eq:states}
\end{equation}
where $\vec{x}_p(t_0)$, $\vec{x}_s(t_0)\in\mathbb{R}^6$ are the two uncorrelated random variables, $\vec{\zeta}_p(t_0)$, $\vec{\zeta}_s(t_0)\in\mathbb{R}^6$ are their mean values and $\vec{C}_p(t_0)$, $\vec{C}_s(t_0)\in\mathbb{R}^{6\times6}$ are their covariance matrices. The control acceleration acting on the primary satellite is $\vec{u}(t)\in\mathbb{R}^3$. 
The state of the two objects can be numerically propagated using an arbitrary dynamics model, generally described as 
\begin{equation}
\begin{aligned}
\quad & \dot{\vec{x}}_p(t) = \vec{f}_p(t,\vec{x}_p(t),\vec{u}(t),\vec{p}_p)\, \mbox{ and } \,  \dot{\vec{x}}_s(t) = \vec{f}_s(t,\vec{x}_s(t),\vec{p}_s),
\label{eq:diffeq}
\end{aligned}
\end{equation}
where $t\in\mathbb{R}_{[t_0, t_f]}$ is the continuous time domain, $\vec{p}_p\in\mathbb{R}^{m_p}$ and $\vec{p}_s\in\mathbb{R}^{m_s}$ are sets of parameters, $\vec{f}_p(\cdot): \mathbb{R}_{[t_0, t_f]}\times\mathbb{R}^6\times\mathbb{R}^3\times\mathbb{R}^{m_p}\rightarrow\mathbb{R}^6$ and $\vec{f}_s(\cdot): \mathbb{R}_{[t_0, t_f]}\times\mathbb{R}^6\times\mathbb{R}^{m_s}\rightarrow\mathbb{R}^6$ are continuous functions.  
In \cref{sec:results}, the \gls{aida} dynamics model is used, which was originally introduced in reference \cite{Morselli2014High} and implements the following acceleration vector
\begin{equation}
\vec{a}_{AIDA} = \vec{a}_{E} + \vec{a}_{S} + \vec{a}_{M} + \vec{a}_{drag} + \vec{a}_{SRP},
\end{equation}
where
\begin{itemize}
   \item[] $\vec{a}_{E}$ is the Earth's gravitational pull computed with a potential model of order up to 15;
    \item[] $\vec{a}_{S}$ and $\vec{a}_{M}$ are the third body accelerations of the Sun and Moon computed with NASA's Spice toolkit \footnote{\url{https://naif.jpl.nasa.gov/naif/documentation.html}};
    \item[] $\vec{a}_{drag}$ is the drag acceleration computed with the NRLMSISE-00 atmospheric density model and assuming a cannonball geometry for the spacecraft;
    \item[] $\vec{a}_{SRP}$ is the \gls{srp} computed with dual-cone shadow model and assuming a cannonball geometry for the spacecraft. 
\end{itemize}

As proven by Baù \textit{et al.} \cite{Bau2021}, the use of modified equinoctial elements in the propagation can preserve the Gaussian nature of the states for a much longer period compared to classical orbital elements or Cartesian coordinates. Nonetheless, when the propagation window is sufficiently short, and the state uncertainty is small, Cartesian elements can still preserve the normality of the distributions \cite{Vittaldev2016Space}. Since we are only considering propagations over one or two orbital periods, the results in \cref{sec:results} are obtained using Cartesian coordinates. Nonetheless, the method is applicable to any representation of the states.
To define the \gls{pc} constraint,  the relative position of the primary with respect to the secondary must be made explicit. The position of the primary in \gls{eci} is a function of the state (and the same is true for the secondary):

\begin{equation}
\begin{aligned}
   \quad & \vec{r}_p(t) = \vec{h}(t,\vec{x}_p(t)),
   \label{eq:toCart}
\end{aligned}
\end{equation}
where $\vec{h}(\cdot):\mathbb{R}\times\mathbb{R}^6\rightarrow\mathbb{R}^3$ is a transformation that  is assumed to be quasi-linear for the uncertainty at hand. This assumption holds when the Gaussian covariances are sufficiently small. In Cartesian coordinates, this transformation simply extracts the three position coordinates of the state, as is the case in \cref{sec:results}.
It follows that $\vec{r}_p(t)\sim \mathcal{N}(\vec{\mu}_p(t),\vec{P}_p(t))$ and $\vec{r}_s(t)\sim \mathcal{N}(\vec{\mu}_s(t),\vec{P}_s(t))$.
The relative position is then simply the subtraction of the two normally distributed random variables $\vec{r}_{rel}(t) = \vec{r}_p(t)-\vec{r}_s(t)$.
Given that the subtraction is a linear transformation, also the relative position is normally distributed

\begin{subequations}
\begin{align}
      \quad &  \vec{r}_{rel}(t)\sim \mathcal{N}(\vec{\mu}(t),\vec{P}(t)), \label{eq:stateRel}\\
      \quad &  \vec{\mu}(t) = \vec{\mu}_p(t) - \vec{\mu}_s(t), \\
      \quad & \vec{P}(t) = \vec{P}_p(t) + \vec{P}_s(t).
\end{align}
\label{eq:rel}
\end{subequations}


\subsection{Collision Avoidance Optimal Control Problem}
The original \gls{ca} \gls{ocp} in the continuous domain is stated as follows
\begin{subequations}
\begin{align}
\min_{\vec{u}} \quad & J = \int_{t_0}^{t_f} u(t) \mathrm{d}t \label{eq:ocpObj}\\
\mbox{s.t.} 
\quad &  \dot{\vec{x}} = \vec{f}(\vec{x}(t),\vec{u}(t),t)  \label{eq:ocpDyn}\\
\quad & \pc(t) \leq \pclim \label{eq:ocpPc}\\
\quad & \vec{x}(t_0) = \vec{x}_0 \label{eq:ocpInit}\\
\quad & u(t) = \sqrt{u_1(t)^2 + u_2(t)^2 + u_3(t)^2}\label{eq:nonconvexU}\\
\quad & u(t) \leq u_{max} \label{eq:ocpUmax}
\end{align}
\label{eq:ocp}
\end{subequations}
where $\vec{u}(t) = [u_1(t)$ \hspace{2pt} $u_2(t)$ \hspace{2pt} $u_3(t)]\transp$ and the state of the primary is indicated with $\vec{x}$; $P_C(\cdot):\mathbb{R}\rightarrow\mathbb{R}$ is the \gls{pc} function which is equal to $P_C^0$ at the starting time and is monotonically increasing during the conjunction; it must always be kept below the threshold $\pclim$. In Problem \eqref{eq:ocp}, \cref{eq:ocpObj} is the fuel minimization objective function, \cref{eq:ocpDyn} is the dynamics constraint, \cref{eq:ocpPc} is the \gls{pc} constraint, \cref{eq:ocpInit} is the initial state bound, \cref{eq:nonconvexU} is a non-convex equality constraint on the control variable and \cref{eq:ocpUmax} is the bound on the maximum value of the control action. The mass loss due to the maneuver is not considered in the equations of motion because it is deemed negligible \cite{Armellin2021}.

\subsection{Discretization of the Dynamics}
The first step towards the \gls{socp} formulation is the discretization of Problem \eqref{eq:ocp}. The continuous time variable $t\in\mathbb{R}_{[t_0, t_f]}$ is substituted by the discrete time variable $t_i\in\{t_0, t_1, ..., t_N\}$, where $N+1$ is the number of equally spaced nodes of the discretization. Following \cref{eq:diffeq} and via the use of an integration scheme   - e.g., Runge-Kutta 7-8 -  one obtains the states of the two spacecraft at node $i+1$, which depend on the state and the control at node $i$:
\begin{subequations}
\begin{align}
    \quad & \vec{x}_{p,i+1} = \vec{f}_{p,i}(t_i,\vec{x}_{p,i},\vec{u}_i,\vec{p}_p) \quad & i\in\{0,\hspace{2pt} ...,\hspace{2pt} N-1\}, \label{eq:dynamicsa} \\
    \quad & \vec{x}_{s,i+1} = \vec{f}_{s,i}(t_i,\vec{x}_{s,i},\vec{p}_s) \quad & i\in\{0,\hspace{2pt} ...,\hspace{2pt} N-1\},  
\end{align}
\label{eq:dynamics} 
\end{subequations}
where $\vec{f}_{p,i}(\cdot): \mathbb{R}_{[t_0,t_f]}\times\mathbb{R}^6\times\mathbb{R}^3\times\mathbb{R}^{m_p}\rightarrow\mathbb{R}^6$  and $\vec{f}_{s,i}(\cdot): \mathbb{R}_{[t_0,t_f]}\times\mathbb{R}^6\times\mathbb{R}^{m_s}\rightarrow\mathbb{R}^6$  are the functions that describe the dynamics at node $i$, $\vec{x}_{p,i} = \vec{x}_p(t_i)$, $\vec{x}_{s,i} = \vec{x}_s(t_i)$ and $\vec{u}_{i} = \vec{u}(t_i)$.  Between two consecutive nodes $i$ and $i+1$ the acceleration is considered constant and equal to $\vec{u}_{i}$. 
\gls{da} is used to introduce perturbations on the primary state ($\vec{x}_{p,i}+\delta\vec{x}_{p,i}$)
and acceleration ($\vec{u}_i+\delta\vec{u}_i$) at each node, effectively expressing \cref{eq:dynamicsa} through Taylor polynomials:
\begin{equation}
\begin{aligned}
\quad & \vec{x}_{p,i+1} = \mathcal{T}^q_{\vec{x}_{p,i+1}}(\vec{x}_{p,i},\vec{u}_i) \quad & i\in\{0,\hspace{2pt} ...,\hspace{2pt} N-1\},
\end{aligned}
\label{eq:perturbations}
\end{equation}
where in general the expression $\mathcal{T}^q_{y}(x)$ indicates the $q^\text{th}$-order Taylor expansion of the variable $y$ as a function of $x$,  around the expansion point $\tilde{x}$ in which the polynomial is computed. The reader can find a detailed explanation of the use of \gls{da} in \cite{Armellin2010}. 

\subsection{Selection of the Risk Metric}
\label{sec:risk}
Probability-based criteria are the most widely employed indicators to assess the likelihood of a collision \cite{Alfriend2000}. Nonetheless, many operators adopt a separation distance strategy that only relies on the objects' mean state, so no information about the uncertainty is used. In this work, the use of three distinct metrics is analyzed, namely the  \gls{ipc}, which will be indicated with $\ipc$, the maximum \gls{ipc}, indicated with $\ipcmax$, and the separation distance  $\dmiss$.  

\begin{figure}[b!]
\centering
\input{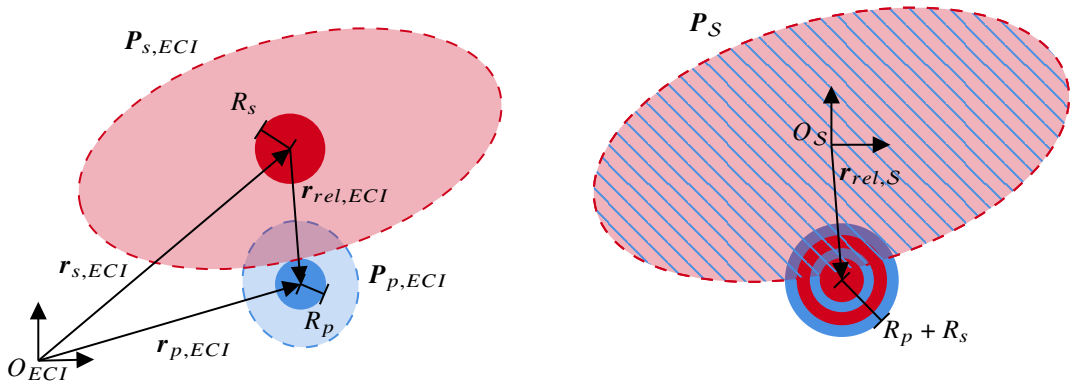}
\caption{Left: KOZ problem in \gls{eci} frame. Right: KOZ problem in $\mathcal{S}$ frame.}
\label{fig:covFrame}
\end{figure}

The most common metric used to formulate the \gls{ca} condition in the short-term \gls{ca} optimization problem is \gls{pc}. Typically an upper limit is set on this variable so that at \gls{tca} $\pc<\pclim$. When considering the long-term problem, the integral formula for the computation of \gls{pc} must consider the uncertainty in the position and velocity of the two space objects \cite{Coppola2012}. Moreover, this metric is highly nonlinear, thus unsuitable for a convex formulation. 
An alternative approach  is  setting a limit on \gls{ipc} rather than on \gls{pc}. \gls{ipc} has no dependence on the velocity uncertainty, making it simpler to compute. Additionally, \gls{ipc} can be further simplified to the \gls{smd}, indicated with $\smd$, transforming the \gls{pc} constraint into an ellipsoidal keep-out zone, as shown in the following.

The relative position of the primary with respect to the secondary is the subtraction of the two multivariate distributions, as in \cref{eq:rel}.  We define the \gls{smd} at time $t$ as: 
\begin{equation}
    \smd(\vec{r}_{rel}(t)) = \vec{\mu}(t)\transp\vec{P}(t)^{-1}\Vec{\mu}(t).
    \label{eq:smd1}
\end{equation}
 With reference to \cref{fig:covFrame}, we now consider a reference frame $\mathcal{S}$  with axes aligned to \gls{eci} and centered in the secondary spacecraft to simplify the calculations.  The two spacecraft are assumed to have a spherical geometry, which is described by the \gls{hbr}, indicated with $R_p$ (for the primary) and $R_s$ (for the secondary).  The combined \gls{hbr} of the two spacecraft is defined as the sum of their individual \glspl{hbr}, $R = R_p + R_s$, and the volume of the two objects is condensed around the primary \cite{Armellin2021}. Also, we assign the whole uncertainty to the secondary spacecraft. So, $\Vec{r}_{rel}(t)$ is shifted by an amount equal to its mean, yielding a multivariate random variable centered  at the position of the secondary, defined as the origin
\begin{equation}
    \Vec{s}(t) = \Vec{r}_{rel}(t)-\Vec{\mu}(t)\sim \mathcal{N}(\vec{0}_3,\vec{P}(t)),
\end{equation} 
Note that in the new reference system, the primary position at time $t$ is deterministic and equal to $\Vec{\mu}(t)$: it will be indicated by the symbol $\vec{r}(t)$ in the following.
The \gls{smd} in \cref{eq:smd1}, then, becomes a measure of the distance of $\vec{r}(t)$ from the normal distribution $\vec{s}(t)$
\begin{equation}
   \smd = \vec{r}\transp\vec{P}^{-1}\vec{r},
    \label{eq:smd}
\end{equation}
where the argument $t$ has been dropped for simplicity.
The value of \gls{ipc} at the time instant $t$ is the integral of the \gls{pdf} over the sphere $\mathbb{S}_\mathrm{HBR}$ centered in $\Vec{r}$ and of radius $R$ \cite{Serra2015}:
\begin{equation}
\ipc=\frac{1}{(2 \pi)^{3 / 2}\mathrm{det}(\vec{P})^{1/2}} \iiint_{\mathbb{S}_\mathrm{HBR}} e^{-\vec{s}\transp\vec{P}^{-1}\vec{s}/2} \mathrm{d} V,
\label{eq:ipceq}
\end{equation}
Analogously to Alfriend and Akella's method for \gls{pc} \cite{Alfriend2000}, \cref{eq:ipceq} can be simplified by neglecting the variation of the \gls{pdf} inside the integration region. Indeed, in typical applications, the ellipsoid associated with the covariance is significantly larger than the hard body sphere \cite{Patera2003Satellite}.
The \gls{pdf}, then, is evaluated only in the central point of $\mathbb{S}_\mathrm{HBR}$ ($\vec{s}=\vec{r}$), obtaining

\begin{equation}
   \ipc = \sqrt{\frac{2}{\pi \mathrm{det}(\vec{P})}}\frac{R^3}{3}\mathrm{e}^{-\smd/2}.
   \label{eq:constipc}
\end{equation}

The covariance of the position might be estimated with a large margin of error; thus we might want a more conservative approach to the estimation of the collision risk. The maximum \acrlong{ipc}  $\ipcmax$  is computed following the same procedure that is found in \cite{Alfriend2000}
\begin{equation}
    \ipcmax = \frac{(\sqrt{2} R)^3 }{3\mathrm{e}^1 \smd \sqrt{\pi \mathrm{det}(\vec{P})}}.
    \label{eq:maxIpc}
\end{equation}

From \cref{eq:constipc} or \cref{eq:maxIpc}, it is possible to approximate a constraint on $\ipc$ or on $\ipcmax$ into a constraint on \gls{smd}, which describes an ellipsoidal keep-out zone.
Given a limit value of \gls{ipc} ($\ipclim$) or  $\ipcmax$ ($\ipcmaxlim$), the corresponding limit of \gls{smd} alternatively becomes 
\begin{subequations}
\begin{align}
    \quad & \smdlim = -2\,\mathrm{ln}\Big(\frac{3\ipclim}{R^3}\sqrt{\frac{\pi \mathrm{det}(\vec{P})}{2}}\Big) \label{eq:dmlim1}, \\
    \quad & \smdlim = \frac{(\sqrt{2} R)^3 }{3\mathrm{e} \ipcmaxlim\sqrt{\pi \mathrm{det}(\vec{P})}}, \label{eq:dmlim2} 
\end{align}
\end{subequations}
and the  \gls{smd}  constraint at any time is stated as 
\begin{equation}
    \smd \geq \smdlim.
    \label{eq:keepout}
\end{equation}

Alternatively to the \gls{smd} constraint, a separation distance  constraint can be employed using the same formalism. Indeed, the separation distance  does not provide information on the uncertainty, so it can be computed as the Mahalanobis distance of the mean state of the relative position with unit covariance $\dmiss=\sqrt{\vec{r}\transp\vec{r}}$. In this case, \cref{eq:keepout} can still be used to impose a keep-out zone condition, where the limit value is given by an arbitrarily defined distance from the secondary.

\begin{figure}[tb!]
    \centering
    \subfloat[Test case 4: \gls{geo}.]{\includegraphics[width = 0.48\textwidth]{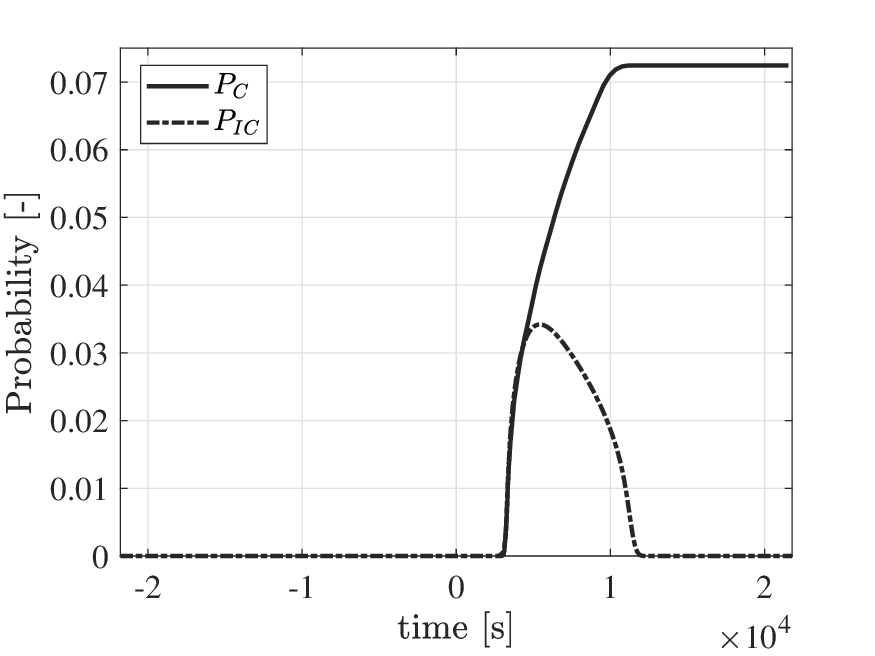}\label{fig:ipcVsPoC4}}
    \subfloat[Test case 9: \gls{heo}.]{\includegraphics[width = 0.48\textwidth]{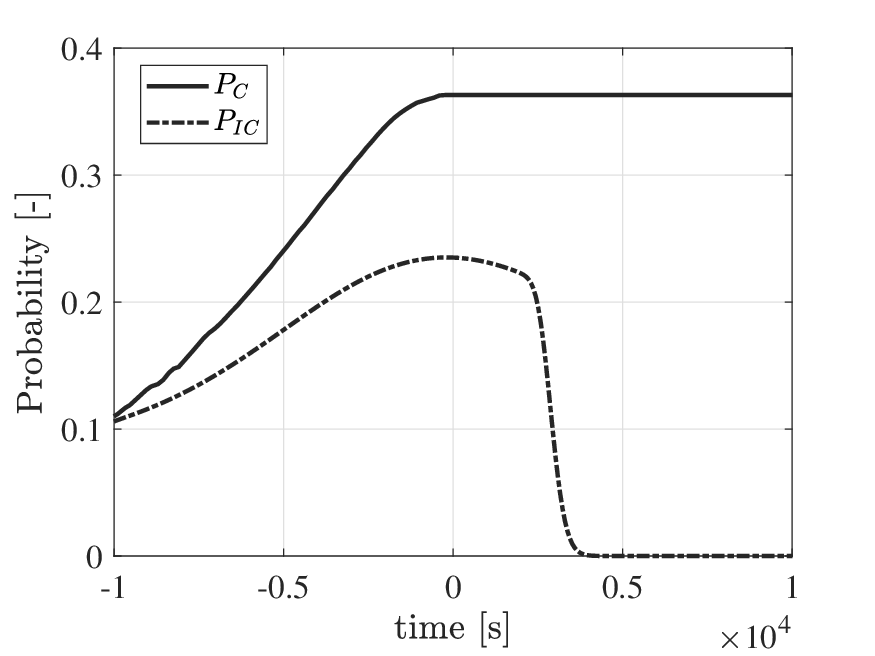}\label{fig:ipcVsPoC9}}  
    \caption{Comparison of \gls{ipc} and \gls{pc} for two test cases from reference \cite{Alfano2009Satellite}.} 
    \label{fig:ipcVsPoC}
\end{figure}

The use of \gls{ipc} as a replacement for \gls{pc} has been proposed in the literature by \cite{Serra2015, NunezGarzon2022}. To confirm that this is a reasonable substitution, in \cref{fig:ipcVsPoC}, two examples of the relationship between \gls{ipc} and \gls{pc} in relevant test cases from reference \cite{Alfano2009Satellite} are shown. Clearly, the evolution of \gls{ipc} follows  the increase in \gls{pc}: when the slope of \gls{pc} is high, \gls{ipc} is non-negligible. For this reason, we can assume that controlling \gls{ipc} can be considered a reasonable alternative to reducing \gls{pc} in the long-term encounter \gls{cam} optimization problem.

\subsubsection{Validity of the IPoC approximation}
\label{sec:validity}
The validity of the \gls{ipc} approximation in \cref{eq:constipc} is determined by the entity of the variation of the \gls{pdf} inside the combined hard body sphere. If the \gls{pdf} changes significantly, the approximation will yield inaccurate results. An important role in this is played by the combined \gls{hbr}: the smaller this value is, the less the \gls{pdf} changes inside it. 
We analyze two cases in which the approximation behaves differently. The first case is taken from reference \cite{NunezGarzon2022}, and the second one from \cite{Alfano2009Satellite}. In the first case, the maximum variation of the \gls{pdf} inside the hard body sphere is around $2\times10^{-11}$ for an unrealistic \gls{hbr} of $320$ \si{m}, $1\times10^{-12}$ for $32$ \si{m}, $1.4\times10^{-13}$ for $3.2$ \si{m}, and $10^{-12}$ for $0.32$ \si{m}. In the second case, instead, the variations are between $0.004$ and $0.04$ for an \gls{hbr} of $0.6$ \si{m} and $12$ \si{m}. 

\begin{figure}[tb!]
    \centering
    \subfloat[Test case from reference \cite{NunezGarzon2022}.]{\includegraphics[width = 0.48\textwidth]{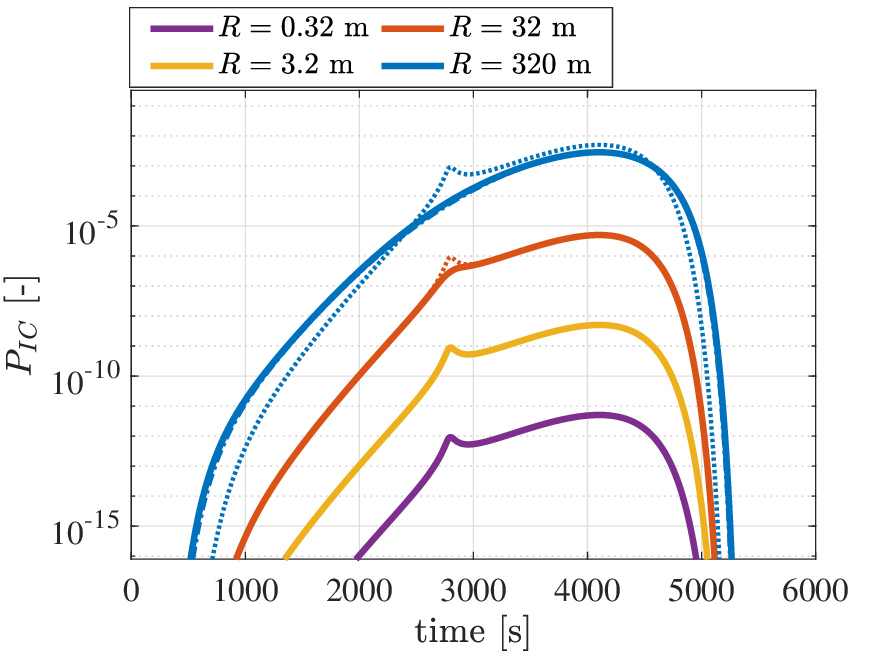}\label{fig:garzonipc}}
    \subfloat[Test case 9 from reference \cite{Alfano2009Satellite}.]{\includegraphics[width = 0.48\textwidth]{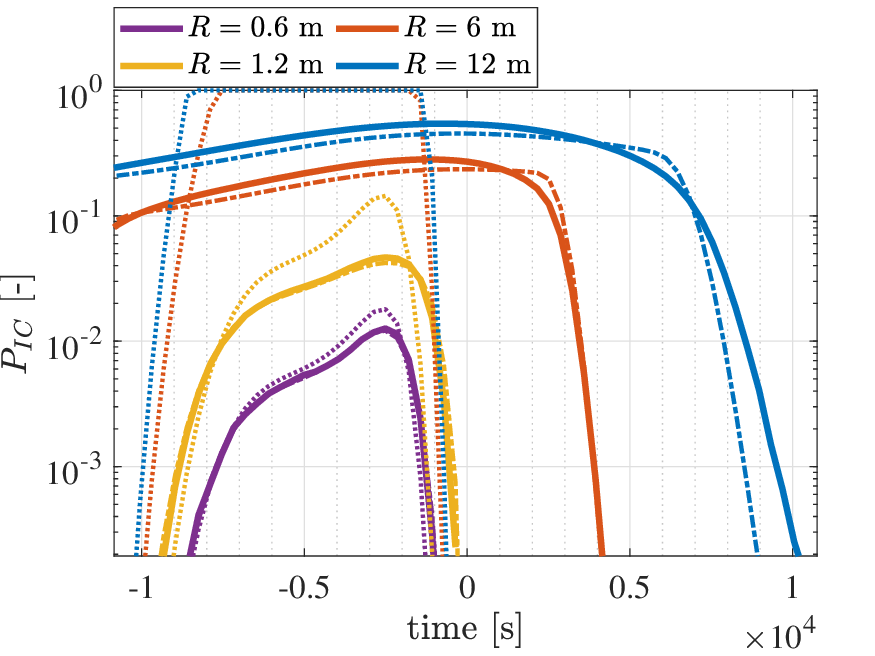}\label{fig:alfanoipc}}  
    \caption{IPoC comparison between Monte Carlo (solid), constant (dotted), and cuboid (dash-dotted) approximations.}
        \label{fig:ipcvalidity}
\end{figure}

In \cref{fig:ipcvalidity}, a comparison is provided between the real (Monte Carlo based) \gls{ipc} and the approximated one for the different values of \gls{hbr} in the two cases. When the models are in accordance with the Monte Carlo result, the corresponding line is not visible. In \cref{fig:garzonipc}, we see that for \gls{hbr} up to $32$ \si{m}, the approximation is very accurate, and the curve starts to separate from the real \gls{ipc} only when \gls{hbr} is exceedingly high.
In the second scenario, represented in \cref{fig:alfanoipc}, the approximation has a maximum relative error of $50$\% in the best case, where the \gls{hbr} is $0.6$ \si{m}, and it goes over $270.4$\% for the nominal case with \gls{hbr} equal to $6$ \si{m}. 
Since in typical applications, the size of the \gls{hbr} is in the order of magnitude of $1-10$ \si{m}, the approximation cannot be considered accurate for demanding scenarios like this one. 

In such situations, we propose to employ a different approximation, which was introduced by Zhang \textit{et al.} \cite{Zhang2020}. 
This method, referred to as "cuboid," transforms the originally spherical integration region into a parallelogram and decouples the integration into three independent one-dimensional integrals. 
As it can be seen from the figure, in the first scenario. this approximation can accurately follow the evolution of the real \gls{ipc}. In the second one, it is better than the constant approximation, reaching maximum errors of $3.9$\%, $9.5$\%, $20.1$\%, $19.6$\%, and $13$\% over the considered time span.

For this reason, we propose an inversion algorithm to obtain a \gls{smd} limit based on the cuboid \gls{ipc} to replace \cref{eq:dmlim1} when the constant \gls{pdf} approximation is not accurate. The algorithm is a simple \gls{nlp}: for a generic node $i$, the \gls{nlp} starts from point $\vec{r} = \vec{r}_i$, used to initialize the solution, to find the optimized point $\hat{r}\in\mathbb{R}^3$
\begin{subequations}
\centering
\begin{align} 
\min_{\hat{r}} \quad & ||\hat{r}-\vec{r}||,														      \label{eq:minDeltar}\\
\text{s.t.}      \quad & \ipc(\vec{r}) = f_{cub}(\vec{P},\hat{r},R) = \ipclim, 						      \label{eq:ipcnlp}   \\
 			     \quad & \vec{r} - \Delta\vec{r}_{max} \leq \hat{r} \leq \vec{r} + \Delta\vec{r}_{max}, \label{eq:limnlp}   
\end{align}
\label{eq:cuboidinversion}
\end{subequations}
where $\Delta\vec{r}\in\mathbb{R}^3$ is a vector used to create the boundaries, and $f_{cub}:\mathbb{R}^3\times\mathbb{R}^{3\times3}\times\mathbb{R}_{[0,1]}:\rightarrow\mathbb{R}$ is the cuboid \gls{ipc} function.
\cref{eq:minDeltar} is the objective function that minimizes the distance between the optimization variable $\hat{r}$ and the relative position $\vec{r}$; \cref{eq:ipcnlp} imposes that the cuboid \gls{ipc} must be equal to the limit value; \cref{eq:limnlp} imposes upper and lower boundaries to the optimization variable. Once a solution has been found, the corresponding \gls{smd} threshold is computed using \cref{eq:smd1}, $\smdlim = \hat{r}\transp\vec{P}^{-1}\hat{r}$. 
This inversion of the \gls{ipc} limit only works in the proximity of the original point, so it performs well when the \gls{ca} maneuver is small, as is the case considered in \cref{fig:alfanoipc}, according to reference \cite{Serra2015}. As there is no guarantee that the cuboid \gls{ipc} is constant over the surface of the ellipsoid defined by $\smdlim$, a check must be included to verify \textit{a posteriori} that the value of \gls{ipc} computed with the cuboid approximation fulfils the required constraint. In the opposite case, a new iteration of the \gls{scp} is carried on until the constraint is met.

\section{Successive Convex Program Formulation}
\label{sec:socpprob}

Maneuvers can be performed at $N$ discretization nodes in the form of a constant acceleration within the segment, $\vec{u}_i$. The last node is an exception since the acceleration would not affect the solution. 

The optimization variables are the states and controls at each node. Here the covariance matrix at each node is independent of the control because there is no feedback. 
Hence, the scalar optimization variables are 9 for each node: the  6  components of the state and the 3 components of the control. In the following, for clarity in the formulas, the state of the primary spacecraft,  that was defined in \cref{eq:states} as $\vec{x}_p$, will be indicated with $\vec{x}$. 
 The full history of state and controls are expressed with the symbols $\mathbb{x}\in\mathbb{R}^{6(N+1)}$ and $\mathbb{u}\in\mathbb{R}^{3N}$

\begin{equation}
    \mathbb{x} = \begin{bmatrix}\vec{x}_0\transp & \vec{x}_1\transp & ...  & \vec{x}_N\transp \end{bmatrix}\transp \, \mbox{ and } \, \mathbb{u} = \begin{bmatrix}\vec{u}_0\transp & \vec{u}_1\transp & ... & \vec{u}_{N-1}\transp\end{bmatrix}\transp.
\end{equation}

Three nonlinearity sources are present in the problem: (i) the dynamics in \cref{eq:perturbations} can include polynomials up to any order; (ii) the constraint in \cref{eq:keepout} is a \gls{koz} constraint, which is a non-convex one; (iii) the minimization of the fuel expenditure requires the use of a non-convex constraint in \cref{eq:nonconvexU}, which must be relaxed into a convex one to build the \gls{socp}. 
Three main steps are then required to formulate the \gls{cam} design as a \gls{socp} \cite{Armellin2021}. 
Firstly, in \cref{sec:linearization} the dynamics are automatically linearized using \gls{da}, and an iterative \gls{scp} method is employed. 
Afterward, in \cref{sec:eqTransf} the objective function and the constraint on the control acceleration magnitude are reformulated by introducing an equivalent transformation.
Lastly, in \cref{sec:projLin} a projection and linearization approach is used to linearize the \gls{koz} constraint. It would be more complicated to handle a constraint on \gls{ipc} since the expression is not a polynomial function of the position, and it would need a more coarse linearization than the one on \gls{smd}. 

\subsection{Linearization of the Dynamics}
\label{sec:linearization}
In general, orbital dynamics are highly non-linear and cannot be directly modeled into a convex optimization framework.  
To address this  issue , the nonlinearities are managed through successive linearizations of the dynamics equations. The original \gls{ocp} is linearized, locally,  into a convex sub-problem in the framework of a \gls{scp}. This process may require numerous iterations, referred to as \textit{major iterations} denoted by index  $j$.
Once the optimization problem of a major iteration $j-1$ is solved, the solution ($\mathbb{x}^{j-1}$, $\mathbb{u}^{j-1}$) becomes available, a column vector comprising the state and control at each node.
Employing \gls{da},  this solution serves as an expansion point for constructing linear dynamics maps for iteration $j$.
The continuity condition is enforced by requiring that the state after the propagation of node $i$ is equal to the state before the propagation of node $i+1$. 
To favor the convergence of the iterations, it is advisable to normalize the control over its maximum value so that $\vec{u}_i \rightarrow \vec{u}_i/u_{max}$ and $0\leq||\vec{u}_i||\leq 1$. In the following, the dynamics equations need to account for the normalized control, i.e., when computing the linear maps, the control must be scaled back into its original dimensions.
The state and acceleration of each node are expanded around the reference points, which are the output of the previous major iteration $\tilde{\Vec{x}}_i^j=\vec{x}_i^{j-1}$ and $\tilde{\Vec{u}}_i^j=\vec{u}_i^{j-1}$ 
\begin{subequations}
\begin{align}
   \quad & \vec{x}_{i}^j = \tilde{\Vec{x}}_i^j + \delta \vec{x}_i^j \quad & i\in\{0,\hspace{2pt} ...,\hspace{2pt} N-1\},
   \label{eq:statePerturbation} \\
   \quad & \vec{u}_{i}^j = \tilde{\Vec{u}}_i^j + \delta \vec{u}_i^j \quad & i\in\{0,\hspace{2pt} ...,\hspace{2pt} N-1\},
\end{align}
\label{eq:exBefore}
\end{subequations} 
The state at the subsequent node is obtained through the propagation of the first-order dynamics, as in \cref{eq:perturbations}

\begin{equation}
\begin{aligned}
 \quad & \vec{x}_{i+1}^j = \mathcal{T}^1_{\vec{x}_{i+1}^j}(\vec{x}_i^j, \vec{u}_i^j) \quad & i\in\{0,\hspace{2pt} ...,\hspace{2pt} N-1\}.
 \label{eq:linMaps}
\end{aligned}
\end{equation}
\cref{eq:linMaps} can be written in matrix form, using the linear maps that establish the first-order dynamics relationship between the perturbations before the propagation of node $i$ and the ones after the propagation, at node $i+1$. These maps are $\vec{A}_{i+1}^j\in\mathbb{R}^{6\times6}$, which is the \gls{stm}, and $\vec{B}_{i+1}^j\in\mathbb{R}^{6\times3}$, which is the control-state transition matrix 
\begin{equation}
\begin{aligned}
    \quad & \vec{x}_{i+1}^j = \bar{\vec{x}}_{i+1}^j + \vec{A}_{i+1}^j \delta \vec{x}_i^j  + \vec{B}_{i+1}^j
    \delta \vec{u}_i^j  \quad & i\in\{0,\hspace{2pt} ...,\hspace{2pt} N-1\},
    \label{eq:propP}
\end{aligned}
\end{equation}
where $\bar{\vec{x}}_{i+1} = \vec{f}_{p,i}(t_i,\tilde{\vec{x}}_i,\tilde{\vec{u}}_i,\vec{p}_p)$ is the constant part of the \gls{da} propagation.
Recalling \cref{eq:exBefore}, the continuity constraint can be written:
\begin{equation}
\begin{aligned}
    \quad & \vec{x}_{i+1}^j = \vec{A}_{i+1}^j\vec{x}_i^j  + \vec{B}_{i+1}^j
    \vec{u}_i^j + \vec{c}_i^j   \quad & i\in\{0,\hspace{2pt} ...,\hspace{2pt} N-1\},
\end{aligned}
\label{eq:forcedDyn}
\end{equation}
where $\vec{c}_i^j$ is the residual of the linearization $\vec{c}_i^j = \overline{\vec{x}}_{i+1} - \vec{A}_{i+1}^j\tilde{\vec{x}}_i^j  - \vec{B}_{i+1}^j
\tilde{\vec{u}}_i^j$.
The initial condition is fixed because the maneuver cannot alter it 
\begin{equation}
    \vec{x}_0^j = \vec{x}_0^0.
    \label{eq:initBound}
\end{equation}

\subsection{Propagation of the Uncertainty}
It is fundamental to propagate the covariance matrices of the two spacecraft throughout the trajectory so that $\vec{C}_{p,i}^j$ and $\vec{C}_{s,i}$ are associated with each node.
 We assume that the uncertainty of the system evolves linearly.  
So, the initial covariance matrices of the two spacecraft are propagated in the discretized time window through the use of the \gls{stm}, obtained with a first-order \gls{da} expansion
\begin{subequations}
\begin{align}
\quad & \vec{C}_{p,i+1}^j = \vec{A}_{p,i+1}^j\vec{C}_{p,i}^j(\vec{A}_{p,i+1}^j)\transp \quad & i\in\{0,\hspace{2pt} ...,\hspace{2pt} N-1\}.
\label{eq:cov} \\
\quad & \vec{C}_{s,i+1} = \vec{A}_{s,i+1}\vec{C}_{s,i}^j\vec{A}_{s,i+1}\transp \quad & i\in\{0,\hspace{2pt} ...,\hspace{2pt} N-1\}.
\label{eq:covSec}
\end{align}
\end{subequations}
In the case of the secondary, the \glspl{stm} are independent of the optimization, and so they are constant throughout the major iterations.
Using a first-order \gls{da} expansion, the Jacobian matrices of the nonlinear transformations from \cref{eq:toCart}, $\vec{H}_{p,i}^j$ and $\vec{H}_{s,i}\in\mathbb{R}^{3\times6}$, are obtained. These are employed to compute the covariance of the Cartesian positions of the two spacecraft
\begin{subequations}
\begin{align}
    \quad & \vec{P}_{p,i}^j = \vec{H}_{p,i}^j \vec{C}_{p,i}^j (\vec{H}_{p,i}^j)\transp \quad & i\in\{0,\hspace{2pt} ...,\hspace{2pt} N\},     \label{eq:covTransf}\\
    \quad & \vec{P}_{s,i} = \vec{H}_{s,i} \vec{C}_{s,i} \vec{H}_{s,i}\transp \quad & i\in\{0,\hspace{2pt} ...,\hspace{2pt} N\}.
\label{eq:covTransfSec}
\end{align}
\end{subequations}
In the case in which the propagation is performed using Cartesian elements, $\vec{H}_{p,i}^j=\vec{H}_{s,i} = [\vec{I}_3, \hspace{2pt} \vec{0}_3; \hspace{2pt} \vec{0}_{3\times6}]$. 
Since the states are assumed to be Gaussian, the covariance of the Cartesian relative position at each node is the sum of the covariances of the Cartesian position of the two spacecraft 
\begin{equation}
    \begin{aligned}
        \quad & \vec{P}_i^j = \vec{P}_{p,i}^j + \vec{P}_{s,i} \quad & i\in\{0,\hspace{2pt} ...,\hspace{2pt} N\}.
    \label{eq:posCovSum}
    \end{aligned}
\end{equation}
This $3\times3$ matrix defines a 3D ellipsoid that evolves in time and is pivotal in the computation of \gls{smd} and in the definition of the \gls{cam} scheme.


We stress the importance of updating the history of the covariance matrices of the primary spacecraft with each major iteration since a specific maneuver could  significantly  change the state transition between two nodes, leading to a different covariance.  In particular, a maneuver could cause a stretch and/or a rotation of the covariance at node $i$ with respect to the same node in the previous major iteration.  According to \cref{eq:smd}, this could cause a \gls{smd} constraint to be violated in the node, thus requiring a different maneuver with respect to the previous major iteration.

\subsection{Lossless Relaxation of the Control Magnitude Constraint}
\label{sec:eqTransf}

Equation \eqref{eq:nonconvexU} is a non-convex equality constraint. Following the work from \cite{Wang2018}, we introduce a lossless relaxation to convexify the constraint: \cref{eq:nonconvexU} is transformed into an inequality constraint and the control magnitude is added to the optimization vector. In this way, \cref{eq:nonconvexU} becomes a second-order cone constraint. The variable $u_i$ is now allowed to take values higher than the norm of the control that acts on the dynamics

\begin{equation}
     u_i \geq \sqrt{u_{i,1}^2 + u_{i,2}^2 + u_{i,3}^2} \quad  i\in\{0,\hspace{2pt} ...,\hspace{2pt} N-1\}, \label{eq:slackConstr}
\end{equation}
The discretized forms of \cref{eq:ocpObj} and \cref{eq:ocpUmax} become respectively
\begin{equation}
J = \sum_{i=0}^{N-1} u_i, \label{eq:obj}
\end{equation}

\begin{equation}
\begin{aligned}
    0 \leq u_i \leq 1 \quad & i\in\{0,\hspace{2pt} ...,\hspace{2pt} N-1\},
    \label{eq:slackbound}
\end{aligned}
\end{equation}
This relaxation is lossless, meaning that the optimal solution for the convexified problem is also optimal for the original problem.

\subsection{Convexification of the Keep-Out-Zone Constraint}
\label{sec:projLin}

\begin{figure}[b!]\centering
\tikzset{every picture/.style={line width=0.75pt}} 

\begin{tikzpicture}[x=0.75pt,y=0.75pt,yscale=-1,xscale=1]

\draw [color={rgb, 255:red, 208; green, 2; blue, 27 }  ,draw opacity=1 ]   (367.92,464.14) -- (378.74,541.23) ;
\draw [shift={(367.64,462.16)}, rotate = 82.01] [fill={rgb, 255:red, 208; green, 2; blue, 27 }  ,fill opacity=1 ][line width=0.08]  [draw opacity=0] (12,-3) -- (0,0) -- (12,3) -- cycle    ;
\draw    (368.2,535.06) -- (377.77,533.7) ;
\draw    (368.2,535.06) -- (369.56,544.63) ;

\draw    (236.69,586.21) -- (244.95,581.19) ;
\draw    (236.69,586.21) -- (241.71,594.47) ;

\draw [color={rgb, 255:red, 65; green, 117; blue, 5 }  ,draw opacity=1 ]   (250.52,588.64) -- (369.61,488.99) ;
\draw [shift={(371.15,487.71)}, rotate = 140.08] [fill={rgb, 255:red, 65; green, 117; blue, 5 }  ,fill opacity=1 ][line width=0.08]  [draw opacity=0] (12,-3) -- (0,0) -- (12,3) -- cycle    ;
\draw [color={rgb, 255:red, 65; green, 117; blue, 5 }  ,draw opacity=1 ]   (379.02,543.42) -- (435.62,513.18) ;
\draw [shift={(437.39,512.24)}, rotate = 151.89] [fill={rgb, 255:red, 65; green, 117; blue, 5 }  ,fill opacity=1 ][line width=0.08]  [draw opacity=0] (12,-3) -- (0,0) -- (12,3) -- cycle    ;
\draw [color={rgb, 255:red, 0; green, 0; blue, 0 }  ,draw opacity=1 ] [dash pattern={on 0.84pt off 2.51pt}]  (371.15,487.71) -- (379.02,543.42) ;
\draw  [color={rgb, 255:red, 208; green, 2; blue, 27 }  ,draw opacity=1 ][fill={rgb, 255:red, 208; green, 2; blue, 27 }  ,fill opacity=0.3 ][line width=0.75]  (177.35,692.36) .. controls (163.6,643.48) and (235.59,580.96) .. (338.16,552.73) .. controls (440.72,524.5) and (535.02,541.24) .. (548.77,590.13) .. controls (562.52,639.02) and (490.52,701.53) .. (387.95,729.76) .. controls (285.39,758) and (191.1,741.25) .. (177.35,692.36) -- cycle ;
\draw  [dash pattern={on 4.5pt off 4.5pt}]  (431.5,477.12) -- (131.75,662.38) ;
\draw  [fill={rgb, 255:red, 0; green, 0; blue, 0 }  ,fill opacity=1 ] (249.4,590.53) .. controls (248.35,589.93) and (248,588.59) .. (248.62,587.54) .. controls (249.24,586.49) and (250.59,586.13) .. (251.64,586.74) .. controls (252.69,587.34) and (253.04,588.68) .. (252.42,589.73) .. controls (251.8,590.78) and (250.45,591.14) .. (249.4,590.53) -- cycle ;
\draw  [dash pattern={on 4.5pt off 4.5pt}]  (539.5,514) -- (279.5,561) ;
\draw  [fill={rgb, 255:red, 0; green, 0; blue, 0 }  ,fill opacity=1 ] (379.99,545.41) .. controls (378.9,545.94) and (377.59,545.48) .. (377.06,544.38) .. controls (376.52,543.28) and (376.97,541.96) .. (378.06,541.43) .. controls (379.14,540.91) and (380.46,541.37) .. (380.99,542.47) .. controls (381.52,543.56) and (381.07,544.88) .. (379.99,545.41) -- cycle ;
\draw [color={rgb, 255:red, 52; green, 104; blue, 163 }  ,draw opacity=1 ]   (363.06,641.25) -- (183.86,483.2) ;
\draw [shift={(182.36,481.88)}, rotate = 41.41] [fill={rgb, 255:red, 52; green, 104; blue, 163 }  ,fill opacity=1 ][line width=0.08]  [draw opacity=0] (12,-3) -- (0,0) -- (12,3) -- cycle    ;
\draw [color={rgb, 255:red, 52; green, 104; blue, 163 }  ,draw opacity=1 ]   (363.06,641.25) -- (371.18,491.89) ;
\draw [shift={(371.29,489.89)}, rotate = 93.11] [fill={rgb, 255:red, 52; green, 104; blue, 163 }  ,fill opacity=1 ][line width=0.08]  [draw opacity=0] (12,-3) -- (0,0) -- (12,3) -- cycle    ;
\draw [color={rgb, 255:red, 52; green, 104; blue, 163 }  ,draw opacity=1 ]   (363.06,641.25) -- (437.04,514.78) ;
\draw [shift={(438.05,513.05)}, rotate = 120.33] [fill={rgb, 255:red, 52; green, 104; blue, 163 }  ,fill opacity=1 ][line width=0.08]  [draw opacity=0] (12,-3) -- (0,0) -- (12,3) -- cycle    ;
\draw  [dash pattern={on 0.84pt off 2.51pt}]  (182.36,481.88) -- (249.35,586.78) ;
\draw  [fill={rgb, 255:red, 52; green, 104; blue, 163 }  ,fill opacity=1 ] (179.32,481.17) .. controls (178.67,480.15) and (178.99,478.8) .. (180.03,478.16) .. controls (181.07,477.52) and (182.43,477.84) .. (183.07,478.86) .. controls (183.72,479.89) and (183.4,481.24) .. (182.36,481.88) .. controls (181.33,482.51) and (179.96,482.2) .. (179.32,481.17) -- cycle ;
\draw  [fill={rgb, 255:red, 52; green, 104; blue, 163 }  ,fill opacity=1 ] (368.94,487.85) .. controls (368.86,486.65) and (369.78,485.61) .. (371,485.53) .. controls (372.22,485.45) and (373.27,486.36) .. (373.35,487.57) .. controls (373.43,488.77) and (372.51,489.81) .. (371.29,489.89) .. controls (370.07,489.97) and (369.02,489.06) .. (368.94,487.85) -- cycle ;
\draw  [fill={rgb, 255:red, 52; green, 104; blue, 163 }  ,fill opacity=1 ] (437.39,512.24) .. controls (436.86,511.15) and (437.33,509.84) .. (438.43,509.32) .. controls (439.54,508.79) and (440.85,509.25) .. (441.37,510.33) .. controls (441.9,511.42) and (441.43,512.73) .. (440.33,513.26) .. controls (439.23,513.79) and (437.91,513.33) .. (437.39,512.24) -- cycle ;
\draw    (363.06,641.25) -- (378.29,547.57) ;
\draw [shift={(378.61,545.6)}, rotate = 99.24] [fill={rgb, 255:red, 0; green, 0; blue, 0 }  ][line width=0.08]  [draw opacity=0] (12,-3) -- (0,0) -- (12,3) -- cycle    ;
\draw    (363.06,641.25) -- (254.23,590.57) ;
\draw [shift={(252.42,589.73)}, rotate = 24.97] [fill={rgb, 255:red, 0; green, 0; blue, 0 }  ][line width=0.08]  [draw opacity=0] (12,-3) -- (0,0) -- (12,3) -- cycle    ;
\draw  [fill={rgb, 255:red, 208; green, 2; blue, 27 }  ,fill opacity=1 ] (361.18,642.4) .. controls (360.54,641.37) and (360.85,640.03) .. (361.89,639.39) .. controls (362.93,638.75) and (364.29,639.07) .. (364.93,640.09) .. controls (365.58,641.12) and (365.26,642.47) .. (364.22,643.1) .. controls (363.19,643.74) and (361.82,643.43) .. (361.18,642.4) -- cycle ;
\draw [color={rgb, 255:red, 208; green, 2; blue, 27 }  ,draw opacity=1 ]   (209.35,524.26) -- (249.35,586.78) ;
\draw [shift={(208.27,522.57)}, rotate = 57.38] [fill={rgb, 255:red, 208; green, 2; blue, 27 }  ,fill opacity=1 ][line width=0.08]  [draw opacity=0] (12,-3) -- (0,0) -- (12,3) -- cycle    ;

\draw (187.12,466.93) node [anchor=north west][inner sep=0.75pt]  [color={rgb, 255:red, 52; green, 104; blue, 163 }  ,opacity=1 ]  {$\vec{r}^{0}$};
\draw (242.06,593.76) node [anchor=north west][inner sep=0.75pt]    {$\vec{z}^{1}$};
\draw (381.28,546.73) node [anchor=north west][inner sep=0.75pt]    {$\vec{z}^{2}$};
\draw (366.59,642.58) node [anchor=north west][inner sep=0.75pt]  [color={rgb, 255:red, 208; green, 2; blue, 27 }  ,opacity=1 ]  {$\vec{r}_{s}$};
\draw (351.76,497.46) node [anchor=north west][inner sep=0.75pt]  [color={rgb, 255:red, 52; green, 104; blue, 163 }  ,opacity=1 ]  {$\vec{r}^{1}$};
\draw (441.83,505.76) node [anchor=north west][inner sep=0.75pt]  [color={rgb, 255:red, 52; green, 104; blue, 163 }  ,opacity=1 ]  {$\vec{r}^{2}$};
\draw (163.6,538.65) node [anchor=north west][inner sep=0.75pt]  [color={rgb, 255:red, 208; green, 2; blue, 27 }  ,opacity=1 ]  {$\vec{\nabla }\left( d_{m}^{2}\right)\Bigl|_{\vec{z}^{1}}$};
\draw (370.72,446.6) node [anchor=north west][inner sep=0.75pt]  [color={rgb, 255:red, 208; green, 2; blue, 27 }  ,opacity=1 ]  {$\vec{\nabla }\left( d_{m}^{2}\right)\Bigl|_{\vec{z}^{2}}$};
\draw (414.42,490.44) node [anchor=north west][inner sep=0.75pt]    {$\textcolor[rgb]{0.25,0.46,0.02}{\vec{r}}\textcolor[rgb]{0.25,0.46,0.02}{^{2}}\textcolor[rgb]{0.25,0.46,0.02}{-}\textcolor[rgb]{0.25,0.46,0.02}{\vec{z}}\textcolor[rgb]{0.25,0.46,0.02}{^{2}}$};
\draw (324.63,474.71) node [anchor=north west][inner sep=0.75pt]    {$\textcolor[rgb]{0.25,0.46,0.02}{\vec{r}}\textcolor[rgb]{0.25,0.46,0.02}{^{1}}\textcolor[rgb]{0.25,0.46,0.02}{-}\textcolor[rgb]{0.25,0.46,0.02}{\vec{z}}\textcolor[rgb]{0.25,0.46,0.02}{^{1}}$};

\end{tikzpicture}
\caption{Simplified 2D version of the projection and linearization process for the generic node $i$.}
\label{fig:minors}
\end{figure}
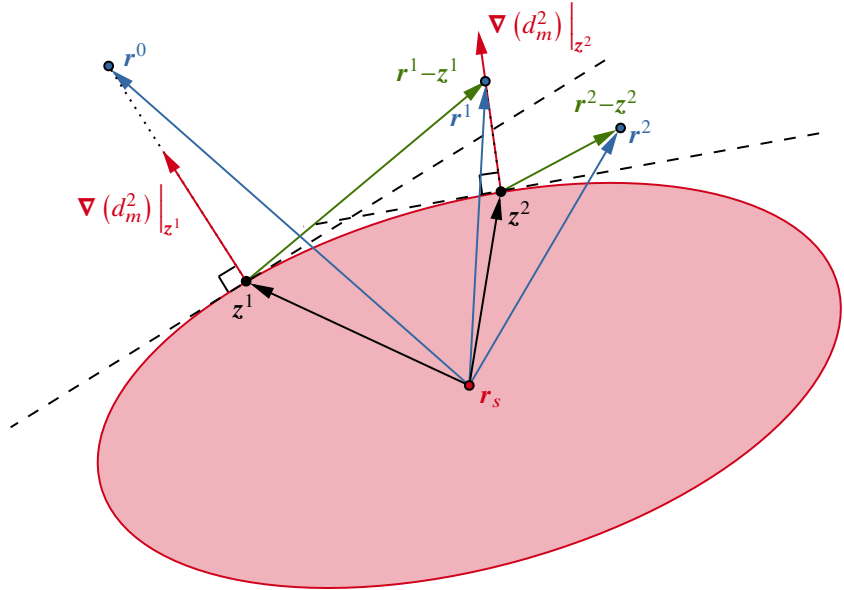

The primary goal of the  maneuver  is to reduce the collision risk by lowering \gls{ipc}. This objective is mathematically formulated using \gls{smd} by leveraging the method proposed by Mao \textit{et al.} \cite{Mao2021} and Armellin \cite{Armellin2021}.  In the following discussion, the indexes $j$ and $i$ are dropped since one the process is repeated multiple times inside the same major iteration and for each node. 
A projection and linearization algorithm is utilized to convexify the nonlinear constraint in \cref{eq:keepout}, as depicted in \cref{fig:minors}. This algorithm operates iteratively within the framework of minor iterations, denoted by the symbol $k$. For each node, the projection convex sub-problem aims to find the point on the surface of the ellipsoid that is closest to the relative position $\vec{r}^{k-1}$ from the previous minor iteration. 

With reference to \cref{fig:minors}, let the ballistic relative position of the primary spacecraft at a generic node $i$ be $\vec{r}^0$. A tangent plane is built on $\vec{z}^1$, which is the point on the surface of the ellipsoid that is closest to $\vec{r}^0$: the admissible region for the optimization, then, is the hemi-space identified by the plane. The optimized relative trajectory in node $i$ after the first minor iteration is $\vec{r}^1$, which is used as a starting point to repeat the projection and linearization process. This process is repeated until a tolerance on the difference in relative position between two consecutive iterations is met, i.e. $||\vec{r}^k-\vec{r}^{k-1}||\leq tol_m$. 
When $k=1$, the value of $\vec{r}^0$ is the value of the last minor iteration of the previous major iteration. Additionally, if $j=1$, $\vec{r}^0$ is the relative position of the ballistic trajectory.

The first part of the minor iterations process consists in the projection of the relative trajectory point $\vec{r}^k$ onto the ellipsoid identified by $\bar{d}_m^2$, i.e., the search for the closest point to $\vec{r}^k$  belonging to the surface of the \gls{koz}. To perform this operation , we solve a quadratic optimization problem. First of all it is convenient to diagonalize the covariance matrix.  
Let the dynamics used in the major convex problem be expressed in the $\mathcal{B}$ reference frame (e.g., an \gls{eci} frame), then $\vec{r}_\mathcal{B}^{k-1}\in\mathbb{R}^3$ and $\vec{P}_\mathcal{B}\in\mathbb{R}^{3\times3}$ are respectively the mean value and the covariance matrix  of the relative position multivariate random variable expressed in this frame. 
Using the covariance's eigenvalues matrix $\vec{D} = \mathrm{diag}([\lambda_1 \hspace{3pt} \lambda_2 \hspace{3pt} \lambda_3])$ and the corresponding eigenvectors matrix $\vec{V}$, the problem is rotated into reference frame $\mathcal{C}$ and scaled so that the covariance is transformed into the identity matrix $\vec{I}_3\in\mathbb{R}^{3\times3}$:
\begin{equation}
    \hat{r}_\mathcal{C}^{k-1} = \vec{D}^{-1}\vec{V}\transp \vec{r}_\mathcal{B}^{k-1}\, \mbox{ and } \,
    \hat{P}_\mathcal{C} = \vec{I}_3.
\end{equation}
Now, a simple quadratic optimization problem is solved in order to find the point $\hat{z}_\mathcal{C}$
\begin{subequations}
\begin{align}
\min_{\hat{z}_\mathcal{C}} \quad &  ||\hat{z}_\mathcal{C}^k - \hat{r}_\mathcal{C}^{k-1}|| \label{eq:objsub}\\
\textrm{s.t.} \quad & (\hat{z}_\mathcal{C}^k)\transp \hat{z}_\mathcal{C}^k \leq \smdlim, \label{eq:consub}
\end{align}
\label{eq:probEllipsoid}
\end{subequations}
The objective \cref{eq:objsub} minimizes the distance between the position of the previous iteration and the optimization variable $\hat{z}_\mathcal{C}$. \cref{eq:consub} imposes a relaxed condition on the optimization variable to be inside the ellipsoid. This relaxed condition is lossless since the minimization of the objective guarantees that the optimization vector is always positioned on the surface of the ellipsoid, maximizing the distance from its center.

Once $\hat{z}_\mathcal{C}$ is determined, the solution is transformed back into the physical space using the equation  
$ \vec{z}_\mathcal{B}^k = \vec{V}\hspace{3pt} \vec{D}\hat{z}_\mathcal{C}^k$.
Now, the linearization of the \gls{koz} constraint is performed on $\vec{z}^k$. 
The dot product between the gradient of \gls{smd} computed on $\vec{z}^k$ and the vector going from $\vec{z}^k$ to the optimized trajectory point $\vec{r}^k$ must be positive.  With the aid of \cref{fig:minors}, the constraint is employed by ensuring that the green red vectors both point outwards from the ellipsoid. The equation of the constraint in iteration $k$ is (including also the indexes $j$ and $i$) 
\begin{equation}
\begin{aligned}
    \nabla(\smd)^{j,k}\Big|_{\vec{z}_i^{j,k}}\cdot (\vec{r}_i^{j,k}-\vec{z}_i^{j,k}) \geq 0 \quad & i\in\{1,\hspace{2pt} ...,\hspace{2pt} N\}.
    \label{eq:caConstr}
\end{aligned}
\end{equation}

As highlighted by Malyuta et al. \cite{Malyuta2021Tutorial}, \gls{scp} algorithms can only converge if the initial trajectory guess is feasible with respect to the convex constraint, even if it does not comply with the nonlinear dynamics. In other words, the method described above only works if $\hat{r}_\mathcal{C}^{k-1}$ lies outside the region occupied by the ellipsoid volume for every node. Otherwise, if the  ballistic relative trajectory  point satisfies the inequality constraint in \cref{eq:consub}, the relaxation fails because the objective function in \cref{eq:objsub} becomes zero when $\hat{z}_\mathcal{C}^k=\hat{r}_\mathcal{C}^{k-1}$: this leads to \cref{eq:caConstr} being undefined since the value of the gradient must be computed on the surface of the ellipsoid for it to be a relevant relaxation of the nonlinear constraint.
So, for the nodes in which the original point is inside the ellipsoid, one should find a starting point (at major iteration $j$ and minor $k=0$) that satisfy \cref{eq:caConstr}, before applying the projection and linearization algorithm. A straightforward approach to address this issue is to use the point of intersection between the surface of the ellipsoid and the line connecting the ellipsoid's origin with the original point as the initial expansion point. In frame $\mathcal{C}$, this point can be computed as follows
\begin{equation}
    \vec{r}_\mathcal{C}'^{j,0} = \Big[\Big(\frac{x}{\lambda_1}\Big)^2 + \Big(\frac{y}{\lambda_2}\Big)^2 + \Big(\frac{z}{\lambda_3}\Big)^2\Big]^{-\frac{1}{2}}\vec{r}_\mathcal{C}^{j,0},
    \label{eq:startpoint}
\end{equation}
where $\vec{r}_\mathcal{C}'^{j,0}$ is the new point that might not satisfy the dynamics, $\vec{r}_\mathcal{C}^{j,0} = [x \hspace{3pt} y \hspace{3pt} z]\transp$ is the original point inside the ellipsoid.

\subsection{Squared Mahalanobis Distance Sensitivity Constraint}
\label{sec:smdSensitivity}
A typical solution of the long-encounter \gls{ca} problem yields a \gls{ipc} profile, presenting at least one local maximum that is usually very close to $\ipclim$.
If the gradient of \gls{smd} for these points is large,  a small deviation from the optimized trajectory (for example, due to errors in the actuation of the maneuver)   may cause the actual value of \gls{smd} to vary significantly. For this reason, it is chosen to introduce a  constraint that bounds the value of the norm of the gradient of \gls{smd} for these nodes.

Let us assume that a solution of a major iteration $j-1$ is available, then $\ipci^{j-1}$ is known for every node. The constraint on the gradient of \gls{smd} is applied at node $i$ only if the following condition is  fulfilled  
\begin{equation}
   \ipci^{j-1} \geq (1-\varepsilon) \ipclim,
   \label{eq:nablalim}
\end{equation}
where $\varepsilon\in\mathbb{R}_{[0,1]}$ is an arbitrary percentage of the \gls{ipc} limit. This constraint bounds the norm of the gradient  of \gls{smd}  to be lower or equal to a certain limit $\lambda$
\begin{equation}
    \begin{aligned}
    \text{if \cref{eq:nablalim} is true,} \hspace{20pt} ||\nabla(\smd)_i|| \leq \gamma_i \quad & & i\in\{1,\hspace{2pt} ...,\hspace{2pt} N\}.
    \end{aligned}
    \label{eq:bcstr}
\end{equation}
The limit value of the gradient $\gamma_i$ is computed according to the following procedure. Let the maximum allowed deviation of \gls{ipc} at a distance $\Delta r$ (e.g. equal to the combined \gls{hbr}) from $\vec{r}$ be equal to or lower than a percentage of $\ipclim$
\begin{equation}
    \Delta P_{IC,max} \leq \rho \ipclim,
    \label{eq:deltaIPC}
\end{equation}
where $\rho\in\mathbb{R}_{[0,1]}$ is a design parameter. The \gls{ipc} variation can be expressed as a function of the derivative of \gls{ipc} with respect to \gls{smd} and of the distance $\mathrm{d}r$ from the relative position $\vec{r}$

\begin{equation}
 \mathrm{d}\ipc = \frac{\partial \ipc}{\partial \smd} \nabla(\smd)\cdot\mathrm{d}\vec{r}.
    \label{eq:ders}
\end{equation}
The derivative of \gls{ipc} is always negative because it is computed according to the approximation in \cref{eq:constipc}
\begin{equation}
    \frac{\partial \ipc}{\partial \smd} = -\frac{\ipc}{2}.
    \label{eq:ipcder}
\end{equation}
To maximize the expression in \cref{eq:ders}, then,  the deviation must be parallel and opposite to $\nabla\smd$ and equal to $\Delta r$ in magnitude: 
\begin{equation}
\mathrm{d}\vec{r} = -\Delta r \frac{\nabla(\smd)}{||\nabla(\smd)||}.
\label{eq:dx}
\end{equation}
Substituting \cref{eq:dx} and \cref{eq:ipcder} into  \cref{eq:ders}, one gets an explicit expression for the limit of the variation of \gls{ipc} 
\begin{equation}
    \Delta P_{IC,max} = \frac{\ipc}{2} ||\nabla(\smd)|| \Delta r.
    \label{eq:deltaipclim}
\end{equation}
 Now, substituting \cref{eq:deltaipclim} into \cref{eq:deltaIPC} and rearranging, the limit imposed to $\nabla\smd$ in \cref{eq:bcstr} eventually depends on the value of  \gls{ipc}
\begin{equation}
    \begin{aligned}
    \quad & \gamma_i = \frac{2\rho}{\Delta r } \frac{\ipclim}{\ipci} \quad & i\in\{1,\hspace{2pt} ...,\hspace{2pt} N\}.
    \label{eq:smdGradLim}
\end{aligned}
\end{equation}
The components of $\nabla(\smd)_i$ are 3 new optimization variables per node.

Since the expression of \gls{smd} is quadratic with respect to the relative position variable $\vec{r}_i$, its gradient is linear, and it reads
\begin{equation}
\begin{aligned}
   \nabla(\smd)_i =  2\vec{r}_i\transp\vec{P}^{-1}_i  \quad & i\in\{1,\hspace{2pt} ...,\hspace{2pt} N\}.
   \label{eq:gradCont}
\end{aligned}
\end{equation}

\section{Station-Keeping Constraints}
\label{sec:sk}
The base \gls{socp} presented in \cref{sec:socpprob} is completed in this section by the translation of \gls{sk} requirements into a linear constraints. 

\subsection{GEO Station-Keeping Constraint}
Maneuvers  in  \gls{geo} orbits must consider the need to respect a \gls{sk} constraint. The \gls{sk} is viewed as a "keep-in box" in two dimensions, the longitude and the latitude of the spacecraft.
The nonlinear dynamics of the evolution of the two geodetic variables are dependent only on the state of the primary spacecraft and on the time variable
\begin{equation}
    \vec{\phi}(t) = \vec{g}(t,\vec{x}(t)),
    \label{eq:funcGeodetic}
\end{equation}
where $\vec{g}(\cdot): \mathbb{R}_{[t_0, t_f]}\times\mathbb{R}^6\rightarrow\mathbb{R}^2$ is a function that first transforms the state into \gls{ecef} and then into geodetic and $\vec{\phi} \in \mathbb{R}^2$ is the vector comprising latitude and longitude.
After discretization, the perturbation in the state is given by \cref{eq:statePerturbation}, where the expansion point is, as usual, the output of the previous major iteration ($\tilde{\Vec{x}}_i^j = \Vec{x}_i^{j-1}$).

Using \gls{da}, \cref{eq:funcGeodetic} is approximated as a first-order Taylor polynomial
\begin{equation}
\begin{aligned}
    \quad & \vec{\phi}_i^j =\mathcal{T}^1_{\vec{\phi}_i^j} (\vec{x}_i^j)  \quad & i\in\{1,\hspace{2pt} ...,\hspace{2pt} N\}.
\end{aligned}
\label{eq:mapsPhi}
\end{equation}
Equation \eqref{eq:mapsPhi} allows for the representation of the perturbation of the geodetic coordinates as a linear transformation of the perturbation of the state
\begin{equation}
    \begin{aligned}
    \quad & \delta \vec{\phi}_i^j =\vec{G}_i^j\delta\vec{x}_i^j  \quad & i\in\{1,\hspace{2pt} ...,\hspace{2pt} N\},
    \end{aligned}
    \label{eq:latlonLin}
\end{equation}
where $\vec{G}_i^j\in\mathbb{R}^{2\times6}$ is the linear map of the geodetic transformation for node $i$ in major iteration $j$.

The \gls{sk} requirement states that at all times, the latitude and longitude need to be inside a rectangular box which is centered on the nominal coordinates $\Vec{\phi}_0\in\mathbb{R}^2$; the sides of the box are the elements of $\Delta\vec{\phi}\in\mathbb{R}^2$
\begin{equation}
    \begin{aligned}
    \quad & \vec{\phi}_0 - \Delta\vec{\phi} \preceq \vec{\phi}_i^j \preceq \vec{\phi}_0 + \Delta\vec{\phi} \quad & i\in\{1,\hspace{2pt} ...,\hspace{2pt} N\},
    \end{aligned}   
    \label{eq:latlonConstr}
\end{equation}
where the symbol $\preceq$ indicates the componentwise inequality.
Thus, rearranging \cref{eq:latlonLin} and \cref{eq:statePerturbation} and splitting the two inequalities, the linearized \gls{sk} constraint can be written as two componentwise inequalities
\begin{subequations}
\begin{align}
   \quad & \vec{G}_i^j\vec{x}_i^{j} \succeq \vec{\phi}_0 -\Delta\vec{\phi} +\vec{d}_i^j  \quad & i\in\{1,\hspace{2pt} ...,\hspace{2pt} N\}, \\
   \quad & \vec{G}_i^j\vec{x}_i^{j} \preceq \vec{\phi}_0 +\Delta\vec{\phi} + \vec{d}_i^j \quad & i\in\{1,\hspace{2pt} ...,\hspace{2pt} N\},
\end{align}
\label{eq:skConstr}
\end{subequations}
where $\vec{d}_i^j=\vec{G}_i^j\vec{x}_i^{j-1}-\overline{\vec{\phi}}_i^j$ is the residual of the linearization.  $\overline{\vec{\phi}}_i^j = \vec{g}(t_i,\tilde{\vec{x}}_i^{j})$ is the constant part of the geodetic transformation, i.e., the nonlinear function evaluation of the expansion point.

\subsection{Station-Keeping State Targeting}

A condition of return to the nominal orbit can be implemented setting the final target state of the optimization to be equal to the final state of the ballistic trajectory.
Thus, a new bound constraint is  introduced  into the original convex problem, which bounds the state at the last node of the optimization:
\begin{equation}
    \vec{x}_N + \Vec{s}_T^\mathrm{+}  - \Vec{s}_T^\mathrm{-} = \vec{x}_T,
    \label{eq:target}
\end{equation}
where the variables $\Vec{s}_T^\mathrm{+}$ and $\Vec{s}_T^\mathrm{-} \in \mathbb{R}^6$ are slack variables used to create a soft constraint, and $\vec{x}_T = \vec{x}_N^0$. A trade-off between the pure \gls{ca} and the pure \gls{sk} maneuvers can be obtained adding a term to the objective function. The weight $\kappa_T$ determines the softness of the constraint
\begin{subequations}
\begin{equation}
    J_T = \kappa_T ||\Vec{s}_T^\mathrm{+}+\Vec{s}_T^\mathrm{-}||_1
    \label{eq:minTarget}
\end{equation}    
\begin{equation}
    \Vec{s}_T^\mathrm{+},\Vec{s}_T^\mathrm{-} \succeq 0,
    \label{eq:slackBoundTarg}
\end{equation}
\end{subequations}
In \gls{leo} scenarios, the return to the nominal orbit can be considered a sufficient station keeping requirement, but the same is not valid for \gls{geo} orbits.

For \gls{geo} orbits, finding the target state $\vec{x}_T$ to minimize the violation of the \gls{sk} box is the typical optimization problem of the periodic \gls{sk} maneuvers. Analytical solutions exist when only the first harmonics of the gravity potential are considered. In contrast, classical nonlinear optimization methods can solve the problem numerically when more perturbations are considered. To overcome the long computational time required by the latter method, a \gls{scp} formulation is presented, where the dynamics are dealt with using the method presented in \cref{sec:linearization}.
The objective of the optimization is to maximize the time spent inside the \gls{sk} box during a period $t\in \mathbb{R}_{[t_f,t_f+y]}$ where $t_f$ is the final time of the \gls{ca} window and $y$ is the length of the time frame before executing another \gls{sk} maneuver. The time window is discretized in the usual way into $M+1$ equally spaced nodes.
After the discretization, it is straightforward to define the dynamics constraint for the state of the satellite in a very similar way to \cref{eq:forcedDyn}. In this case, the initial state must remain unconstrained and no control is acting
\begin{equation}
\begin{aligned}
\vec{x}_{i+1}^{j} - \vec{A}_{i+1}^j \vec{x}_i^j = \vec{c}_i^j \quad & i \ in \{0,\hspace{2pt} ..., \hspace{2pt} M\},
\end{aligned}
\label{eq:forceDynTarget}
\end{equation}
Where $\vec{A}_{i+1}^j\in\mathbb{R}^{6\times6}$ is the \gls{stm} and $\vec{c}_i^j=\overline{\vec{x}}_{i+1}^j - \vec{A}_{i+1}^j \tilde{\vec{x}}_i^j$ is the residual of the linearization.
The \gls{sk} constraint is enforced similarly to that of \cref{eq:skConstr}. In this problem the control is not available in every node to adjust the position of the spacecraft, so it is unavoidable that, in a long period of propagation (e.g., two weeks), the orbital perturbations make the satellite violate the \gls{sk} box. The constraint is then enforced as a soft constraint via the introduction of $2\times (M+1)$ vector slack variables (each comprising two elements, which influence longitude and latitude respectively), $\vec{\chi}_i^\mathrm{+}$ and $\vec{\chi}_i^\mathrm{-}\in\mathbb{R}^2$. These variables quantify the entity of the violation of the \gls{sk} box requirement over each node
\begin{subequations}
\begin{align}
   \quad & \vec{G}_i^j\vec{r}_i^{j} + \vec{\chi}_i^\mathrm{+} \succeq\vec{\phi}_0 + \vec{d}_i^j-\Delta\vec{\phi}  \quad & i \ in \{0,\hspace{2pt} ..., \hspace{2pt} M\}, \\
   \quad & \vec{G}_i^j\vec{r}_i^{j} - \vec{\chi}_i^\mathrm{-} \preceq\vec{\phi}_0 + \vec{d}_i^j +\Delta\vec{\phi}  \quad & i \ in \{0,\hspace{2pt} ..., \hspace{2pt} M\},
\end{align}
\label{eq:skConstrTarget}
\end{subequations}
where $\vec{d}_i^j = \vec{G}_i^j\vec{r}_i^{j-1}-\overline{\vec{\phi}}_i^j$ is the usual residual of the linearization.
The slack variables need to be non-negative
\begin{equation}
    \begin{aligned}
    \quad & \vec{\chi}_i^\mathrm{+} \text{, }\vec{\chi}_i^\mathrm{-}\succeq 0 \quad &  i \ in \{0,\hspace{2pt} ..., \hspace{2pt} M\},
    \end{aligned}
\label{eq:auxBound}
\end{equation}

The optimization minimizes these violations by acting only on the initial state of the propagation. Thus the objective function is
\begin{equation}
    J_T = \sum_{i=0}^M ||\vec{\chi}_i^\mathrm{+} + \vec{\chi}_i^\mathrm{-}||_1.
    \label{eq:objTarget}
\end{equation}
The optimization problem can be summarized as

\begin{equation}
\begin{aligned}
\min_{\vec{x}_0} \quad & \text{\cref{eq:objTarget},} \quad & \mbox{s.t.}  \quad & \text{\cref{eq:forceDynTarget,eq:skConstrTarget,eq:auxBound}}
\end{aligned}
\label{eq:opt}
\end{equation}

\section{Finalization of the SOCP}
\label{sec:finalize}
In order to finalize the \gls{socp}, a trust region algorithm combined with virtual controls is introduced.

\subsection{Trust Region Constraint}
The use of a trust region algorithm is of pivotal importance when dealing with a complexly nonlinear problem that has been linearized. The classic approach introduces a node-wise constraint that bounds the norm of the maximum deviation allowed to the state variables with respect to the linearization point. The radius of the trust region is usually proportional to some performance index that is related to how well the linearized dynamics represent the actual non-linear problem \cite{Malyuta2021Tutorial}. Here we introduce a methodology that adjusts the trust region radius for every single variable according to a measure of its nonlinearity and uses a limited number of parameters to update the radius. 

The idea for this algorithm stems from the work of Losacco \textit{et al.} on the \gls{nli} \cite{Losacco2023} and the one of Bernardini \textit{et al.} on trust region \cite{Bernardini2023}: a second-order Taylor expansion of the nonlinear dynamics is computed using \gls{da}. This allows one to obtain the first-order expansion of the Jacobian of the constraints with respect to the optimization variables (both the states and controls):
\begin{equation}
\begin{aligned}
   \quad & \vec{J}_i = \bar{\Vec{J}}_i+\delta\vec{J}_i
   \quad & i\in\{1,\hspace{2pt} ...,\hspace{2pt} N\},
\end{aligned}
\end{equation}
where $\vec{J}_i\in\mathbb{R}^{6\times 9}$. From now on, in this section, the index $i$ will be dropped for readability; still, the equations are valid for each node.

The deviation of the Jacobian is a first-order function of the deviation of the optimization variables

\begin{equation}
\begin{aligned}
    \quad & J_{uv} = \bar{J}_{uv} + \delta {J}_{uv} =  \bar{J}_{uv} + \sum_{w=1}^9 a_{uvw} \delta x_w
    \quad & u\in\{1,\hspace{2pt} ..., \hspace{2pt} 6\}
    \quad & v\in\{1,\hspace{2pt} ..., \hspace{2pt} 9\},
\end{aligned}
\end{equation}
where $\delta x_w = x_w - \tilde{x}_w$ is the componentwise deviation from the reference variable, which comprises both the state and the control: $w \in \{1,\hspace{2pt} ..., \hspace{2pt} 6\}$ indicates the state components, $w\in\{7,\hspace{2pt} ..., \hspace{2pt} 9\}$ the control.

The component-wise \gls{nli} is defined as 
\begin{equation}
    \nu_w = \sqrt{\frac{\sum_{u=1}^6 \sum_{v=1}^9 a_{uvw}^2 }{\sum_{u=1}^6 \sum_{v=1}^9 \bar{J}_{uv}^2}} |\delta x_w| = \xi_w|\delta x_w|.
    \label{eq:nli}
\end{equation}
The variable $\xi_w\in\mathbb{R}_{\geq0}$ is a measure of the nonlinearity of the reference solution; $\nu_w\in\mathbb{R}_{\geq0}$ is an indicator of the nonlinearities for a variation $\delta x_w$ from the reference. To allow variations within the accuracy of the linearizations, $\nu_w$ is bounded by a maximum value $\overline{\nu}$:
\begin{equation}
\begin{aligned}
\quad & \nu_w =  \xi_w|\delta x_w|\leq \overline{\nu} \quad & w\{1,\hspace{2pt} ..., \hspace{2pt} 9\},
\end{aligned}
\label{eq:xik}
\end{equation}

Making the absolute value explicit and bringing the optimization variables to the left, \cref{eq:xik} translates into two trust region constraints
\footnote{It is important to have $\xi_{iw}$ in the numerator, otherwise when $\xi_{iw} = 0$ (linear dynamics) the constraint would become undefined.}

\begin{subequations}
    \begin{align}
        \vec{\xi}_{i} \odot [\vec{x}_{i}\transp \hspace{2pt} \vec{u}_{i}\transp]\transp \preceq \vec{\xi}_{i} \odot [\tilde{\vec{x}}_i\transp \hspace{2pt} \tilde{\vec{u}}_{i}\transp]\transp + \overline{\nu}\cdot\mathbf{1} 
        \quad & i\in\{1,\hspace{2pt} ...,\hspace{2pt} N\}, \\
        \vec{\xi}_{i} \odot[\vec{x}_{i}\transp \hspace{2pt} \vec{u}_{i}\transp]\transp \succeq \vec{\xi}_{i}\odot[\tilde{\vec{x}}_{i}\transp \hspace{2pt} \tilde{\vec{u}}_{i}\transp]\transp - \overline{\nu}\cdot\mathbf{1} 
        \quad & i\in\{1,\hspace{2pt} ...,\hspace{2pt} N\},
    \end{align}
    \label{eq:tr}
\end{subequations}
where $\vec{x}_{i}$ and $\vec{u}_{i}$ are the state and control to be optimized, $\tilde{\vec{x}}_{i}$ and $\tilde{\vec{u}}_{i}$ are the expansion points,  $\vec{\xi}_{i}=[\xi_{i1}, \hspace{2pt} \xi_{i2}, \hspace{2pt} ..., \hspace{2pt} \xi_{i9}]\transp$, $\mathbf{1}\in\mathbb{R}^9$ is a vector of ones, and the symbol $\odot$ indicates the Hadamard product; the index $i$ is used again to indicate the node-wise constraints. For $i=0$ no deviation is allowed so the constraint is not set as it would be redundant with \cref{eq:initBound}. 

\subsubsection{Virtual Controls}
The introduction of the trust region constraint can cause artificial infeasibility, so virtual controls, and virtual buffers are added to the dynamics, \gls{ca}, \gls{sk}, and $\nabla\smd$ constraints, which improve the convergence of the method. The new constraints with virtual controls become
\begin{subequations}
\begin{align}
     \quad & \vec{x}_{i+1}^j - \vec{A}_{i+1}\vec{x}_i^j  - \vec{B}_{i+1}^j \vec{u}_i^j + \Vec{\upsilon}_{dyn,i+1}^j = \vec{c}_i^j   \quad & i\in\{0,\hspace{2pt} ...,\hspace{2pt} N-1\}, \label{eq:forcedDynVc} \\
    \quad & \nabla(\smd)^{j,k}\Big|_{\vec{z}_i^{j,k}}\cdot (\vec{r}_i^{j,k}-\vec{z}_i^{j,k}) + \upsilon_{ca,i}^j \geq 0 \quad & i\in\{1,\hspace{2pt} ...,\hspace{2pt} N\},    \label{eq:caConstrVc} \\
    \quad & \vec{G}_i^j\vec{x}_i^{j}  +\Vec{\upsilon}_{sk,i}^j \succeq \vec{\phi}_0 -\Delta\vec{\phi} +\vec{d}_i^j  \quad & i\in\{1,\hspace{2pt} ...,\hspace{2pt} N\},
    \label{eq:skConstrVc1}\\
    \quad & \vec{G}_i^j\vec{x}_i^{j} +\Vec{\upsilon}_{sk,i}^j\preceq \vec{\phi}_0 +\Delta\vec{\phi} + \vec{d}_i^j \quad & i\in\{1,\hspace{2pt} ...,\hspace{2pt} N\},     \label{eq:skConstrVc2}\\
    \quad & ||\nabla(\smd)_i|| + \upsilon_{s,i}^j \leq \gamma_i \quad & i\in\{1,\hspace{2pt} ...,\hspace{2pt} N\},
    \label{eq:smdGradConstrCa}
\end{align}
\label{vcConstraints}
\end{subequations}
where $\Vec{\upsilon}_{dyn,i}^j\in \mathbb{R}^6$ is the virtual control vector, $\upsilon_{ca,i}^j\in \mathbb{R}$ is the virtual buffer for the \gls{smd} constraint, $\Vec{\upsilon}_{sk,i}^j\in \mathbb{R}^2$ is the virtual buffer for the station keeping constraint, and $\Vec{\upsilon}_{s,i}^j\in \mathbb{R}$ is the virtual buffer for the \gls{smd} sensitivity constraint.

\begin{figure}[tb!]
\centering
\includegraphics[width = \textwidth]{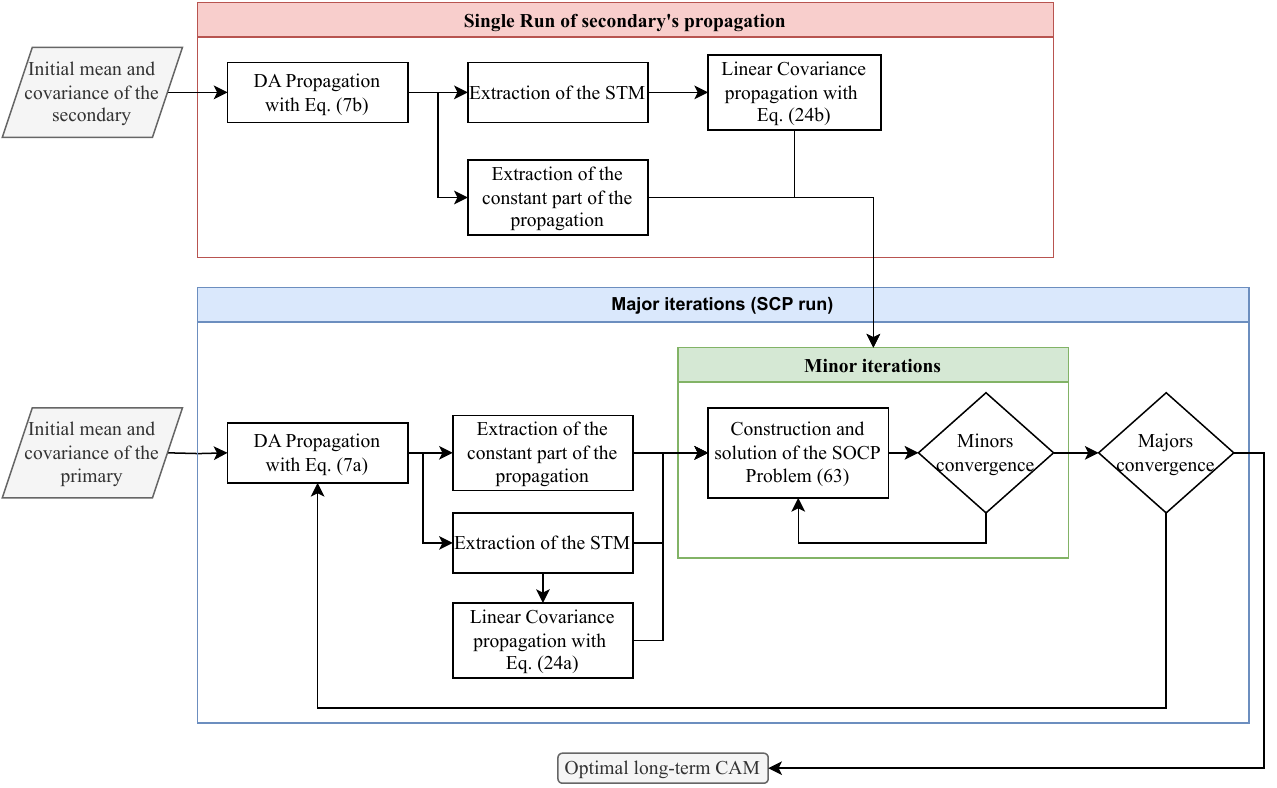}
\caption{High-level flowchart of the \gls{scp} iterative process.}
\label{fig:flowchart}
\end{figure}

Ideally, the virtual controls should all go to $0$ at convergence, so a second-order cone constraint is built on each node, and in the objective function, a term is added which is proportional to it; the value of the weight in the objective function should be orders of magnitude higher than the one on the control, e.g., $10^4$, to favor the penalization of the virtual control variables

\begin{subequations}
    \begin{align}
    \quad & \upsilon_i \leq \sqrt{(\upsilon_{s,i}^j)^2 + (\upsilon_{ca,i}^j)^2 + ||\vec{\upsilon}_{dyn,i}^j|| + ||\vec{\upsilon}_{sk,i}^j||} \quad & i\in\{1,\hspace{2pt} ...,\hspace{2pt} N\}, \label{eq:vcCone}\\
    \quad & J_{vc} = \kappa_{vc}\sum_{i=0}^N \upsilon_i,  \label{eq:vcCost}
    \end{align}
\end{subequations}

\subsection{Final Form of the SOCP}
The final objective function is given by the sum of \cref{eq:obj}, \cref{eq:minTarget}, and \cref{eq:vcCost}:
\begin{equation}
    J = \kappa_T ||\Vec{s}_T^\mathrm{+} + \Vec{s}_T^\mathrm{-}||_1 + \kappa_{vc}\sum_{i=0}^N \upsilon_i + \sum_{i=0}^{N-1} u_i.
    \label{eq:totObj}
\end{equation}
The full convex optimization problem with the novel constraints (\gls{sk} and \gls{smd} sensitivity) is summarized in Problem \eqref{eq:optFinal}.
\begin{equation}
\begin{aligned}
\min_{\mathbb{x},\mathbb{u}} \quad & \text{\cref{eq:totObj}}\\
\mbox{s.t.} 
\quad & \text{\cref{eq:forcedDynVc,eq:caConstrVc,eq:skConstrVc1,eq:gradCont,eq:slackConstr,eq:initBound,eq:target,eq:slackbound,eq:slackBoundTarg,eq:smdGradConstrCa,eq:tr,eq:vcCone}}\\
\end{aligned}
\label{eq:optFinal}
\end{equation}
In \cref{fig:flowchart} and in \cref{alg:scp}, a schematics and the algorithmic representation of the method are reported, respectively.

\section{Results}
\label{sec:results}
In this section, we study a \gls{leo}, a \gls{geo}  and a \gls{heo}  scenario: the effect of different parameters and constraints is analyzed to determine how they affect the computed optimal \gls{cam} thrust profile. In particular, we will analyze the following factors: risk metric, maximum thrust, and inclusion of operational constraints (\gls{sk} and $\nabla\smd$).
The simulations are run with MATLAB r2022b on AMD Ryzen 9 6900HS @ 3.3GHz. The optimization is performed using MOSEK 10.0.24, which implements a state-of-the-art primal-dual interior point solver. 
 In all the considered test-cases, the nominal \gls{tca} is assumed to happen the 2nd of March 2015 at 06:00:00 A.M. .

Despite the algorithm working with a low-thrust formulation, i.e., integrating the control acceleration, it was chosen to display the corresponding $\dv$ in the figures, as it is an easier-to-understand quantity, and the mass loss due to the propulsion is neglected. The $\dv$ is obtained by integrating the acceleration over the time step: $\Delta\vec{v}_i = \Vec{u}_i\Delta t$.

\subsection{LEO Scenario}
The \gls{leo} scenario is based on  the set of test cases   reported in \cite{NunezGarzon2022}. In \cref{tab:paramsLeo}, the physical properties of the spacecraft are shown, and in \cref{tab:initialConditionsLeo} we report the orbital parameters at \gls{tca}. The simulations are run for two orbital periods with \gls{tca} as the median point of the propagation; the time nodes previous to \gls{tca} are negative, and the ones after are positive. Each orbital period is discretized into $60$ nodes, granting a minimum thrust arc of $3$ \si{deg} and a total of $120$ available thrusting opportunities.  The propulsion system is chemical, and can achieve a maximum thrust of $1$ \si{N}, which corresponds to a maximum acceleration of $5$ \si{mm/s^2}  for a mass of $200$ \si{kg}. In the following, the simulations in which the maximum acceleration is $5$ \si{mm/s^2} will be referred to as \textit{high-thrust}, as opposed to the \textit{low-thrust} ones, in which the maximum acceleration is much lower. In both cases, the maneuvers are modeled as finite burns.  

\begin{table}[b!]
\centering
\caption{\gls{leo} scenario: physical properties of the spacecraft.}
\label{tab:paramsLeo}
    \begin{tabular}{lllllll}
    \Xhline{4\arrayrulewidth}
    \textbf{Spacecraft} & $m$ [\si{kg}] & $A_{drag}$ [\si{m^2}]  & $C_D$ [-] & $A_{SRP}$ [\si{m^2}] & $C_r$ [-] & \gls{hbr} [\si{m}]  \\
    \Xhline{3\arrayrulewidth}
    Primary   & $200$ & $1$     & $2.2$ & $1$    & $1.31$ & $25$\\ \hline 
    Secondary & $50$  & $0.05$  & $2$   & $0.05$ & $1.31$ & $7$\\
    \Xhline{4\arrayrulewidth}
    \end{tabular}
\end{table}

\begin{table}[b!]
\centering
\caption{\gls{leo} scenario: orbit parameters at \gls{tca}.}
\label{tab:initialConditionsLeo}
    \begin{tabular}{lllllll}
    \Xhline{4\arrayrulewidth}
    \textbf{Spacecraft} & $\vec{a}$ {[\si{km}]} & $\vec{e}$ {[-]}  & $\vec{i}$ {[\si{deg}]} & $\vec{\omega}$ {[\si{deg}]} & $\vec{\Omega}$ {[\si{deg}]} &  $\vec{\theta}$ {[\si{deg}]}   \\
    \Xhline{3\arrayrulewidth}
    Primary   & $6800$ & $0$                 & $0$                 & $0$      & $0$ & $0$        \\ \hline 
    Secondary & $6802$ & $4.42\times10^{-4}$ & $8.4\times10^{-5}$  & $1.9103$ & $0$ & $-1.9103$  \\
    \Xhline{4\arrayrulewidth}
    \end{tabular}
\end{table}

In \cref{tab:covarianceLeo}, the covariance at \gls{tca} is expressed in the \gls{rtn} reference frame of the two spacecraft. Before summing them, the two covariance matrices must be expressed in a common reference frame, e.g., \gls{eci}.

\begin{table}[tb!]
\centering
\caption{\gls{leo} scenario: diagonal elements of the covariance at \gls{tca}.}
\label{tab:covarianceLeo}
    \begin{tabular}{lllllll}
    \Xhline{4\arrayrulewidth}
   \textbf{Spacecraft} & $C_{rr}$ [\si{m^2}] & $C_{tt}$ [\si{m^2}] & $C_{nn}$ [\si{m^2}] & $C_{\dot{r}\dot{r}}$ [\si{m^2/s^2}] & $C_{\dot{t}\dot{t}}$ [\si{m^2/s^2}] & $C_{\dot{n}\dot{n}}$ [\si{m^2/s^2}] \\
    \Xhline{3\arrayrulewidth}
    Primary    & $0.625$ & $10$ & $3.025$ & $0.00625$ & $0.05625$ & $0.00225$ \\ \hline 
    Secondary  & $5.625$ &$ 90$ & $27.225$ & $0.05625$ & $0.50625$ & $0.02025$ \\ \hline 
    \Xhline{4\arrayrulewidth}
    \end{tabular}
\end{table}

\begin{figure}[tb!]
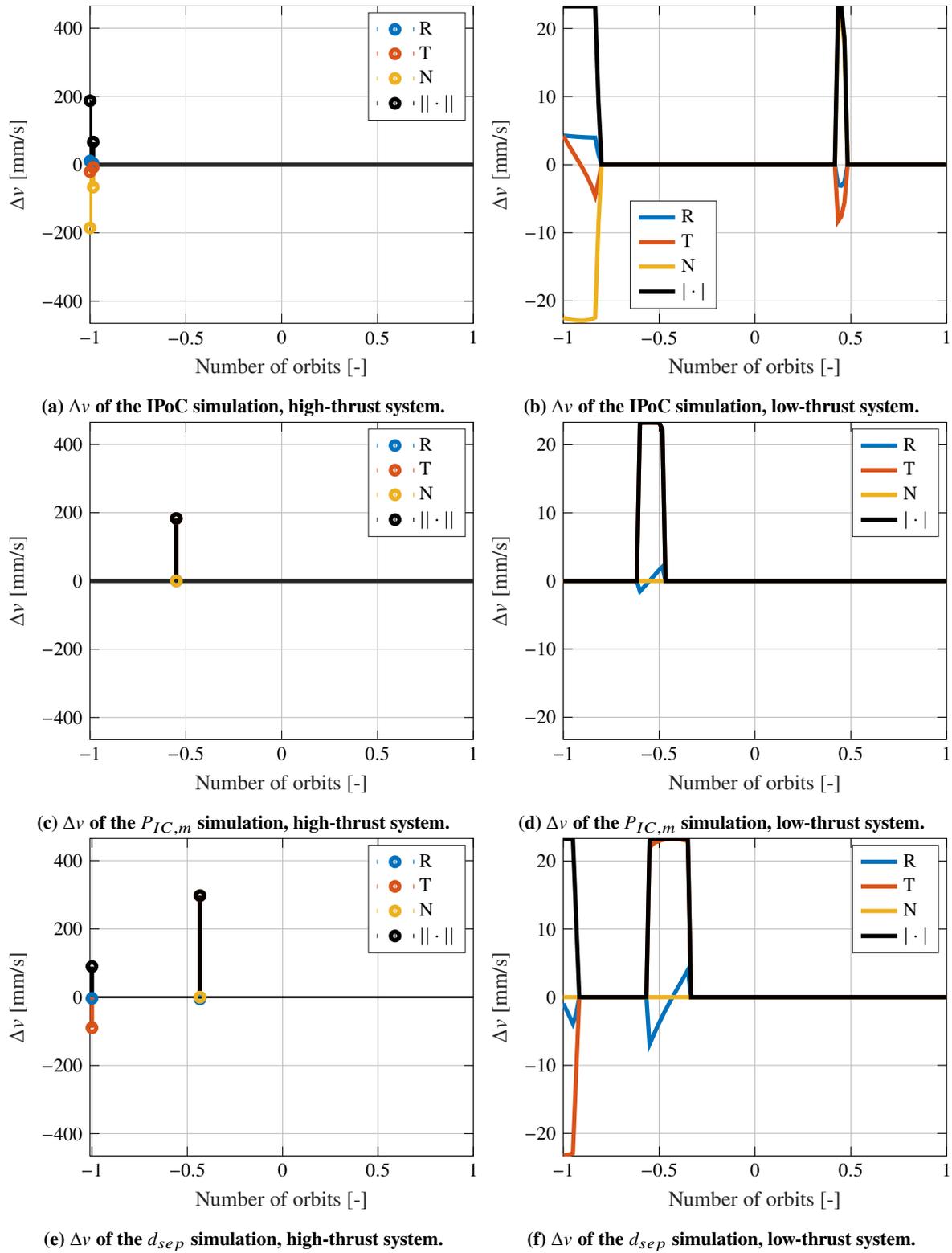

    \centering
    \subfloat[$\dv$ of the \gls{ipc} simulation, high-thrust system.]{
%
%
\definecolor{mycolor1}{rgb}{0.00000,0.44700,0.74100}%
\definecolor{mycolor2}{rgb}{0.85000,0.32500,0.09800}%
\definecolor{mycolor3}{rgb}{0.92900,0.69400,0.12500}%
\begin{tikzpicture}

\begin{axis}[%
at={(0\textwidth,0\textwidth)},
width=0.48\textwidth,
xmin=-1,
xmax=1,
xlabel style={font=\color{white!15!black}},
xlabel={Number of orbits [-]},
ymin=-465.042991335137,
ymax=465.042991335137,
ylabel style={font=\color{white!15!black}},
ylabel={$\dv$ [mm/s]},
axis background/.style={fill=white},
xmajorgrids,
ymajorgrids,
legend style={legend cell align=left, align=left, draw=white!15!black}
]
\addplot[ycomb, color=mycolor1, line width=2.0pt, mark=o, mark options={solid, mycolor1}] table[row sep=crcr] {%
-1	10.7731393621985\\
-0.983333333333333	3.53658472581218\\
};
\addlegendentry{R}

\addplot[ycomb, color=mycolor2, line width=2.0pt, mark=o, mark options={solid, mycolor2}] table[row sep=crcr] {%
-1	-20.3514490984198\\
-0.983333333333333	-7.92563584938701\\
};
\addlegendentry{T}

\addplot[ycomb, color=mycolor3, line width=2.0pt, mark=o, mark options={solid, mycolor3}] table[row sep=crcr] {%
-1	-185.699913467796\\
-0.983333333333333	-65.5541210307546\\
};
\addlegendentry{N}

\addplot[ycomb, color=black, line width=2.0pt, mark=o, mark options={solid, black}] table[row sep=crcr] {%
-1	187.122152280456\\
-0.983333333333333	66.1261364307243\\
};
\addplot[forget plot, color=white!15!black, line width=2.0pt] table[row sep=crcr] {%
-1	0\\
1	0\\
};
\addlegendentry{$||\cdot||$}
\end{axis}

\end{tikzpicture}%
\label{fig:dv1}}
    \subfloat[$\dv$ of the \gls{ipc} simulation, low-thrust system.]{\input{dv5.tex}\label{fig:dv5}} \\
    \subfloat[$\dv$ of the $\ipcmax$ simulation, high-thrust system.]{
%
%
\definecolor{mycolor1}{rgb}{0.00000,0.44700,0.74100}%
\definecolor{mycolor2}{rgb}{0.85000,0.32500,0.09800}%
\definecolor{mycolor3}{rgb}{0.92900,0.69400,0.12500}%
\begin{tikzpicture}

\begin{axis}[%
at={(0\textwidth,0\textwidth)},
width=0.48\textwidth,
xmin=-1,
xmax=1,
xlabel style={font=\color{white!15!black}},
xlabel={Number of orbits [-]},
ymin=-465.042991335137,
ymax=465.042991335137,
ylabel style={font=\color{white!15!black}},
ylabel={$\dv$ [mm/s]},
axis background/.style={fill=white},
xmajorgrids,
ymajorgrids,
legend style={legend cell align=left, align=left, draw=white!15!black}
]
\addplot[ycomb, color=mycolor1, line width=2.0pt, mark=o, mark options={solid, mycolor1}] table[row sep=crcr] {%
-0.55	0.341027183534404\\
};
\addlegendentry{R}

\addplot[ycomb, color=mycolor2, line width=2.0pt, mark=o, mark options={solid, mycolor2}] table[row sep=crcr] {%
-0.55	183.024828226903\\
};
\addlegendentry{T}

\addplot[ycomb, color=mycolor3, line width=2.0pt, mark=o, mark options={solid, mycolor3}] table[row sep=crcr] {%
-0.55 0 \\
};
\addlegendentry{N}

\addplot[ycomb, color=black, line width=2.0pt, mark=o, mark options={solid, black}] table[row sep=crcr] {%
-0.55	183.025145941828\\
};
\addplot[forget plot, color=white!15!black, line width=2.0pt] table[row sep=crcr] {%
-1	0\\
1	0\\
};
\addlegendentry{$||\cdot||$}

\end{axis}
\end{tikzpicture}%
\label{fig:dv2}}
    \subfloat[$\dv$ of the $\ipcmax$ simulation, low-thrust system.]{\input{dv6.tex}\label{fig:dv6}} \\
    \subfloat[$\dv$ of the $\dmiss$ simulation, high-thrust system.]{
%
%
\definecolor{mycolor1}{rgb}{0.00000,0.44700,0.74100}%
\definecolor{mycolor2}{rgb}{0.85000,0.32500,0.09800}%
\definecolor{mycolor3}{rgb}{0.92900,0.69400,0.12500}%
\begin{tikzpicture}

\begin{axis}[%
at={(0\textwidth,0\textwidth)},
width=0.48\textwidth,
xmin=-1.01,
xmax=1,
xlabel style={font=\color{white!15!black}},
xlabel={Number of orbits [-]},
ymin=-465.042991335137,
ymax=465.042991335137,
ylabel style={font=\color{white!15!black}},
ylabel={$\dv$ [mm/s]},
axis background/.style={fill=white},
xmajorgrids,
ymajorgrids,
legend style={legend cell align=left, align=left, draw=white!15!black}
]
\addplot[ycomb, color=mycolor1, line width=2.0pt, mark=o, mark options={solid, mycolor1}] table[row sep=crcr] {%
-1	-3.55420365635211\\
-0.433333333333333	-5.04595156540997\\
};

\addlegendentry{R}

\addplot[ycomb, color=mycolor2, line width=2.0pt, mark=o, mark options={solid, mycolor2}] table[row sep=crcr] {%
-1	-89.5973391287857\\
-0.433333333333333	297.642034634423\\
};
\addlegendentry{T}

\addplot[ycomb, color=mycolor3, line width=2.0pt, mark=o, mark options={solid, mycolor3}] table[row sep=crcr] {%
-0.433333333333333	2.04294566360943e-02\\
};
\addlegendentry{N}

\addplot[ycomb, color=black, line width=2.0pt, mark=o, mark options={solid, black}] table[row sep=crcr] {%
-1	89.667806611902\\
-0.433333333333333	297.684804492742\\
};

\addlegendentry{$||\cdot||$}

\addplot [color=black, line width=1.0pt, forget plot]
  table[row sep=crcr]{%
-0.983333333333333	0\\
1	0\\
};
\end{axis}
\end{tikzpicture}%
\label{fig:dv3}}
    \subfloat[$\dv$ of the $\dmiss$ simulation, low-thrust system.]{\input{dv7.tex}\label{fig:dv7}} \\
    \caption{\gls{leo} scenario: $\dv$ for different collision metrics and propulsion systems.}
    \label{fig:dv123567}
\end{figure}

\subsubsection{Comparison of Risk Metrics}
The selection of different risk metrics determines different optimal maneuvers. The chosen risk thresholds are $\ipclim = 10^{-6}$, $\ipcmaxlim = 10^{-4}$, and $\overline{d}_{miss} = 2$ \si{km}. 
\cref{fig:dv123567} makes it clear that when the \gls{ipc} metric is used, the computed maneuver is mostly out-of-plane, whereas it is tangential in the other two cases. 
The less restrictive metric in terms of total $\dv$ is $P_{IC,m}$, which requires $183$ \si{mm/s}. 
The evolution of $\ipcmax$ of the ballistic trajectory envelops \gls{ipc}, as shown in \cref{fig:ipc_sim12}, guaranteeing that the former is a very conservative estimate of the latter. The most demanding metric, instead, is $\dmiss$, with a two-firings $\dv$ of $387$ \si{mm/s}: to avoid the sphere of $2$ \si{km} centered in the secondary spacecraft, the tangential maneuver is the most efficient, as expected from the short-term problem \cite{Bombardelli2015,DeVittori2022}. Most notably, as already suggested in \cite{NunezGarzon2022}, the $\dmiss$ metric in \cref{fig:dMiss_sim3} exhibits an opposite behavior with respect to \gls{ipc} and $\ipcmax$: the evolution of the covariance in the probability-based criteria determines a high risk in the period in which the relative distance between the two objects is higher.
\begin{figure}[tb!]
    \centering
    \subfloat[$\ipc$ and $\ipcmax$ of the ballistic and optimized trajectories.]{\input{ipc_sim12.tex}\label{fig:ipc_sim12}}
    \subfloat[$\dmiss$ of the ballistic and optimized trajectories.]{
%
%
\definecolor{mycolor1}{rgb}{0.00000,0.44700,0.74100}%
\definecolor{mycolor2}{rgb}{0.85000,0.32500,0.09800}%
\begin{tikzpicture}

\begin{axis}[%
at={(0\textwidth,0\textwidth)},
width=0.48\textwidth,
xmin=-1,
xmax=1,
xlabel style={font=\color{white!15!black}},
xlabel={Number of orbits [-]},
ymin=0.100498815257142,
ymax=25.0904880893328,
ylabel style={font=\color{white!15!black}},
ylabel={$d_{sep}$ [\si{km}] },
axis background/.style={fill=white},
xmajorgrids,
ymajorgrids,
legend style={at={(0.41,0.683)}, anchor=south west, legend cell align=left, align=left, draw=white!15!black}
]
\addplot [color=mycolor1, line width=1.0pt]
  table[row sep=crcr]{%
-1	18.711152385567\\
-0.983333333333333	19.0254705864957\\
-0.966666666666667	19.3342141501873\\
-0.95	19.6308212243134\\
-0.933333333333333	19.9089879530322\\
-0.916666666666667	20.1627122196747\\
-0.9	20.3863401582484\\
-0.883333333333333	20.574599235605\\
-0.866666666666667	20.7226424288824\\
-0.85	20.8260793743969\\
-0.833333333333333	20.881011489778\\
-0.816666666666667	20.8840666311635\\
-0.8	20.8324319381914\\
-0.783333333333333	20.7238804830425\\
-0.766666666666667	20.5567935613873\\
-0.75	20.3301692510899\\
-0.733333333333333	20.0436303675938\\
-0.716666666666667	19.6974339978827\\
-0.7	19.2924714405769\\
-0.683333333333333	18.8302656477192\\
-0.666666666666667	18.3129583838074\\
-0.65	17.7432878824399\\
-0.633333333333333	17.1245648691707\\
-0.616666666666667	16.4606404947504\\
-0.6	15.7558761739501\\
-0.583333333333333	15.015111915089\\
-0.566666666666667	14.243620210402\\
-0.55	13.4470558500933\\
-0.533333333333333	12.6314023106826\\
-0.516666666666667	11.8029214871503\\
-0.5	10.9681095763158\\
-0.483333333333333	10.1336632782199\\
-0.466666666666667	9.30644106683137\\
-0.45	8.49344230395308\\
-0.433333333333333	7.70178543055662\\
-0.416666666666667	6.93870370451851\\
-0.4	6.2115557693456\\
-0.383333333333333	5.52784476021853\\
-0.366666666666667	4.89522221059292\\
-0.35	4.32142557813262\\
-0.333333333333333	3.81403719269354\\
-0.316666666666667	3.37989601108872\\
-0.3	3.02395631162615\\
-0.283333333333333	2.7475540778047\\
-0.266666666666667	2.54656541992923\\
-0.25	2.41055800292888\\
-0.233333333333333	2.32392553079705\\
-0.216666666666667	2.26874716987132\\
-0.2	2.22792940769606\\
-0.183333333333333	2.18732602534049\\
-0.166666666666667	2.1365459443846\\
-0.15	2.06887887875034\\
-0.133333333333333	1.98085016783198\\
-0.116666666666667	1.8717508401224\\
-0.1	1.74331478842613\\
-0.0833333333333333	1.59963137677167\\
-0.0666666666666667	1.44735656895608\\
-0.05	1.29628761360222\\
-0.0333333333333333	1.16014982633744\\
-0.0166666666666667	1.05674116593406\\
0	1.00498815257142\\
0.0166666666666667	1.01684713520725\\
0.0333333333333333	1.08941339903236\\
0.05	1.20703347386001\\
0.0666666666666667	1.35052437110113\\
0.0833333333333333	1.5034636728311\\
0.1	1.65363833979257\\
0.116666666666667	1.79247944939647\\
0.133333333333333	1.91430543068119\\
0.15	2.01588654330959\\
0.166666666666667	2.09635053194777\\
0.183333333333333	2.15736440263845\\
0.2	2.20352563108553\\
0.216666666666667	2.24284547086211\\
0.233333333333333	2.28711102388931\\
0.25	2.3517084181819\\
0.266666666666667	2.45435983917649\\
0.283333333333333	2.61254154074193\\
0.3	2.84025135573644\\
0.316666666666667	3.14564700086733\\
0.333333333333333	3.53068807851892\\
0.35	3.99249410778345\\
0.366666666666667	4.52524991030971\\
0.383333333333333	5.12174312653523\\
0.4	5.77428795619141\\
0.416666666666667	6.47516280640487\\
0.433333333333333	7.21676782305687\\
0.45	7.99165294482102\\
0.466666666666667	8.79250322303423\\
0.483333333333333	9.61211970610888\\
0.5	10.4434196781902\\
0.516666666666667	11.2794492194796\\
0.533333333333333	12.1134024903461\\
0.55	12.9386524757557\\
0.566666666666667	13.7487836965775\\
0.583333333333333	14.5376321623306\\
0.6	15.2993323616315\\
0.616666666666667	16.0283691620026\\
0.633333333333333	16.71962479543\\
0.65	17.3684314627545\\
0.666666666666667	17.9706042762664\\
0.683333333333333	18.5224805350931\\
0.7	19.0209516114047\\
0.716666666666667	19.4634922670272\\
0.733333333333333	19.848184501539\\
0.75	20.1737347926472\\
0.766666666666667	20.4394794207312\\
0.783333333333333	20.6453940123747\\
0.8	20.7920835079671\\
0.816666666666667	20.8807782230909\\
0.833333333333333	20.9133210326768\\
0.85	20.892150792735\\
0.866666666666667	20.8202815109746\\
0.883333333333333	20.701277890737\\
0.9	20.5392223384904\\
0.916666666666667	20.3386831363924\\
0.933333333333333	20.1046746954549\\
0.95	19.8426125211137\\
0.966666666666667	19.5582759364684\\
0.983333333333333	19.2577697000685\\
1	18.9474870276375\\
};
\addlegendentry{Ballistic}

\addplot [color=mycolor2, line width=1.0pt]
  table[row sep=crcr]{%
-1	18.9049321208929\\
-0.983333333333333	19.2213241908468\\
-0.966666666666667	19.5362137227924\\
-0.95	19.8382499996041\\
-0.933333333333334	20.1207807168027\\
-0.916666666666667	20.3774805192875\\
-0.9	20.6023939561772\\
-0.883333333333334	20.789976833535\\
-0.866666666666667	20.9351362054767\\
-0.85	21.0332688661339\\
-0.833333333333333	21.0802979471831\\
-0.816666666666667	21.0727070697727\\
-0.8	21.0075714249775\\
-0.783333333333333	20.8825851444054\\
-0.766666666666667	20.6960843542226\\
-0.75	20.4470653749935\\
-0.733333333333333	20.1351976441043\\
-0.716666666666667	19.7608310456807\\
-0.7	19.3249970305803\\
-0.683333333333333	18.8294041677721\\
-0.666666666666667	18.2764277337881\\
-0.65	17.6690936633255\\
-0.633333333333333	17.0110573272857\\
-0.616666666666667	16.3065778082947\\
-0.6	15.5604886083952\\
-0.583333333333333	14.7781680580704\\
-0.566666666666667	13.9655103920123\\
-0.55	13.1288837570091\\
-0.533333333333333	12.275108930176\\
-0.516666666666667	11.4114415001347\\
-0.5	10.5455662402516\\
-0.483333333333333	9.68561009787948\\
-0.466666666666667	8.8401806579026\\
-0.45	8.01843678420089\\
-0.433333333333333	7.23019448450017\\
-0.416666666666667	6.47667662319409\\
-0.4	5.76978474441247\\
-0.383333333333333	5.13183253109009\\
-0.366666666666667	4.57677051678642\\
-0.35	4.11777779167062\\
-0.333333333333333	3.76470057621006\\
-0.316666666666667	3.52036674055923\\
-0.3	3.37737283983407\\
-0.283333333333333	3.31784440886661\\
-0.266666666666667	3.31707474074582\\
-0.25	3.34895494936904\\
-0.233333333333333	3.39028413914546\\
-0.216666666666667	3.42278562157574\\
-0.2	3.43335441664697\\
-0.183333333333333	3.41348163351484\\
-0.166666666666667	3.35850346152202\\
-0.15	3.26697130028791\\
-0.133333333333333	3.14024109548855\\
-0.116666666666667	2.98230670447662\\
-0.1	2.79989152776874\\
-0.0833333333333333	2.60281511298212\\
-0.0666666666666667	2.40460727445369\\
-0.05	2.22312797439079\\
-0.0333333333333333	2.08037614072445\\
-0.0166666666666667	1.9998100618391\\
0	1.99980435487287\\
0.0166666666666667	2.08586840468339\\
0.0333333333333333	2.24835142496608\\
0.05	2.46787507661471\\
0.0666666666666667	2.72281759800964\\
0.0833333333333333	2.99377734380679\\
0.1	3.26484045403987\\
0.116666666666667	3.52333661838285\\
0.133333333333333	3.75924192164616\\
0.15	3.9646700509125\\
0.166666666666667	4.13354589701436\\
0.183333333333333	4.26144773668599\\
0.2	4.34558923021248\\
0.216666666666667	4.38492452131337\\
0.233333333333333	4.38037694834343\\
0.25	4.33520770693863\\
0.266666666666667	4.25555007279802\\
0.283333333333333	4.15112447158129\\
0.3	4.03608643811385\\
0.316666666666667	3.92978463235273\\
0.333333333333333	3.85685353675438\\
0.35	3.8456415044312\\
0.366666666666667	3.92406603108029\\
0.383333333333333	4.11349635040675\\
0.4	4.4236540344933\\
0.416666666666667	4.85174777157523\\
0.433333333333333	5.38583214793364\\
0.45	6.0095003683144\\
0.466666666666667	6.70543275253818\\
0.483333333333333	7.45721532227112\\
0.5	8.24993906054711\\
0.516666666666667	9.07021730721263\\
0.533333333333333	9.90598057206849\\
0.55	10.7462549656519\\
0.566666666666667	11.5809828478703\\
0.583333333333333	12.400898057003\\
0.6	13.1974492197183\\
0.616666666666667	13.9627599434271\\
0.633333333333333	14.6896151749286\\
0.65	15.371464913817\\
0.666666666666667	16.0024384626285\\
0.683333333333333	16.5773645630969\\
0.7	17.0917929358845\\
0.716666666666667	17.5420148619257\\
0.733333333333333	17.9250807288971\\
0.75	18.2388130466455\\
0.766666666666667	18.4818139047531\\
0.783333333333333	18.6534661991189\\
0.8	18.7539282465162\\
0.816666666666667	18.7841216564666\\
0.833333333333333	18.7457125497999\\
0.85	18.6410864225472\\
0.866666666666667	18.4733171478642\\
0.883333333333334	18.2461308088702\\
0.9	17.9638652550297\\
0.916666666666667	17.6314264797638\\
0.933333333333334	17.2542431221654\\
0.95	16.8382205902103\\
0.966666666666667	16.3896964643039\\
0.983333333333333	15.9153989279771\\
1	15.4224099207073\\
};
\addlegendentry{Optimized}

\addplot [color=black, dashed, line width=1.0pt]
  table[row sep=crcr]{%
-1	2\\
1	2\\
};
\addlegendentry{Limit}

\end{axis}

\end{tikzpicture}%
\label{fig:dMiss_sim3}} 
    \caption{\gls{leo} scenario: collision metrics comparison.} 
    \label{fig:ipc_sim13}
\end{figure}

\subsubsection{Comparison of Thrust Systems}
 The same scenarios of the previous section are analyzed using a low-thrust system with a maximum thrust of $50$ \si{mN}, corresponding to a  maximum acceleration of $0.25$ \si{mm/s^2}.   In \cref{fig:dv123567} the maneuvers of the different cases are shown. In the \gls{ipc} case, the total $\dv$ required is $330$ \si{mm/s}, in the $\ipcmax$ case it is $185$ \si{mm/s} and in the $\dmiss$ case $407$ \si{mm/s}. The number of total nodes where thrust is active increases in the three cases from the 2 of the high-thrust system to 13, 7, and 15 for the low-thrust system.  As expected from a fuel-optimal low-thrust solution, the optimizer achieves a bang-bang profile, and the firing windows overlap with the short firings produced by the high-thrust system.

\subsubsection{Return to the Nominal Orbit}
\begin{figure}[b!]
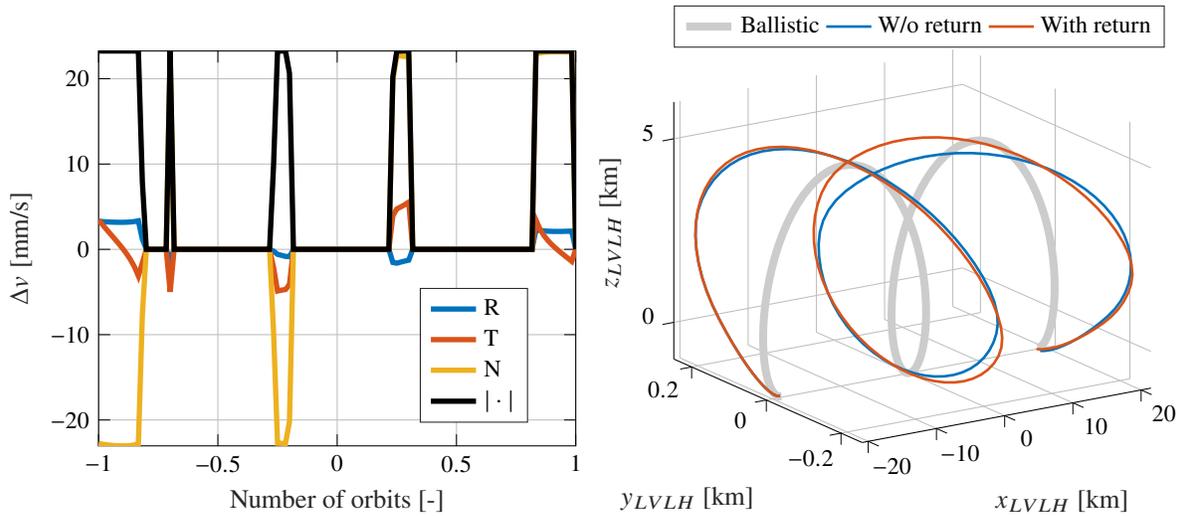

    \centering
    \subfloat[$\dv$ of the return trajectory.]{\input{dvReturn.tex}\label{fig:dvreturn}}
    \subfloat[Relative trajectory without and with return constraint.]{\input{relTrajReturn}\label{fig:reltrajReturn}} 
    \caption{\gls{leo} scenario: case with the return to the nominal orbit constraint.} 
    \label{fig:return}
\end{figure}

After completing reducing the collision risk, it might be required to the primary spacecraft to return to its nominal orbit. This is obtained by constraining the optimized final state to match the final state of the ballistic trajectory. So, in \cref{eq:target}, $\vec{x}_T = \vec{x}_N^0$. The two trajectories, with and without return constraints, are shown in \cref{fig:reltrajReturn}. The thrust profile in \cref{fig:dvreturn} should be compared with the one in \cref{fig:dv5}, which is the corresponding scenario with no return constraint. The first firing is substantially the same, with only a slight modification in the direction of the thrust (radial and along-track components). All of the following firings are used to reroute the spacecraft toward the desired final state. The total $\dv$ increases to  $733$ \si{mm/s}. 

\subsubsection{Squared Mahalanobis Distance Sensitivity Constraint}
\begin{figure}[b!]
    \centering
    \subfloat[\gls{ipc} evolution.]{
%
%
\definecolor{mycolor1}{rgb}{0.00000,0.44700,0.74100}%
\definecolor{mycolor2}{rgb}{0.92941,0.69412,0.12549}%
\definecolor{mycolor3}{rgb}{0.85098,0.32549,0.09804}%
\definecolor{mycolor4}{rgb}{0.49412,0.18431,0.55686}%
\definecolor{mycolor5}{rgb}{1.00000,0.41176,0.16078}%
\begin{tikzpicture}

\begin{axis}[%
at={(0\textwidth,0\textwidth)},
width=0.48\textwidth,
height = 0.36\textwidth,
xmin=0.55,
xmax=0.85,
xlabel style={font=\color{white!15!black}},
xlabel={Number of orbits [-]},
ymode=log,
ymin=4.59914743811156e-08,
ymax=6.81875601133947e-06,
yminorticks=true,
ylabel style={font=\color{white!15!black}},
ylabel={$\ipc$ [-]},
axis background/.style={fill=white},
xmajorgrids,
ymajorgrids,
legend style={at={(0,1)}, anchor=south west, legend cell align=left, align=left, draw=white!15!black},
legend columns = 3,
]
\addplot [color=mycolor1, line width=1.0pt]
  table[row sep=crcr]{%
0.55	7.41370069335787e-07\\
0.566666666666667	8.90357538600221e-07\\
0.583333333333333	1.10821298534829e-06\\
0.6	1.39864588584212e-06\\
0.616666666666667	1.76721786553597e-06\\
0.633333333333333	2.21617270446659e-06\\
0.65	2.73943495220807e-06\\
0.666666666666667	3.31681614541167e-06\\
0.683333333333333	3.90777433552177e-06\\
0.7	4.44635165101515e-06\\
0.716666666666667	4.84062686013225e-06\\
0.733333333333333	4.98191899764521e-06\\
0.75	4.76972629151028e-06\\
0.766666666666667	4.15491828160462e-06\\
0.783333333333333	3.19158637373257e-06\\
0.8	2.06597189545988e-06\\
0.816666666666667	1.05330668492498e-06\\
0.833333333333333	3.81163281394103e-07\\
0.85	8.29028489686564e-08\\
};
\addlegendentry{Ballistic}

\addplot [color=white!80!black, line width=3.0pt]
  table[row sep=crcr]{%
0.55	2.37484306079831e-07\\
0.566666666666667	2.55357642915062e-07\\
0.583333333333333	2.95416200667202e-07\\
0.6	3.53414646531328e-07\\
0.616666666666667	4.28166992813818e-07\\
0.633333333333333	5.18537118227785e-07\\
0.65	6.21882855338475e-07\\
0.666666666666667	7.32779110926303e-07\\
0.683333333333333	8.41888147473576e-07\\
0.7	9.35264758873196e-07\\
0.716666666666667	9.94735706651716e-07\\
0.733333333333333	1.00028705980239e-06\\
0.75	9.35371351746215e-07\\
0.766666666666667	7.9515048962956e-07\\
0.783333333333333	5.95247472475828e-07\\
0.8	3.74769208866714e-07\\
0.816666666666667	1.85337177851589e-07\\
0.833333333333333	6.48170864822225e-08\\
0.85	1.35572122112483e-08\\
};
\addlegendentry{No constraint}

\addplot [color=mycolor2, line width=1.0pt]
  table[row sep=crcr]{%
0.55	2.42918471166469e-07\\
0.566666666666667	2.62569783618006e-07\\
0.583333333333333	3.04296945585846e-07\\
0.6	3.63986863462644e-07\\
0.616666666666667	4.40379958125428e-07\\
0.633333333333333	5.32152832780357e-07\\
0.65	6.36382249447681e-07\\
0.666666666666667	7.47290718674719e-07\\
0.683333333333333	8.55169638663214e-07\\
0.7	9.45785933595679e-07\\
0.716666666666667	1.00091718401608e-06\\
0.733333333333333	1.00092407148761e-06\\
0.75	9.30190662179655e-07\\
0.766666666666667	7.8529670791536e-07\\
0.783333333333333	5.83323582088153e-07\\
0.8	3.64052161460285e-07\\
0.816666666666667	1.78245978639014e-07\\
0.833333333333333	6.1625804448426e-08\\
0.85	1.27201676703671e-08\\
};
\addlegendentry{$\rho = 0.4$}

\addplot [color=mycolor3, line width=1.0pt]
  table[row sep=crcr]{%
0.55	2.16600689185473e-07\\
0.566666666666667	2.43462197158456e-07\\
0.583333333333333	2.89039161282101e-07\\
0.6	3.51346786353818e-07\\
0.616666666666667	4.29907743296269e-07\\
0.633333333333333	5.2372868423679e-07\\
0.65	6.30000627350389e-07\\
0.666666666666667	7.42916253843272e-07\\
0.683333333333333	8.52622591930326e-07\\
0.7	9.4466208996025e-07\\
0.716666666666667	1.00057319003681e-06\\
0.733333333333333	1.00056997798207e-06\\
0.75	9.2911250595311e-07\\
0.766666666666667	7.8316353487584e-07\\
0.783333333333333	5.80418698298114e-07\\
0.8	3.61187415413822e-07\\
0.816666666666667	1.76248839067751e-07\\
0.833333333333333	6.07281807451458e-08\\
0.85	1.25044584957418e-08\\
};
\addlegendentry{$\rho = 0.3$}

\addplot [color=mycolor4, line width=1.0pt]
  table[row sep=crcr]{%
0.55	1.13987761296253e-07\\
0.566666666666667	1.36574764607525e-07\\
0.583333333333333	1.7158453538965e-07\\
0.6	2.20076690449015e-07\\
0.616666666666667	2.83885290558137e-07\\
0.633333333333333	3.6468188168559e-07\\
0.65	4.630603053075e-07\\
0.666666666666667	5.77358627098096e-07\\
0.683333333333333	7.02175466632595e-07\\
0.7	8.26772583686116e-07\\
0.716666666666667	9.33930282569671e-07\\
0.733333333333333	1.00036415777959e-06\\
0.75	1.00034612226217e-06\\
0.766666666666667	9.14090523816552e-07\\
0.783333333333333	7.40536190043549e-07\\
0.8	5.09099540660848e-07\\
0.816666666666667	2.78244256109927e-07\\
0.833333333333333	1.09359606691193e-07\\
0.85	2.6339643346847e-08\\
};
\addlegendentry{$\rho = 0.2$}

\addplot [color=mycolor5, line width=1.0pt]
  table[row sep=crcr]{%
0.55	1.07148153022666e-07\\
0.566666666666667	1.31100564630246e-07\\
0.583333333333333	1.66851088560594e-07\\
0.6	2.15848929673024e-07\\
0.616666666666667	2.80098250402061e-07\\
0.633333333333333	3.61360778865039e-07\\
0.65	4.60278392085481e-07\\
0.666666666666667	5.75202042348394e-07\\
0.683333333333333	7.00704282447177e-07\\
0.7	8.2597740411583e-07\\
0.716666666666667	9.33693379647809e-07\\
0.733333333333333	1.00044515237486e-06\\
0.75	1.00041974654144e-06\\
0.766666666666667	9.13855291214092e-07\\
0.783333333333333	7.39870528934724e-07\\
0.8	5.08167026565462e-07\\
0.816666666666667	2.77407760240552e-07\\
0.833333333333333	1.0889043288927e-07\\
0.85	2.61990033152183e-08\\
};
\addlegendentry{$\rho = 0.1$}

\addplot [color=black, dashed, line width=1.0pt]
  table[row sep=crcr]{%
0.733333	1e-08\\
0.733333	1e-05\\
};
\addplot [color=black, dotted, line width=1.0pt]
  table[row sep=crcr]{%
0	1e-06\\
1	1e-06\\
};

\end{axis}

\end{tikzpicture}%
\label{fig:ipcgrad}}
    \subfloat[$\Delta\ipc$ at a distance of \gls{hbr} from the relative position.]{
%
%
\definecolor{mycolor1}{rgb}{0.92941,0.69412,0.12549}%
\definecolor{mycolor2}{rgb}{0.85098,0.32549,0.09804}%
\definecolor{mycolor3}{rgb}{0.49412,0.18431,0.55686}%
\definecolor{mycolor4}{rgb}{1.00000,0.41176,0.16078}%
\begin{tikzpicture}

\begin{axis}[%
at={(0\textwidth,0\textwidth)},
width=0.48\textwidth,
xmin=0.55,
xmax=0.85,
xlabel style={font=\color{white!15!black}},
xlabel={Number of orbits [-]},
ymode=log,
ymin=3.25313356521552e-08,
ymax=5.65473797619583e-07,
yminorticks=true,
ylabel style={font=\color{white!15!black}},
ylabel={$\Delta\ipc$ [-]},
axis background/.style={fill=white},
xmajorgrids,
ymajorgrids,
yminorgrids,
]
\addplot [color=white!80!black, line width=3.0pt]
  table[row sep=crcr]{%
0.55	2.00125232152333e-07\\
0.566666666666667	1.81190511484855e-07\\
0.583333333333333	1.83820603778636e-07\\
0.6	1.98747006376341e-07\\
0.616666666666667	2.21911724255263e-07\\
0.633333333333333	2.5173460899641e-07\\
0.65	2.86566888350916e-07\\
0.666666666666667	3.24255986826553e-07\\
0.683333333333333	3.61685765513359e-07\\
0.7	3.94481875976822e-07\\
0.716666666666667	4.16955226412045e-07\\
0.733333333333333	4.22465988762156e-07\\
0.75	4.04485420037258e-07\\
0.766666666666667	3.58668052754806e-07\\
0.783333333333333	2.86015236976903e-07\\
0.8	1.96224578673877e-07\\
0.816666666666667	1.0818997282162e-07\\
0.833333333333333	4.30716129381852e-08\\
0.85	1.04140619995071e-08\\
};

\addplot [color=mycolor1, line width=1.0pt]
  table[row sep=crcr]{%
0.55	1.68603787229458e-07\\
0.566666666666667	1.55311640338511e-07\\
0.583333333333333	1.59934301909267e-07\\
0.6	1.74320890555e-07\\
0.616666666666667	1.96854584288336e-07\\
0.633333333333333	2.25331732380225e-07\\
0.65	2.58721881719285e-07\\
0.666666666666667	2.95236111836839e-07\\
0.683333333333333	3.32177372029243e-07\\
0.7	3.65649493949504e-07\\
0.716666666666667	3.9045415745016e-07\\
0.733333333333333	4.00331553888092e-07\\
0.75	3.88793073271048e-07\\
0.766666666666667	3.50873694794921e-07\\
0.783333333333333	2.86045763260558e-07\\
0.8	2.01792503666446e-07\\
0.816666666666667	1.15255845430921e-07\\
0.833333333333333	4.79947176138493e-08\\
0.85	1.23006099161136e-08\\
};

\addplot [color=mycolor2, line width=1.0pt]
  table[row sep=crcr]{%
0.55	9.20043732103685e-08\\
0.566666666666667	9.03297402673516e-08\\
0.583333333333333	9.71330066089194e-08\\
0.6	1.09892378252489e-07\\
0.616666666666667	1.27621952643798e-07\\
0.633333333333333	1.49785885854164e-07\\
0.65	1.75753008935292e-07\\
0.666666666666667	2.04471623701474e-07\\
0.683333333333333	2.34231581248671e-07\\
0.7	2.62508175403225e-07\\
0.716666666666667	2.85922692371092e-07\\
0.733333333333333	3.00349199603613e-07\\
0.75	3.01179367359784e-07\\
0.766666666666667	2.83852123489424e-07\\
0.783333333333333	2.45144860612816e-07\\
0.8	1.86102562448611e-07\\
0.816666666666667	1.1618129374763e-07\\
0.833333333333333	5.37000164872166e-08\\
0.85	1.55440839799084e-08\\
\\
};

\addplot [color=mycolor3, line width=1.0pt]
  table[row sep=crcr]{%
0.55	1.0830216407688e-07\\
0.566666666666667	1.05588148413413e-07\\
0.583333333333333	1.10399793633651e-07\\
0.6	1.20756594062112e-07\\
0.616666666666667	1.33948726310335e-07\\
0.633333333333333	1.48980871703754e-07\\
0.65	1.64505726370775e-07\\
0.666666666666667	1.79035368689658e-07\\
0.683333333333333	1.91029040915239e-07\\
0.7	1.99132268965373e-07\\
0.716666666666667	2.02427485888643e-07\\
0.733333333333333	2.00375280260739e-07\\
0.75	1.92113219971291e-07\\
0.766666666666667	1.75493212634589e-07\\
0.783333333333333	1.47570197658176e-07\\
0.8	1.08057556549322e-07\\
0.816666666666667	6.36276314348147e-08\\
0.833333333333333	2.68733800373011e-08\\
0.85	6.83821203366459e-09\\
};

\addplot [color=mycolor4, line width=1.0pt]
  table[row sep=crcr]{%
0.55	3.65241668115065e-08\\
0.566666666666667	3.97069103551188e-08\\
0.583333333333333	4.55999182119301e-08\\
0.6	5.36726005325071e-08\\
0.616666666666667	6.32029105102053e-08\\
0.633333333333333	7.35186023919008e-08\\
0.65	8.41657542192163e-08\\
0.666666666666667	9.3329377898865e-08\\
0.683333333333333	9.96174472631316e-08\\
0.7	1.01767790421406e-07\\
0.716666666666667	9.99820656783944e-08\\
0.733333333333333	9.7555632065978e-08\\
0.75	1.00342778581716e-07\\
0.766666666666667	1.09596558142825e-07\\
0.783333333333333	1.15992106810042e-07\\
0.8	1.06800818357048e-07\\
0.816666666666667	7.8031828531133e-08\\
0.833333333333333	4.06761227817042e-08\\
0.85	1.28913610576833e-08\\
};

\addplot [color=black, dotted, line width=1.0pt]
  table[row sep=crcr]{%
0	4e-07\\
1	4e-07\\
};

\addplot [color=black, dotted, line width=1.0pt]
  table[row sep=crcr]{%
-0	3e-07\\
1	3e-07\\
};

\addplot [color=black, dotted, line width=1.0pt]
  table[row sep=crcr]{%
0	2e-07\\
1	2e-07\\
};
\addplot [color=black, dotted, line width=1.0pt]
  table[row sep=crcr]{%
-0	1e-07\\
1	1e-07\\
};

\addplot [color=black, dashed, line width=1.0pt]
  table[row sep=crcr]{%
0.733333	1e-12\\
0.733333	0.0001\\
};

\end{axis}
\end{tikzpicture}%
\label{fig:dIpc}} 
    \caption{\gls{leo} scenario: collision metrics for cases with the sensitivity constraint.} 
    \label{fig:ipcgradsmd}
\end{figure}
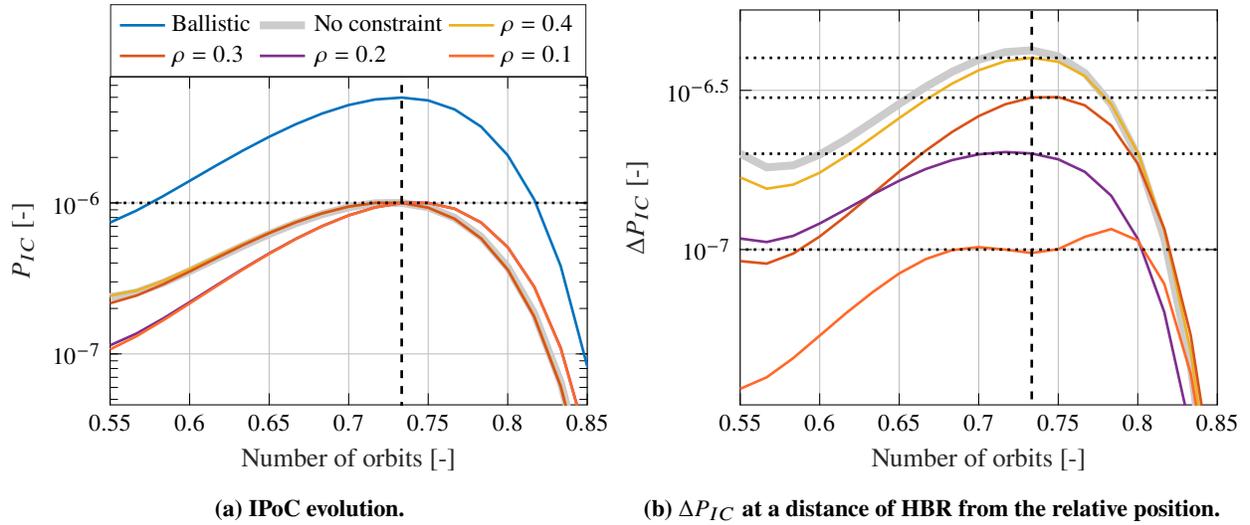

A scenario starting from \gls{tca} and spanning the period of 1 orbit is used to analyze the influence of the \gls{smd} sensitivity constraint on the maneuver.
In \cref{fig:ipcgrad}, \gls{ipc} for the cases with and without the \gls{smd} sensitivity constraint are shown. 
In \cref{fig:ipcgrad} a zoom of the region in which the \gls{ipc} of the ballistic trajectory is over the limit is shown. The maximum value of \gls{ipc} happens at $0.73333$ orbits from \gls{tca}. All four maneuvers have the effect of lowering the \gls{ipc} maximum value to $10^{-6}$; the increasingly constraining value of $\overline{\Delta P}_{IC}$, though, bounds $\Delta\ipc$ to increasingly lower values. Note that the \gls{smd} sensitivity constraint is applied only at $0.73333$ orbits, since it is the only node in which \gls{ipc} is high enough to trigger its activation. Indeed, in \cref{fig:dIpc}, $\Delta\ipc$ computed in the direction opposite to $\nabla\smd$ respects the limit in all cases. This limit is respectively $4\times 10^{-7}$, $3\times 10^{-7}$, $2\times 10^{-7}$, and $10^{-7}$.

The total $\dv$ required with the constraint's introduction increases when the safety margin increases. From the unconstrained case to the case with $\rho=0.1$, the $\dv$ goes from $257$ to $265$, $387$, $507$, and $580$ \si{mm/s}. The maneuver is always a single-firing applied at the first node and the tangential component becomes more and more dominant.

\subsection{GEO Scenario}
A \gls{geo} test case is used to demonstrate the effectiveness of the \gls{scp} algorithm in another orbital regime. Simulations are performed considering a high thrust of $2.5$ \si{N}, which corresponds to a maximum acceleration of $5$ \si{mm/s^2} for a $500$ \si{kg} spacecraft. The orbital period is discretized into $60$ nodes, so the maximum impulsive $\dv$ between successive nodes is $7.18$ \si{m/s}.
For the low-thrust system the maximum thrust is $2.5$ \si{mN}, i.e. a maximum acceleration of $5$ \si{\mu m/s^2}.

In \cref{tab:initialConditionsGeo}, the orbital parameters of the two spacecraft at \gls{tca} are represented. The two orbits are close to a perfect \gls{geo}, with a slight inclination and a difference in the semi-major axis of $1.5$ \si{km}. The covariance of the two spacecraft is the same as in \cref{tab:covarianceLeo}. In \cref{tab:paramsGeo} the physical properties of the spacecraft are shown. The only relevant perturbations in the \gls{geo} regime are higher order gravitational harmonics, \gls{srp}, and third body attraction, so no information is required on the equivalent drag surface area and $C_D$ coefficient.

\begin{figure}[tb!]
    \centering
    \subfloat[\gls{ipc} and $\ipcmax$ of the ballistic and optimized trajectories.]{\input{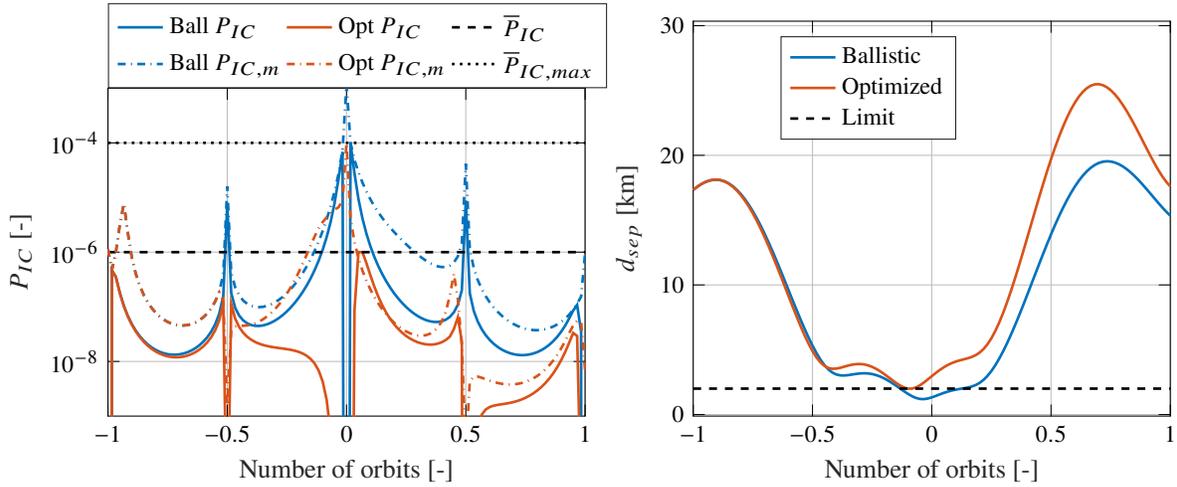}\label{fig:ipc_geo}}
    \subfloat[$\dmiss$ of the ballistic and optimized trajectories.]{
%
%
\definecolor{mycolor1}{rgb}{0.00000,0.44700,0.74100}%
\definecolor{mycolor2}{rgb}{0.85000,0.32500,0.09800}%
\begin{tikzpicture}

\begin{axis}[%
at={(0\textwidth,0\textwidth)},
width=0.48\textwidth,
xmin=-1,
xmax=1,
xlabel style={font=\color{white!15!black}},
xlabel={Number of orbits [-]},
ymin=-0.118364261953803,
ymax=30.3477776124905,
ylabel style={font=\color{white!15!black}},
ylabel={$\dmiss$ [\si{km}]},
axis background/.style={fill=white},
xmajorgrids,
ymajorgrids,
legend style={at={(0.183,0.7)}, anchor=south west, legend cell align=left, align=left, draw=white!15!black}
]
\addplot [color=mycolor1, line width=1.0pt]
  table[row sep=crcr]{%
-1	17.3292426766842\\
-0.983333333333333	17.5677336518869\\
-0.966666666666667	17.7689686820162\\
-0.95	17.9284933510307\\
-0.933333333333333	18.0423168316301\\
-0.916666666666667	18.1069452699559\\
-0.9	18.1194098558482\\
-0.883333333333333	18.0772989332755\\
-0.866666666666667	17.9787775871096\\
-0.85	17.8226099995322\\
-0.833333333333333	17.6081739979356\\
-0.816666666666667	17.3354677584486\\
-0.8	17.0051162248195\\
-0.783333333333334	16.6183693899028\\
-0.766666666666667	16.1770905852965\\
-0.75	15.6837490288386\\
-0.733333333333333	15.1414152226246\\
-0.716666666666667	14.5537146975364\\
-0.7	13.9248123766975\\
-0.683333333333333	13.259385512146\\
-0.666666666666667	12.5625951810453\\
-0.65	11.8400611778789\\
-0.633333333333333	11.0978333758081\\
-0.616666666666667	10.3423756637649\\
-0.6	9.58055701233769\\
-0.583333333333333	8.81965516425145\\
-0.566666666666667	8.06738579892241\\
-0.55	7.33195638697598\\
-0.533333333333333	6.62214965784426\\
-0.516666666666667	5.94743931266923\\
-0.5	5.31810531578705\\
-0.483333333333333	4.74527150615185\\
-0.466666666666667	4.24069318530038\\
-0.45	3.81596303991713\\
-0.433333333333333	3.48074682766053\\
-0.416666666666667	3.23995162735795\\
-0.4	3.09071413956742\\
-0.383333333333333	3.0211381639419\\
-0.366666666666667	3.01218925059545\\
-0.35	3.04187321149047\\
-0.333333333333333	3.08931100290698\\
-0.316666666666667	3.13713282169036\\
-0.3	3.17217907343256\\
-0.283333333333333	3.18521192260553\\
-0.266666666666667	3.17028402555519\\
-0.25	3.12409496735652\\
-0.233333333333333	3.04545864381819\\
-0.216666666666667	2.93491280518032\\
-0.2	2.79444415431869\\
-0.183333333333333	2.62732320077629\\
-0.166666666666667	2.43804288918934\\
-0.15	2.23236315515978\\
-0.133333333333333	2.01749635244148\\
-0.116666666666667	1.80246857111332\\
-0.1	1.59864250142342\\
-0.0833333333333333	1.4201750597002\\
-0.0666666666666667	1.28351830633334\\
-0.05	1.20415082958388\\
-0.0333333333333333	1.18980304211897\\
-0.0166666666666667	1.23488598178208\\
0	1.32288417323933\\
0.0166666666666667	1.43430492114707\\
0.0333333333333333	1.55256629467374\\
0.05	1.66577354406203\\
0.0666666666666667	1.76645280858621\\
0.0833333333333333	1.85096662255576\\
0.1	1.91920548497527\\
0.116666666666667	1.97461748707899\\
0.133333333333333	2.02445029053575\\
0.15	2.07991011621881\\
0.166666666666667	2.15578500365967\\
0.183333333333333	2.26899626732444\\
0.2	2.43592383636208\\
0.216666666666667	2.66931089656183\\
0.233333333333333	2.97622139620675\\
0.25	3.35794365406652\\
0.266666666666667	3.8114821313421\\
0.283333333333333	4.33127292262916\\
0.3	4.91050363521082\\
0.316666666666667	5.54192674388833\\
0.333333333333333	6.21822723117743\\
0.35	6.93215936745747\\
0.366666666666667	7.67657425390331\\
0.383333333333333	8.44441900264657\\
0.4	9.22872790452218\\
0.416666666666667	10.0226248209436\\
0.433333333333333	10.819339958926\\
0.45	11.6122331782025\\
0.466666666666667	12.3948282072752\\
0.483333333333333	13.1608556801245\\
0.5	13.9042955158156\\
0.516666666666667	14.6194237787107\\
0.533333333333333	15.3008628641348\\
0.55	15.9436242037815\\
0.566666666666667	16.5431555533664\\
0.583333333333333	17.0953825409729\\
0.6	17.5967435261611\\
0.616666666666667	18.0442267358983\\
0.633333333333333	18.4353955427846\\
0.65	18.768411056255\\
0.666666666666667	19.0420514882191\\
0.683333333333333	19.2557203383642\\
0.7	19.4094510375332\\
0.716666666666667	19.5039081131712\\
0.733333333333333	19.5403772378789\\
0.75	19.5207522026119\\
0.766666666666667	19.4475195024245\\
0.783333333333334	19.3237309215539\\
0.8	19.1529806007248\\
0.816666666666667	18.9393713892853\\
0.833333333333333	18.6874813348737\\
0.85	18.4023307744906\\
0.866666666666667	18.0893429431088\\
0.883333333333333	17.7543064627645\\
0.9	17.4033400141297\\
0.916666666666667	17.0428513120603\\
0.933333333333333	16.6794974849938\\
0.95	16.3201456920123\\
0.966666666666667	15.9718215417337\\
0.983333333333333	15.6416578367493\\
1	15.3368236983996\\
};
\addlegendentry{Ballistic}

\addplot [color=mycolor2, line width=1.0pt]
  table[row sep=crcr]{%
-1	17.3292426766842\\
-0.983333333333333	17.5687678637495\\
-0.966666666666667	17.772464077413\\
-0.95	17.9347365316244\\
-0.933333333333333	18.0503915070785\\
-0.916666666666667	18.1146847275065\\
-0.9	18.1238197365777\\
-0.883333333333333	18.0751852563707\\
-0.866666666666667	17.9669053594656\\
-0.85	17.7977452185154\\
-0.833333333333333	17.5671283343874\\
-0.816666666666667	17.2751459185121\\
-0.8	16.9225660785661\\
-0.783333333333334	16.5108351054998\\
-0.766666666666667	16.042068948497\\
-0.75	15.5190498951626\\
-0.733333333333333	14.9452285435925\\
-0.716666666666667	14.3246839595225\\
-0.7	13.6621174147531\\
-0.683333333333333	12.9628391329309\\
-0.666666666666667	12.2327586185398\\
-0.65	11.4783858871758\\
-0.633333333333333	10.7045827882405\\
-0.616666666666667	9.91706773390753\\
-0.6	9.12478913190459\\
-0.583333333333333	8.33754575842769\\
-0.566666666666667	7.56610303651211\\
-0.55	6.82235564106881\\
-0.533333333333333	6.1195170222876\\
-0.516666666666667	5.47226045278861\\
-0.5	4.89659192313291\\
-0.483333333333333	4.40902244274601\\
-0.466666666666667	4.02441764135276\\
-0.45	3.75217097153202\\
-0.433333333333333	3.59169854392227\\
-0.416666666666667	3.53002626015699\\
-0.4	3.54459813211506\\
-0.383333333333333	3.60818125404554\\
-0.366666666666667	3.69417375639136\\
-0.35	3.78052865495252\\
-0.333333333333333	3.85061203713808\\
-0.316666666666667	3.89276970807845\\
-0.3	3.89948955592057\\
-0.283333333333333	3.86659150453352\\
-0.266666666666667	3.79260364275447\\
-0.25	3.67833420635926\\
-0.233333333333333	3.52662202269815\\
-0.216666666666667	3.34226023947847\\
-0.2	3.13208263781556\\
-0.183333333333333	2.90523655609389\\
-0.166666666666667	2.67364352547662\\
-0.15	2.44831700253445\\
-0.133333333333333	2.24484454203942\\
-0.116666666666667	2.08724258539763\\
-0.1	1.99990968463266\\
-0.0833333333333333	1.99982685882968\\
-0.0666666666666667	2.08851750017959\\
-0.05	2.25150042832506\\
-0.0333333333333333	2.46585974473327\\
-0.0166666666666667	2.70842778620667\\
0	2.95986995467374\\
0.0166666666666667	3.2054897455749\\
0.0333333333333333	3.43479414540747\\
0.05	3.64093001231616\\
0.0666666666666667	3.81864443397282\\
0.0833333333333333	3.96511033473769\\
0.1	4.0823940624617\\
0.116666666666667	4.17628292980722\\
0.133333333333333	4.25655182125016\\
0.15	4.33710892057835\\
0.166666666666667	4.43572486146189\\
0.183333333333333	4.57297659443763\\
0.2	4.77016474981177\\
0.216666666666667	5.04643304447676\\
0.233333333333333	5.41591964075777\\
0.25	5.88601486426607\\
0.266666666666667	6.45738657477275\\
0.283333333333333	7.12521925516107\\
0.3	7.88096504763054\\
0.316666666666667	8.71396973749345\\
0.333333333333333	9.61260628074085\\
0.35	10.5649584395247\\
0.366666666666667	11.559179041986\\
0.383333333333333	12.5836210423042\\
0.4	13.6272395190517\\
0.416666666666667	14.679282256347\\
0.433333333333333	15.7292003596753\\
0.45	16.7667095546195\\
0.466666666666667	17.7818567028694\\
0.483333333333333	18.7650922389469\\
0.5	19.7073401855712\\
0.516666666666667	20.6000713446923\\
0.533333333333333	21.4353783217639\\
0.55	22.2060410971499\\
0.566666666666667	22.9058269568068\\
0.583333333333333	23.5293090083801\\
0.6	24.0718027648632\\
0.616666666666667	24.5295849814646\\
0.633333333333333	24.8999196886553\\
0.65	25.1810794002758\\
0.666666666666667	25.3723607443035\\
0.683333333333333	25.474086520692\\
0.7	25.4876018155468\\
0.716666666666667	25.4152643126882\\
0.733333333333333	25.2604214075361\\
0.75	25.0273823393957\\
0.766666666666667	24.7213862781885\\
0.783333333333334	24.3485571295103\\
0.8	23.915861585303\\
0.816666666666667	23.4310554831505\\
0.833333333333333	22.902629072044\\
0.85	22.3397511341018\\
0.866666666666667	21.7522153327005\\
0.883333333333333	21.1507632736234\\
0.9	20.5473692916333\\
0.916666666666667	19.9545511033273\\
0.933333333333333	19.3851191571728\\
0.95	18.8519891405411\\
0.966666666666667	18.3679267411092\\
0.983333333333333	17.945232982865\\
1	17.5953570543064\\
};
\addlegendentry{Optimized}

\addplot [color=black, dashed, line width=1.0pt]
  table[row sep=crcr]{%
-1	2\\
1	2\\
};
\addlegendentry{Limit}

\end{axis}

\end{tikzpicture}%
\label{fig:dMiss_geo}} 
    \caption{\gls{geo} scenario: collision metrics comparison.} 
    \label{fig:metrics_geo}
\end{figure}

In \cref{fig:metrics_geo}, the evolution of the three collision metrics for the considered scenario are presented. Similarly to the \gls{leo} scenario, $\ipcmax$ always provides an upper boundary to \gls{ipc}. Contrarily to the \gls{leo} scenario, in this case \gls{ipc} is close to the maximum when $\dmiss$ is close to the minimum. In general, all the computed maneuvers can easily lower the required risk metric.

\begin{table}[tb!]
\centering
\caption{\gls{geo} scenario: orbit parameters at \gls{tca}.}
\label{tab:initialConditionsGeo}
    \begin{tabular}{lllllll}
    \Xhline{4\arrayrulewidth}
    \textbf{Spacecraft} & ${a}$ {[\si{km}]} & ${e}$ {[-]}  & ${i}$ {[\si{deg}]} & ${\omega}$ {[\si{deg}]} & ${\Omega}$ {[\si{deg}]} &  ${\theta}$ {[\si{deg}]}   \\
    \Xhline{3\arrayrulewidth}
    Primary   & $42166.025929$ & $5.6\times10^{-5}$                 & $0.118864$                 & $241.585377$      & $76.181310$ & $119.443522$        \\ \hline 
    Secondary & $42167.765854$ & $1.118\times10^{-4}$ & $0.118860$  & $268.823470$ & $76.181269$ & $92.207487$  \\
    \Xhline{4\arrayrulewidth}
    \end{tabular}
\end{table}

\begin{table}[tb!]
\centering
\caption{\gls{geo} scenario: physical properties of the spacecraft.}
\label{tab:paramsGeo}
    \begin{tabular}{lllll}
    \Xhline{4\arrayrulewidth}
   \textbf{Spacecraft} & ${m}$ {[\si{kg}]} & ${A_{SRP}}$ {[\si{m^2}]} & ${C_r}$ {[-]} & \gls{hbr} [\si{m}] \\
    \Xhline{3\arrayrulewidth}
    Primary   & $500$ & $1$    & $1.31$ & $35$ \\ \hline 
    Secondary & $200$ & $1.2$ & $1.31$ & $10$\\
    \Xhline{4\arrayrulewidth}
    \end{tabular}
\end{table}

\subsubsection{Station-Keeping Constraint and Targeting}
\begin{figure}
    \centering
    \input{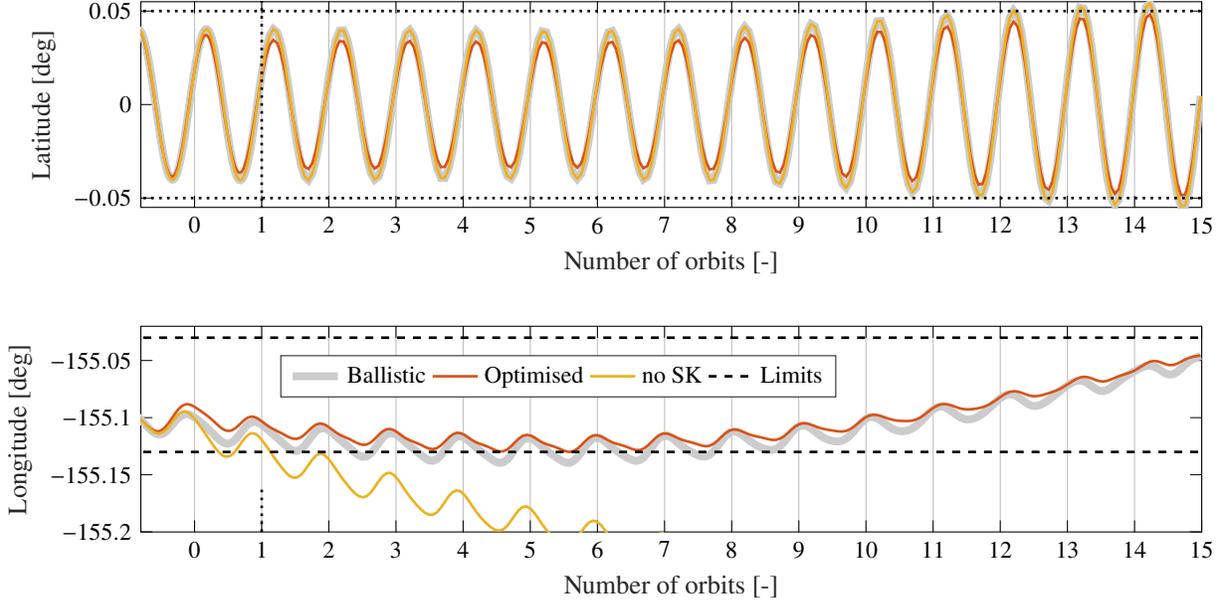}
    \caption{\gls{geo} scenario: latitude and longitude evolution with and without the \gls{sk} requirement.}
    \label{fig:skGeo}
\end{figure}

The primary spacecraft is commanded to always stay inside the latitude-longitude \gls{sk} square box of side $\Delta\phi = 0.05$ \si{deg} around the nominal values of latitude ($0$ \si{deg}) and longitude ($-155.08$ \si{deg}). Moreover, the optimal final state target is computed so that the \gls{sk} requirement is respected by the natural motion of the satellite for the following $14$ days. This requirement is respected as shown in \cref{fig:skGeo}. For comparison, a scenario in which the \gls{sk} constraint is not enforced is also reported in \cref{fig:skGeo}; in that case the maneuver causes an even more serious drift on the longitude than that of the ballistic trajectory. 

\begin{figure}[tb!]
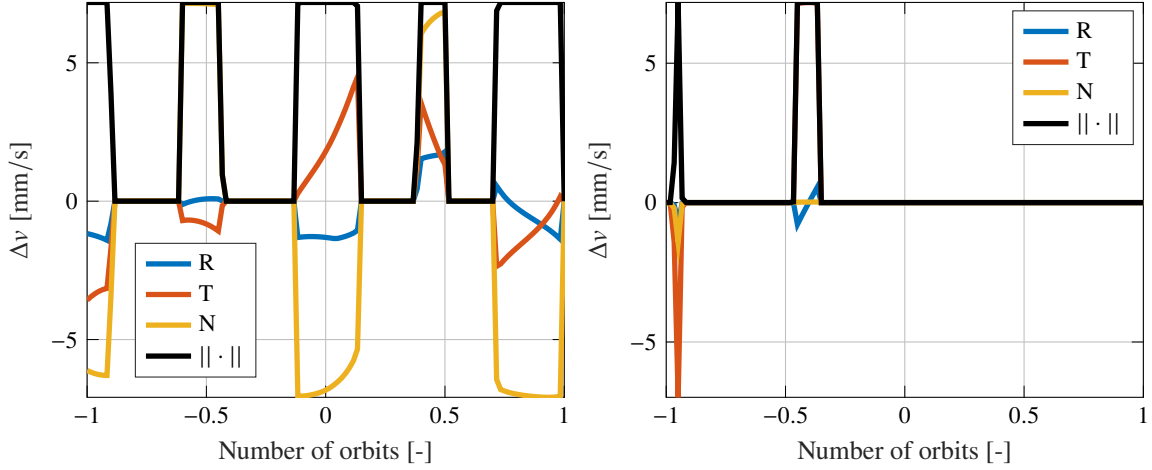
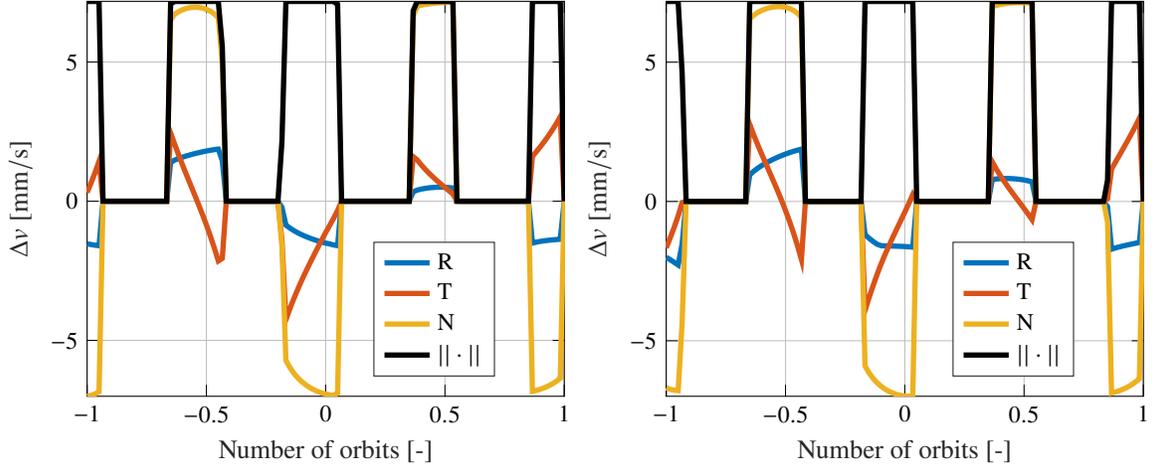

    \centering
    \subfloat[$\dv$ of the \gls{ipc} simulation, low-thrust system.]{\input{dv_sim18.tex}\label{fig:dv18}}
    \subfloat[$\dv$ of the \gls{ipc} simulation with no \gls{sk}, low-thrust system.]{\input{dv_sim25.tex}\label{fig:dv25}} \\
    \subfloat[$\dv$ of the $\ipcmax$ simulation, low-thrust system.]{\input{dv_sim23.tex}\label{fig:dv23}} 
    \subfloat[$\dv$ of the $\dmiss$ simulation, high-thrust system.]{\input{dv_sim24.tex}\label{fig:dv24}} 
    \caption{\gls{geo} Scenario: $\dv$ for different collision metrics and propulsion systems.}
    \label{fig:dvGeo}
\end{figure}

The \gls{sk} constraint completely changes the direction of the maneuver. In fact, the \gls{sk} maneuver is mostly an out-of-plane correction of the inclination, which allows keeping the latitude inside the box \cite{Pavanello2023Long}. The pure \gls{cam} in \cref{fig:dv25}, instead, is an almost purely tangential maneuver. Also, the entity of the total $\dv$ is significantly changed by the constraint, going from $52$ \si{mm/s} in the pure \gls{cam} case to $409$ \si{mm/s} in the case with \gls{sk}. Given the predominance of the \gls{sk} constraint, the maneuver is only slightly modified by the use of different collision metrics: \cref{fig:dv23} and \cref{fig:dv24} are very similar to \cref{fig:dv18}; the total $\dv$ for the case with $\ipcmax$ is $366$ \si{mm/s} and for the case with $\dmiss$ is $364$ \si{mm/s}.

\subsubsection{Squared Mahalanobis Distance Sensitivity Constraint}
The \gls{smd} sensitivity constraint is applied to a \gls{geo} case with no \gls{sk} target. 
The inclusion of the constraint completely shifts the maneuver. While in 
The \gls{socp} does not guarantee to find the global optimum of the original \gls{ocp} because the found solution could fall in a local optimum well. In fact, counter-intuitively the $\dv$ required by the robust maneuver is lower than the one where the constraint was not applied ($37$ \si{mm/s} vs $51$ \si{mm/s}). As pointed out in \cite{Armellin2021}, the \gls{socp} can only recover a local optimum solution depending on the initial reference solution. In other words, if a higher-cost solution falls closer to the initial ballistic trajectory, the solver is likely to find it and miss the global optimum. 
Nonetheless, the solver can still recover the optimal solution that is found in the case where the sensitivity constraint is employed. If the solution of the case with the sensitivity constraint is taken as reference and the \gls{socp} is run without the constraint, the solution found is close to the reference, with a slightly lower $\dv$, $34$ \si{mm/s}.


\subsection{HEO scenario}
\begin{figure}[tb!]
    \centering
%
%
\definecolor{mycolor1}{rgb}{0.00000,0.44700,0.74100}%
\definecolor{mycolor2}{rgb}{0.85000,0.32500,0.09800}%
\begin{tikzpicture}

\begin{axis}[%
at={(0\textwidth,1.7in)},
width=0.95\textwidth,
height = 2in,
xmin=-0.5,
xmax=0.5,
xlabel style={font=\color{white!15!black}},
xlabel={Number of orbits [-]},
ymode=log,
ymin=1e-07,
ymax=0.5,
yminorticks=true,
ylabel style={font=\color{white!15!black}},
ylabel={$P_{IC}$ [-]},
axis background/.style={fill=white},
xmajorgrids,
ymajorgrids,
yminorgrids,
legend style={at={(0.8,0.4)}, anchor=south west, legend cell align=left, align=left, draw=white!15!black}
]
\addplot [color=mycolor1, line width=1.0pt]
  table[row sep=crcr]{%
-0.5	0\\
-0.491666666666667	0\\
-0.483333333333333	0\\
-0.475	0\\
-0.466666666666667	0\\
-0.458333333333333	0\\
-0.45	0\\
-0.441666666666667	0\\
-0.433333333333333	0\\
-0.425	0\\
-0.416666666666667	0\\
-0.408333333333333	0\\
-0.4	0\\
-0.391666666666667	0\\
-0.383333333333333	0\\
-0.375	0\\
-0.366666666666667	0\\
-0.358333333333333	0\\
-0.35	0\\
-0.341666666666667	6.66133814775094e-16\\
-0.333333333333333	2.53752574508326e-12\\
-0.325	1.76638881299596e-09\\
-0.316666666666667	2.90729272323098e-07\\
-0.308333333333333	1.46455542204826e-05\\
-0.3	0.000279344065763909\\
-0.291666666666667	0.00240981249745631\\
-0.283333333333333	0.0109419760669391\\
-0.275	0.0298824338863424\\
-0.266666666666667	0.0554331274974753\\
-0.258333333333333	0.0782164852795216\\
-0.25	0.0929462567855986\\
-0.241666666666667	0.100901620290191\\
-0.233333333333333	0.105490830125809\\
-0.225	0.109034112237251\\
-0.216666666666667	0.112508514750006\\
-0.208333333333333	0.116218709804023\\
-0.2	0.120245735257468\\
-0.191666666666667	0.124612363457463\\
-0.183333333333333	0.129325788694065\\
-0.175	0.134385453741692\\
-0.166666666666667	0.139783381404415\\
-0.158333333333333	0.145503228298416\\
-0.15	0.15151922634417\\
-0.141666666666667	0.157795272513025\\
-0.133333333333333	0.164284314156456\\
-0.125	0.170928175158651\\
-0.116666666666667	0.177657952365928\\
-0.108333333333333	0.184395078788401\\
-0.1	0.191053090408289\\
-0.0916666666666667	0.197540061673198\\
-0.0833333333333333	0.203761598056296\\
-0.075	0.209624195836909\\
-0.0666666666666667	0.215038721505195\\
-0.0583333333333333	0.219923734421175\\
-0.05	0.224208380278481\\
-0.0416666666666667	0.227834631725475\\
-0.0333333333333333	0.230758721256679\\
-0.025	0.232951711644275\\
-0.0166666666666667	0.234399236264267\\
-0.00833333333333333	0.235100530365647\\
0	0.23506692531292\\
0.00833333333333333	0.234320007499904\\
0.0166666666666667	0.232889641254042\\
0.025	0.230812022935704\\
0.0333333333333333	0.228127892677737\\
0.0416666666666667	0.224872342257087\\
0.05	0.219986594222112\\
0.0583333333333333	0.195352263945311\\
0.0666666666666667	0.116457315764729\\
0.075	0.036915613628998\\
0.0833333333333333	0.00636229240109165\\
0.0916666666666667	0.000680356445311392\\
0.1	5.17798298831362e-05\\
0.108333333333333	3.13875683710663e-06\\
0.116666666666667	1.64970688221189e-07\\
0.125	7.99413391039394e-09\\
0.133333333333333	3.72899711109653e-10\\
0.141666666666667	1.72484249105764e-11\\
0.15	8.07021116600026e-13\\
0.158333333333333	3.8746783559418e-14\\
0.166666666666667	1.88737914186277e-15\\
0.175	1.11022302462516e-16\\
0.183333333333333	0\\
0.191666666666667	0\\
0.2	0\\
0.208333333333333	0\\
0.216666666666667	0\\
0.225	0\\
0.233333333333333	0\\
0.241666666666667	0\\
0.25	0\\
0.258333333333333	0\\
0.266666666666667	0\\
0.275	0\\
0.283333333333333	0\\
0.291666666666667	0\\
0.3	0\\
0.308333333333333	0\\
0.316666666666667	0\\
0.325	0\\
0.333333333333333	0\\
0.341666666666667	0\\
0.35	0\\
0.358333333333333	0\\
0.366666666666667	0\\
0.375	0\\
0.383333333333333	0\\
0.391666666666667	0\\
0.4	0\\
0.408333333333333	0\\
0.416666666666667	0\\
0.425	0\\
0.433333333333333	0\\
0.441666666666667	0\\
0.45	0\\
0.458333333333333	0\\
0.466666666666667	0\\
0.475	0\\
0.483333333333333	0\\
0.491666666666667	0\\
0.5	0\\
};
\addlegendentry{Ballistic}

\addplot [color=mycolor2, line width=1.0pt]
  table[row sep=crcr]{%
-0.5	0\\
-0.491666666666667	0\\
-0.483333333333333	0\\
-0.475	0\\
-0.466666666666667	0\\
-0.458333333333333	0\\
-0.45	0\\
-0.441666666666667	0\\
-0.433333333333333	0\\
-0.425	0\\
-0.416666666666667	0\\
-0.408333333333333	0\\
-0.4	0\\
-0.391666666666667	0\\
-0.383333333333333	0\\
-0.375	0\\
-0.366666666666667	0\\
-0.358333333333333	0\\
-0.35	0\\
-0.341666666666667	0\\
-0.333333333333333	0\\
-0.325	0\\
-0.316666666666667	0\\
-0.308333333333333	0\\
-0.3	3.21964677141295e-15\\
-0.291666666666667	2.53019827312073e-13\\
-0.283333333333333	9.90985071780415e-12\\
-0.275	2.15297113470569e-10\\
-0.266666666666667	2.84864398736318e-09\\
-0.258333333333333	2.48006000180823e-08\\
-0.25	1.5146542042821e-07\\
-0.241666666666667	6.84194967792706e-07\\
-0.233333333333333	2.38784545125714e-06\\
-0.225	6.67270283838395e-06\\
-0.216666666666667	1.53677592519275e-05\\
-0.208333333333333	2.98444144657406e-05\\
-0.2	4.97304508717544e-05\\
-0.191666666666667	7.19825723701772e-05\\
-0.183333333333333	9.11649574639739e-05\\
-0.175	0.000101226231408069\\
-0.166666666666667	9.81641859951754e-05\\
-0.158333333333333	8.22433435849756e-05\\
-0.15	5.83589199674162e-05\\
-0.141666666666667	3.39513501511624e-05\\
-0.133333333333333	1.53764552563995e-05\\
-0.125	4.98257002556901e-06\\
-0.116666666666667	9.97800234436852e-07\\
-0.108333333333333	9.3302056414224e-08\\
-0.1	2.33686092698804e-09\\
-0.0916666666666667	3.83366671741214e-11\\
-0.0833333333333333	6.94371331189103e-06\\
-0.075	6.09940690843302e-07\\
-0.0666666666666667	9.30933663489952e-10\\
-0.0583333333333333	7.41295913542217e-13\\
-0.05	1.11022302462516e-16\\
-0.0416666666666667	0\\
-0.0333333333333333	0\\
-0.025	0\\
-0.0166666666666667	0\\
-0.00833333333333333	0\\
0	0\\
0.00833333333333333	0\\
0.0166666666666667	0\\
0.025	0\\
0.0333333333333333	0\\
0.0416666666666667	0\\
0.05	0\\
0.0583333333333333	0\\
0.0666666666666667	0\\
0.075	0\\
0.0833333333333333	0\\
0.0916666666666667	0\\
0.1	0\\
0.108333333333333	0\\
0.116666666666667	0\\
0.125	0\\
0.133333333333333	0\\
0.141666666666667	0\\
0.15	0\\
0.158333333333333	0\\
0.166666666666667	0\\
0.175	0\\
0.183333333333333	0\\
0.191666666666667	0\\
0.2	0\\
0.208333333333333	0\\
0.216666666666667	0\\
0.225	0\\
0.233333333333333	0\\
0.241666666666667	0\\
0.25	0\\
0.258333333333333	0\\
0.266666666666667	0\\
0.275	0\\
0.283333333333333	0\\
0.291666666666667	0\\
0.3	0\\
0.308333333333333	0\\
0.316666666666667	0\\
0.325	0\\
0.333333333333333	0\\
0.341666666666667	0\\
0.35	0\\
0.358333333333333	0\\
0.366666666666667	0\\
0.375	0\\
0.383333333333333	0\\
0.391666666666667	0\\
0.4	0\\
0.408333333333333	0\\
0.416666666666667	0\\
0.425	0\\
0.433333333333333	0\\
0.441666666666667	0\\
0.45	0\\
0.458333333333333	0\\
0.466666666666667	0\\
0.475	0\\
0.483333333333333	0\\
0.491666666666667	0\\
0.5	0\\
};
\addlegendentry{Optimized}

\addplot [color=black, dashed, line width=1.0pt]
  table[row sep=crcr]{%
-0.5	0.0001\\
0.5	0.0001\\
};
\addlegendentry{$\text{P}_{\text{IC}}\text{ limit}$}

\end{axis}
\end{tikzpicture}%
    \caption{\gls{heo} Scenario: \gls{ipc} profile computed with the cuboid approximation.}
    \label{fig:caseSerraSolution}
\end{figure}
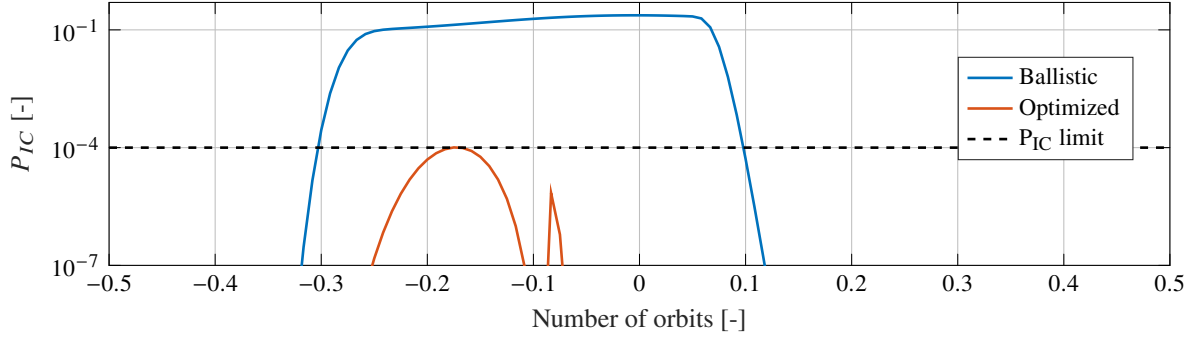

The last test case that we analyze is the \gls{heo} scenario that was proposed as test case 9 in reference \cite{Alfano2009Satellite}. To validate the method, the same dynamical model used in \cite{Serra2015}, so we employ purely Keplerian dynamics. We refer to the original article for the data of the conjunction. This test case is used to validate the method against the solution obtained by Serra \textit{et al.} \cite{Serra2015} who impose a \gls{ipc} threshold of $10^{-4}$. They obtain a four-impulses solution, with a total $\dv$ of $4.380$ \si{mm/s}. Our method recovers a solution that employs a single firing in the first node with \gls{rtn} components equal to $\Delta\vec{v} = [-0.250,\hspace{2pt} -0.128,\hspace{2pt} 0.018]\transp$  \si{mm/s}, and magnitude $\dv = 0.282$ \si{mm/s}. So, our method can obtain a single-firing solution which is twenty times lower than the reference. 
Note that to compute this maneuver, the \gls{smd} threshold method based on the cuboid \gls{ipc} approximation is used, because the constant \gls{ipc} approximation is not accurate, as shown in \cref{sec:validity}.
In \cref{fig:caseSerraSolution}, the evolution of the ballistic and optimized \gls{ipc} profile is shown: The optimized maneuver can  lower \gls{ipc} according to the requirement.

\subsection{Convergence and Analysis of the Solutions}

\begin{table}[tb!]
\centering
\caption{Convergence properties of the simulations.}
\label{tab:convergence}
    \begin{tabular}{llllll|llll}
    \Xhline{4\arrayrulewidth}
    sim n & $n_{orb}$ [-] & orbit & SK & Thrust & $\nabla\smd$ & $n_{maj}$ [-]&  $n_{min}$ [-] & $e$ [\si{mm}] & $\tau$ [\si{s}] \\
    \Xhline{3\arrayrulewidth}
    \textbf{1}   & 2 & \gls{leo} & No & High & No & $3$ & $11$ & $1.28$ & $15.09$ \\ \hline 
    \textbf{2}   & 2 & \gls{leo} & Yes & High & No & $3$ & $9$ & $0.31$ & $14.29$ \\ \hline 
    \textbf{3}   & 2 & \gls{leo} & No & Low & No & $3$ & $9$ & $17.54$ & $13.92$ \\ \hline 
    \textbf{4}   & 2 & \gls{leo} & Yes & Low & No & $3$ & $9$ & $1.426$ & $13.69$ \\ \hline 
    \textbf{5}   & 1 & \gls{leo} & No & High & No & $3$ & $8$ & $0.61$ & $7.09$ \\ \hline 
    \textbf{6}   & 1 & \gls{leo} & No & High & Yes & $4$ & $12$ & $11.88$ & $9.11$ \\ \hline 
    \textbf{7}   & 2 & \gls{geo} & No & High & No & $2$ & $6$ & $7.36$ & $2.84$ \\ \hline 
    \textbf{8}   & 2 & \gls{geo} & No & Low & No & $2$ & $5$ & $3.33$ & $2.59$ \\ \hline 
    \textbf{9}   & 2 & \gls{geo} & Yes & High & No & $2$ & $9$ & $2.66$ & $18.31$ \\ \hline 
    \textbf{10}   & 2 & \gls{geo} & Yes & Low & No & $3$ & $6$ & $56.43$ & $18.03$ \\ \hline 
    \textbf{11}   & 2 & \gls{geo} & Yes & High & Yes & $7$ & $40$ & $71.12$ & $23.6$ \\ \hline 
    \textbf{12}  & 1 & \gls{geo} & No & High & No & $2$ & $6$ & $0.129$ & $3.62$ \\ \hline 
    \textbf{13}  & 1 & \gls{geo} & No & High & Yes & $6$ & $51$ & $21.32$ & $31.07$ \\\hline
    \textbf{14}  &  1 &  \gls{heo} &   No &  Low &   No &  $3$ &  $5$ &  $0.168$ &  $2.3$ \\ 
    \Xhline{4\arrayrulewidth}   
    \end{tabular}
\end{table}

This section analyzes the convergence properties of the \gls{scp}.
In all the simulations presented in this work, the value of $\overline{\nu}$ was $10^{-4}$ for all scenarios. This value is low because we expect the \gls{cam} to deviate from the original orbit by a small amount.
For all the simulations, the tolerance for the convergence of the major iterations $tol_M = 10^{-3}$ and the one for the minor iterations is $tol_m = 10^{-6}$. The major iterations error is computed as the maximum difference between the control action of iteration $j$ and that of iteration $j-1$. The control action is always normalized with respect to the maximum control so that $0\leq||\vec{u}_i||\leq 1$. In this way, the entity of the error is independent of the maximum control and the major iteration tolerance does not need to be adjusted as a function of $u_{max}$. The minor iteration error is the maximum difference between the optimized relative position of two consecutive minor iterations.  A complementary condition to reach convergence is the minimization of the sum of the virtual controls below a threshold of $10^{-7}$, which is achieved in all simulations before the third major iteration.

In \cref{tab:convergence}, the convergence results for the test cases analyzed are reported. The number of major iterations used to linearize the dynamics is always kept below 7, showing that the executed maneuvers are typically small and the deviation from the ballistic trajectory is almost negligible on an orbital scale. The trend indicates that the number of iterations is proportional to the length of the propagation window. Lastly, in LEO, the use of the \gls{sk} targeting constraint improves the convergence because the first reference solution already satisfies the constraint, whereas the same cannot be observed for \gls{geo} cases. The execution time $\tau$ is proportional to the number of minor iterations. More than $89\%$ of the run time is required by the linearization of the dynamics and the building of the linear maps; on average, the solution of the convex problem only requires around $10\%$ of the total run time.  It is worth noticing that the accurate models greatly increase the computational time required by the propagation. For example, if the atmospheric model is not considered, the run time can be brought down to $2$ \si{s} for the simulations in \gls{leo}. Indeed, in the \gls{geo} simulations, where the atmosphere model is not needed, the run times are typically shorter. They only get longer when the computation of the optimal target state is performed using Problem \eqref{eq:opt}.  As a last remark, we note that the implementation of the code is not optimized for speed, and it is the authors' opinion that the time to find a solution could be reduced by one order of magnitude at least if a more suitable implementation was used.

\section{Conclusions}
\label{sec:conclusion}

A \acrfull{scp} was developed to design fuel-optimal collision avoidance maneuvers in long-term encounters. The original non-convex \acrlong{ocp} is locally approximated into a \acrfull{socp}, and the solution is found iteratively.
The collision risk is estimated either via the \acrfull{ipc}, the maximum instantaneous probability of collision, or the separation distance . This allows for using \acrfull{smd} to formulate the collision avoidance constraint as an ellipsoidal keep-out zone.  The dynamics are automatically linearized via \acrlong{da}, allowing for any dynamical model in the \gls{scp}, e.g., high-order gravitational harmonics, atmospheric drag, solar radiation pressure, and third body attraction.  The uncertainty of the system is assumed to evolve linearly, so it is propagated using the state transition matrix.  Moreover, the \acrfull{sk} operational requirement is introduced as a convex constraint, and a sensitivity constraint on the \gls{smd} is used to improve the robustness of the maneuver against modeling and actuation errors. A new trust region algorithm based on the nonlinearity index is introduced to avoid artificial unboundedness due to the linearization. 

The algorithm is tested on realistic scenarios in \gls{geo} and in \gls{leo}. The optimizer can recover an optimal solution in all the test cases considered without prior knowledge of the thrust arc structure or thrust direction. It is shown that the computed maneuver can be different when different collision metrics are selected. 
The algorithm's flexibility allows it to find solutions in high and low-thrust scenarios. The introduction of the novel \gls{smd} sensitivity constraint can modify the maneuver significantly, as it constraints the spacecraft on the surface of the keep-out zone where the gradient of \gls{smd} is lower. 
The proposed method is proven reliable and efficient, resulting in a promising step towards autonomous collision avoidance maneuver computation. 

 Our approach to \gls{cam} design is based on the assumption that the state uncertainties are Gaussian during the entire window of interest. Since the propagation windows are relatively short (e.g., one or two orbital periods) and the initial uncertainties are limited this assumption is not a strong one. If we were to consider longer propagation windows or larger uncertainties, nonlinear uncertainty propagation methods, like \glspl{gmm}, should be considered. Alternatively, different representations of the states, like generalized equinoctial elements, can be used to preserve the normality of the distribution. 

\section*{Funding Sources}
This material is based upon work supported by the Air Force Office of Scientific Research under award number FA2386-21-1-4115.

\bibliography{references}

\begin{thebibliography}{45}
\newcommand{\enquote}[1]{``#1''}
\providecommand{\natexlab}[1]{#1}
\providecommand{\url}[1]{\texttt{#1}}
\providecommand{\urlprefix}{URL }
\expandafter\ifx\csname urlstyle\endcsname\relax
  \providecommand{\doi}[1]{\discretionary{}{}{}https://doi.org/#1}\else
  \providecommand{\doi}[1]{\discretionary{}{}{}\urlstyle{rm}\url{https://doi.org/#1}}\fi

\bibitem[{ESOC(2023)}]{ESA2023}
ESOC, \enquote{{ESA's Annual Space Environment Report},} Tech. rep., ESA, 2023.

\bibitem[{Letizia et~al.(2015)Letizia, Colombo, and Lewis}]{Letizia2015}
Letizia, F., Colombo, C., and Lewis, H.~G., \enquote{{Collision Probability Due
  to Space Debris Clouds Through a Continuum Approach},} \emph{Journal of
  Guidance, Control, and Dynamics}, Vol.~39, No.~10, 2015, pp. 2240--2249.
\newblock \doi{10.2514/1.G001382}.

\bibitem[{Zhang et~al.(2022{\natexlab{a}})Zhang, Li, Wang, Zhang, and
  Wang}]{Zhang2022}
Zhang, H., Li, Z., Wang, W., Zhang, Y., and Wang, H., \enquote{{Geostationary
  Orbital Debris Collision Hazard after a Collision},} \emph{Aerospace},
  Vol.~9, No. 258, 2022{\natexlab{a}}.
\newblock \doi{10.3390/aerospace9050258},
  \urlprefix\url{https://doi.org/10.3390/aerospace9050258}.

\bibitem[{Zhang et~al.(2022{\natexlab{b}})Zhang, Li, Liu, and
  Sang}]{Zhang2022Analysis}
Zhang, Y., Li, B., Liu, H., and Sang, J., \enquote{{An analysis of close
  approaches and probability of collisions between LEO resident space objects
  and mega constellations},} \emph{Geo-spatial Information Science}, Vol.~25,
  No.~1, 2022{\natexlab{b}}, pp. 104--120.
\newblock \doi{10.1080/10095020.2022.2031313}.

\bibitem[{Browns(2010)}]{Browns2010}
Browns, A.~C., \enquote{{Human Spaceflight Recent Conjunctions of Interest
  Human Spaceflight Screening and Notification},} \emph{Proceedings of the
  USSTRATCOM Conjunction Summary Message Workshop}, 2010, p.~3.

\bibitem[{Hernando-Ayuso and Bombardelli(2021)}]{Hernando-Ayuso2020}
Hernando-Ayuso, J., and Bombardelli, C., \enquote{{Low-thrust collision
  avoidance in circular orbits},} \emph{Journal of Guidance, Control, and
  Dynamics}, Vol.~44, No.~5, 2021, pp. 983--995.
\newblock \doi{10.2514/1.G005547}.

\bibitem[{Armellin(2021)}]{Armellin2021}
Armellin, R., \enquote{{Collision avoidance maneuver optimization with a
  multiple-impulse convex formulation},} \emph{Acta Astronautica}, Vol. 186,
  2021, pp. 347--362.
\newblock \doi{10.1016/j.actaastro.2021.05.046}.

\bibitem[{{De Vittori} et~al.(2022){De Vittori}, Palermo, {Di Lizia}, and
  Armellin}]{DeVittori2022}
{De Vittori}, A., Palermo, M.~F., {Di Lizia}, P., and Armellin, R.,
  \enquote{{Low-Thrust Collision Avoidance Maneuver Optimization},}
  \emph{Journal of Guidance, Control, and Dynamics}, Vol.~45, No.~10, 2022, pp.
  1815--1829.
\newblock \doi{10.2514/1.G006630}.

\bibitem[{{N{\'{u}}{\~{n}}ez Garz{\'{o}}n} and
  Lightsey(2022)}]{NunezGarzon2022}
{N{\'{u}}{\~{n}}ez Garz{\'{o}}n}, U.~E., and Lightsey, E.~G.,
  \enquote{{Relating Collision Probability and Separation Indicators in
  Spacecraft Formation Collision Risk Analysis},} \emph{Journal of Guidance,
  Control, and Dynamics}, Vol.~45, No.~3, 2022, pp. 517--532.
\newblock \doi{10.2514/1.G005744}.

\bibitem[{Patera(2002)}]{Patera2003Satellite}
Patera, R., \enquote{{Satellite Collision Probability for Non-Linear Relative
  Motion},} \emph{AIAA/AAS Astrodynamics Specialist Conference and Exhibit},
  Vol.~26, American Institute of Aeronautics and Astronautics, Reston,
  Virigina, 2002.
\newblock \doi{10.2514/6.2002-4632}.

\bibitem[{Coppola(2012)}]{Coppola2012}
Coppola, V.~T., \enquote{{Including velocity uncertainty in the probability of
  collision between space objects},} \emph{AIAA/AAS Astrodynamics Specialist
  Conference 2014}, Vol. 143, San Diego, California, 2012, pp. 2159--2178.

\bibitem[{Alfano(2006)}]{AlfanoSpace}
Alfano, S., \enquote{{Addressing Nonlinear Relative Motion For Spacecraft
  Collision Probability},} \emph{AIAA/AAS Astrodynamics Specialist Conference
  and Exhibit}, American Institute of Aeronautics and Astronautics, Reston,
  Virigina, 2006, pp. 1--10.
\newblock \doi{10.2514/6.2006-6760}.

\bibitem[{Alfano(2014)}]{Alfano2014}
Alfano, S., \enquote{{Eliminating Assumptions Regarding Satellite Conjunction
  Analysis},} \emph{The Journal of the Astronautical Sciences}, Vol.~59, 2014,
  pp. 676--705.
\newblock \doi{10.1007/s40295-014-0002-4}.

\bibitem[{Xu and Xiong(2011)}]{Xu2011Research}
Xu, X.-l., and Xiong, Y.-q., \enquote{{A Research on the Collision Probability
  Calculation of Space Debris for Nonlinear Relative Motions†},}
  \emph{Chinese Astronomy and Astrophysics}, Vol.~35, No.~3, 2011, pp.
  304--317.
\newblock \doi{10.1016/j.chinastron.2011.07.008}.

\bibitem[{Wen and Qiao(2022)}]{Wen2022}
Wen, C., and Qiao, D., \enquote{{Calculating collision probability for
  long-term satellite encounters through the reachable domain method},}
  \emph{Astrodynamics}, Vol.~6, No.~2, 2022, pp. 141--159.
\newblock \doi{10.1007/s42064-021-0119-8}.

\bibitem[{Chan(2008)}]{chan2008Spacecraft}
Chan, K., \emph{{Spacecraft Collision Probability}}, The Aerospace Press, El
  Segundo, USA, 2008.

\bibitem[{Jones and Doostan(2013)}]{Jones2013Satellite}
Jones, B.~A., and Doostan, A., \enquote{{Satellite collision probability
  estimation using polynomial chaos expansions},} \emph{Advances in Space
  Research}, Vol.~52, No.~11, 2013, pp. 1860--1875.
\newblock \doi{10.1016/j.asr.2013.08.027},
  \urlprefix\url{http://dx.doi.org/10.1016/j.asr.2013.08.027}.

\bibitem[{Adurthi and Singla(2015)}]{Adurthi2015Conjugate}
Adurthi, N., and Singla, P., \enquote{{Conjugate unscented transformation-based
  approach for accurate conjunction analysis},} \emph{Journal of Guidance,
  Control, and Dynamics}, Vol.~38, No.~9, 2015, pp. 1642--1658.
\newblock \doi{10.2514/1.G001027}.

\bibitem[{Zhang et~al.(2020)Zhang, Fu, Chen, and Cao}]{Zhang2020}
Zhang, S., Fu, T., Chen, D., and Cao, H., \enquote{{Satellite instantaneous
  collision probability computation using equivalent volume cuboids},}
  \emph{Journal of Guidance, Control, and Dynamics}, Vol.~43, No.~9, 2020, pp.
  1757--1763.
\newblock \doi{10.2514/1.G004711}.

\bibitem[{Mueller(2009)}]{Mueller2009}
Mueller, J., \enquote{{Onboard Planning of Collision Avoidance Maneuvers Using
  Robust Optimization},} \emph{AIAA Infotech@Aerospace Conference}, American
  Institute of Aeronautics and Astronautics, Reston, Virigina, 2009.
\newblock \doi{10.2514/6.2009-2051},
  \urlprefix\url{https://arc.aiaa.org/doi/10.2514/6.2009-2051}.

\bibitem[{Serra et~al.(2015)Serra, Arzelier, Joldes, and
  Rondepierre}]{Serra2015}
Serra, R., Arzelier, D., Joldes, M., and Rondepierre, A.,
  \enquote{{Probabilistic Collision Avoidance for Long-term Space Encounters
  via Risk Selection},} \emph{Advances in Aerospace Guidance, Navigation and
  Control}, Springer International Publishing, 2015, pp. 679--698.
\newblock \doi{10.1007/978-3-319-17518-8_39}.

\bibitem[{Liu et~al.(2017)Liu, Lu, and Pan}]{Liu2017}
Liu, X., Lu, P., and Pan, B., \enquote{{Survey of convex optimization for
  aerospace applications},} \emph{Astrodynamics}, Vol.~1, No.~1, 2017, pp.
  23--40.
\newblock \doi{10.1007/s42064-017-0003-8}.

\bibitem[{Dutta and Misra(2022)}]{Dutta2022}
Dutta, S., and Misra, A.~K., \enquote{{Convex optimization of collision
  avoidance maneuvers in the presence of uncertainty},} \emph{Acta
  Astronautica}, Vol. 197, 2022, pp. 257--268.
\newblock \doi{10.1016/j.actaastro.2022.05.038}.

\bibitem[{Pinson and Lu(2018)}]{Pinson2018}
Pinson, R., and Lu, P., \enquote{{Trajectory design employing convex
  optimization for landing on irregularly shaped asteroids},} \emph{Journal of
  Guidance, Control, and Dynamics}, Vol.~41, No.~6, 2018, pp. 1243--1256.
\newblock \doi{10.2514/1.G003045}.

\bibitem[{Alonso-Mora et~al.(2019)Alonso-Mora, Montijano, N{\"{a}}geli,
  Hilliges, Schwager, and Rus}]{Alonso-Mora2019}
Alonso-Mora, J., Montijano, E., N{\"{a}}geli, T., Hilliges, O., Schwager, M.,
  and Rus, D., \enquote{{Distributed multi-robot formation control in dynamic
  environments},} \emph{Autonomous Robots}, Vol.~43, No.~5, 2019, pp.
  1079--1100.
\newblock \doi{10.1007/s10514-018-9783-9}.

\bibitem[{Pirovano and Armellin(2024)}]{Pirovano2024}
Pirovano, L., and Armellin, R., \enquote{{Detection and estimation of
  spacecraft maneuvers for catalog maintenance},} \emph{Acta Astronautica},
  Vol. 215, No. December 2023, 2024, pp. 387--397.
\newblock \doi{10.1016/j.actaastro.2023.12.016}.

\bibitem[{Boyd and Vandenberghe(2004)}]{Boyd2}
Boyd, S., and Vandenberghe, L., \emph{{Convex Optimization}}, Cambridge
  University Press, Cambridge, NY, 2004.

\bibitem[{Lew et~al.(2020)Lew, Bonalli, and Pavone}]{Lew2020}
Lew, T., Bonalli, R., and Pavone, M., \enquote{{Chance-Constrained Sequential
  Convex Programming for Robust Trajectory Optimization},} \emph{2020 European
  Control Conference (ECC)}, IEEE, 2020, pp. 1871--1878.
\newblock \doi{10.23919/ECC51009.2020.9143595}.

\bibitem[{Ridderhof et~al.(2020)Ridderhof, Pilipovsky, and
  Tsiotras}]{Ridderhof2021}
Ridderhof, J., Pilipovsky, J., and Tsiotras, P., \enquote{{Chance-Constrained
  Covariance Control for Low-Thrust Minimum-Fuel Trajectory Optimization},}
  \emph{2020 AAS/AIAA Astrodynamics Specialist Conference}, South Lake Tahoe,
  CA, 2020, pp. 9--13.

\bibitem[{Benedikter et~al.(2022{\natexlab{a}})Benedikter, Zavoli, Wang,
  Pizzurro, and Cavallini}]{Benedikter2022Covariance}
Benedikter, B., Zavoli, A., Wang, Z., Pizzurro, S., and Cavallini, E.,
  \enquote{{Covariance Control for Stochastic Low-Thrust Trajectory
  Optimization},} \emph{AIAA SciTech 2022 Forum}, American Institute of
  Aeronautics and Astronautics, Reston, Virginia, 2022{\natexlab{a}}.
\newblock \doi{10.2514/6.2022-2474}.

\bibitem[{Benedikter et~al.(2022{\natexlab{b}})Benedikter, Zavoli, Wang,
  Pizzurro, and Cavallini}]{Benedikter2022Convex}
Benedikter, B., Zavoli, A., Wang, Z., Pizzurro, S., and Cavallini, E.,
  \enquote{{Convex Approach to Covariance Control with Application to
  Stochastic Low-Thrust Trajectory Optimization},} \emph{Journal of Guidance,
  Control, and Dynamics}, Vol.~45, No.~11, 2022{\natexlab{b}}, pp. 2061--2075.
\newblock \doi{10.2514/1.G006806}.

\bibitem[{Mueller et~al.(2013)Mueller, Griesemer, and Thomas}]{Mueller2013}
Mueller, J.~B., Griesemer, P.~R., and Thomas, S.~J., \enquote{{Avoidance
  maneuver planning incorporating station-keeping constraints and automatic
  relaxation},} \emph{Journal of Aerospace Information Systems}, Vol.~10,
  No.~6, 2013, pp. 306--322.
\newblock \doi{10.2514/1.54971}.

\bibitem[{Malyuta et~al.(2022)Malyuta, Reynolds, Szmuk, Lew, Bonalli, Pavone,
  and A{\c{c}}ıkmeşe}]{Malyuta2021Tutorial}
Malyuta, D., Reynolds, T.~P., Szmuk, M., Lew, T., Bonalli, R., Pavone, M., and
  A{\c{c}}ıkmeşe, B., \enquote{{Convex Optimization for Trajectory
  Generation: A Tutorial on Generating Dynamically Feasible Trajectories
  Reliably and Efficiently},} \emph{IEEE Control Systems}, Vol.~42, No.~5,
  2022, pp. 40--113.
\newblock \doi{10.1109/MCS.2022.3187542},
  \urlprefix\url{https://ieeexplore.ieee.org/document/9905530/}.

\bibitem[{Losacco et~al.(2024)Losacco, Foss{\`{a}}, and Armellin}]{Losacco2023}
Losacco, M., Foss{\`{a}}, A., and Armellin, R., \enquote{{Low-Order Automatic
  Domain Splitting Approach for Nonlinear Uncertainty Mapping},} \emph{Journal
  of Guidance, Control, and Dynamics}, 2024, pp. 1--20.
\newblock \doi{10.2514/1.G007271}.

\bibitem[{Bernardini et~al.(2023)Bernardini, Wijayatunga, Baresi, and
  Armellin}]{Bernardini2023}
Bernardini, N., Wijayatunga, M.~C., Baresi, N., and Armellin, R.,
  \enquote{{State-Dependent Trust Region for Successive Convex Optimization of
  Spacecraft Trajectories},} \emph{AAS/AIAA Space Flight Mechanics Meeting},
  Austin, TX, 2023, pp. 1--20.

\bibitem[{Morselli et~al.(2014)Morselli, Armellin, {Di Lizia}, and {Bernelli
  Zazzera}}]{Morselli2014High}
Morselli, A., Armellin, R., {Di Lizia}, P., and {Bernelli Zazzera}, F.,
  \enquote{{A high order method for orbital conjunctions analysis: Sensitivity
  to initial uncertainties},} \emph{Advances in Space Research}, Vol.~53,
  No.~3, 2014, pp. 490--508.
\newblock \doi{10.1016/j.asr.2013.11.038}.

\bibitem[{Ba{\`{u}} et~al.(2021)Ba{\`{u}}, Hernando-Ayuso, and
  Bombardelli}]{Bau2021}
Ba{\`{u}}, G., Hernando-Ayuso, J., and Bombardelli, C., \enquote{{A
  generalization of the equinoctial orbital elements},} \emph{Celestial
  Mechanics and Dynamical Astronomy}, Vol. 133, No. 11-12, 2021.
\newblock \doi{10.1007/s10569-021-10049-1}.

\bibitem[{Vittaldev and Russell(2016)}]{Vittaldev2016Space}
Vittaldev, V., and Russell, R.~P., \enquote{{Space object collision probability
  using multidirectional Gaussian mixture models},} \emph{Journal of Guidance,
  Control, and Dynamics}, Vol.~39, No.~9, 2016, pp. 2161--2167.
\newblock \doi{10.2514/1.G001610}.

\bibitem[{Armellin et~al.(2010)Armellin, {Di Lizia}, Bernelli-Zazzera, and
  Berz}]{Armellin2010}
Armellin, R., {Di Lizia}, P., Bernelli-Zazzera, F., and Berz, M.,
  \enquote{{Asteroid close encounters characterization using differential
  algebra: The case of Apophis},} \emph{Celestial Mechanics and Dynamical
  Astronomy}, Vol. 107, No.~4, 2010, pp. 451--470.
\newblock \doi{10.1007/s10569-010-9283-5}.

\bibitem[{Alfriend et~al.(1999)Alfriend, Akella, Frisbee, Foster, Lee, and
  Wilkins}]{Alfriend2000}
Alfriend, K.~T., Akella, M.~R., Frisbee, J., Foster, J.~L., Lee, D.-J., and
  Wilkins, M., \enquote{{Probability of collision error analysis},} \emph{Space
  Debris}, 1999.
\newblock \doi{https://doi.org/10.1023/A:1010056509803}.

\bibitem[{Alfano(2009)}]{Alfano2009Satellite}
Alfano, S., \enquote{{Satellite conjunction Monte Carlo analysis},} \emph{19th
  AAS/AIAA Space Flight Mechanics Meeting}, Vol. 134, American Astronautical
  Society, Savannah, Georgia, 2009, pp. 2007--2024.

\bibitem[{Wang and Grant(2018)}]{Wang2018}
Wang, Z., and Grant, M.~J., \enquote{{Minimum-Fuel Low-Thrust Transfers for
  Spacecraft: A Convex Approach},} \emph{IEEE Transactions on Aerospace and
  Electronic Systems}, Vol.~54, No.~5, 2018, pp. 2274--2290.
\newblock \doi{10.1109/TAES.2018.2812558}.

\bibitem[{Mao and Acikmese(2021)}]{Mao2021}
Mao, Y., and Acikmese, B., \enquote{{SCvx-fast: A Superlinearly Convergent
  Algorithm for A Class of Non-Convex Optimal Control Problems},} \emph{ArXiv},
  Vol. 2112, 2021.
\newblock \urlprefix\url{http://arxiv.org/abs/2112.00108}.

\bibitem[{Bombardelli and Hernando-Ayuso(2015)}]{Bombardelli2015}
Bombardelli, C., and Hernando-Ayuso, J., \enquote{{Optimal impulsive collision
  avoidance in low earth orbit},} \emph{Journal of Guidance, Control, and
  Dynamics}, Vol.~38, AIAA International, 2015, pp. 217--225.
\newblock \doi{10.2514/1.G000742}.

\bibitem[{Pavanello et~al.(2023)Pavanello, Pirovano, and
  Armellin}]{Pavanello2023Long}
Pavanello, Z., Pirovano, L., and Armellin, R., \enquote{{Long-Term Encounters
  Collision Avoidance Maneuver Optimization with a Convex Formulation},}
  \emph{AAS/AIAA Space Flight Mechanics Meeting}, Austin, TX, 2023, pp. 1--20.

\end{thebibliography}
\pagebreak
\appendix
\section{High level flow of the \gls{scp}}
In \cref{alg:scp} the algorithm of the \gls{scp} is shown.
\begin{algorithm}
\label{alg:scp}
\begin{algorithmic}[1]
\State {Get inputs for the spacecraft: $HBR$, $A_{drag}$, $C_D$, $A_{SRP}$, $C_r$ $\vec{x}_{0}$, $\vec{C}_0$} 
\State {Assign $t_0$, $\Delta t$, $N, u_{max}$, $u_{min}$, $\ipclim$ (or $\ipcmaxlim$ or $\overline{d}_{miss})$, $\rho$, $\varepsilon$, $\overline{\nu}$, $\mathrm{tol}_M$, $\mathrm{tol}_m$, $j_{max}$, $k_{max}$}
\State $t \gets t_0 : \Delta t : t_0 + N\Delta t$
\State {Perform a first-order \gls{da} propagation of the secondary trajectory starting from $\Vec{x}_{s,0}$.}
\For {$i = 1:N$}
\State{$\overline{\vec{x}}_{s,i} \gets $ constant part of the propagation}
 
\State{$\vec{A}_{s,i} \gets$ linear part of the propagation}
\State {$\vec{C}_{s,i} \gets$ \cref{eq:covSec}}
\State {$\vec{P}_{s,i} \gets$ \cref{eq:covTransfSec}}
\EndFor
\State {$j \gets 0$}
\While {$||\mathbb{u}^j-\mathbb{u}^{j-1}||_\infty > \mathrm{tol}_M \land j < j_{max}$}
    \State {$j \gets j + 1$}
    \State {$k \gets 0$}
    \If{$j>1$} 
        \State {Perform a second-order \gls{da} propagation with control history  $\mathbb{u}^{j-1}$ and expansion points from $\mathbb{x}^{j-1}$.}
    \Else
        \State {Perform a second-order forward \gls{da} propagation with no control.}
    \EndIf
    \For {$i = 0:N$}
        \State{$\overline{\vec{x}}_i^j,\overline{\vec{\phi}}_i^j \gets $ constant part of the propagation geodetic coordinates}
        \State{$\vec{A}_i^j,\vec{B}_i^j,\vec{G}_i^j \gets$ linear part of the propagation and geodetic coordinates transformation}
        \State {$\vec{C}_{p,i}^j \gets$ \cref{eq:cov}}
        \State {$\vec{P}_{p,i}^j \gets$ \cref{eq:covTransf}} 
        \State $\vec{P}_i^j \gets$ \cref{eq:posCovSum}
        \State $\vec{r}_i^j \gets \vec{r}_{p,i}^j - \vec{r}_{s,i}$
        \State {$(\smdlim)_i^j \gets$ \cref{eq:dmlim1}, \cref{eq:dmlim2}, or Problem \eqref{eq:cuboidinversion}}  
        \State {$(\smd)_i^j \gets$  \cref{eq:smd}}
        \State {$(\ipc)_i^j \gets$  \cref{eq:constipc}}
        \State {$\xi_i \gets$  \cref{eq:nli}}
        \If {$(P_{IC,i}^j>(1-\varepsilon)\ipclim$}
            \State{$\gamma_i \gets$ \cref{eq:smdGradLim}}
        \EndIf
    \EndFor
    \For{$i=1:N \land (\smd)_i^j < (\smdlim)_i^j$}
        \State {Find starting point on the ellipsoid's surface using \cref{eq:startpoint}}
    \EndFor
    \While{$(||\mathbb{r}^{j,k}-\mathbb{r}^{j,k-1}||_\infty > \mathrm{tol}_m \land k < k_{max})$}
        \State{$k \gets k+1$}
        \State{Solve \acrshort{socp} Problem \eqref{eq:optFinal}}
    \EndWhile
    \State{$\mathbb{x}^j \gets \mathbb{x}^{j,k}$}
    \State{$\mathbb{u}^j \gets \mathbb{u}^{j,k}$}
\EndWhile
\State {Validation: Propagate forward from the initial state using the control history from the last iteration and check the highest position error with respect to the last major iteration.}
\end{algorithmic}
\end{algorithm}

\end{document}